\documentclass[11pt]{article}
\renewcommand{\theequation}{\arabic{section}.\arabic{equation}}  

\newcommand{\beq}{\begin{equation}}   
\newcommand{\eeq}{\end{equation}}    
\newcommand{\bea}{\begin{eqnarray}}   
\newcommand{\eea}{\end{eqnarray}}    
\newcommand{\ber}{\begin{array}}      \newcommand{\eer}{\end{array}}    
  
\newcommand{\dms}{\displaystyle}      \newcommand{\fp}{\Phi\Pi}  
\newcommand{\dx}{d^{4}x}              \newcommand{\de}{\partial}

\newcommand{\alp}{\alpha}                 
\newcommand{\nabh}{\hat{\nabla}}          
\newcommand{\hg}{\hat{\gamma}}            
              \newcommand{\Om}{\Omega}    
        \newcommand{\gam}{\gamma}      
\newcommand{\bet}{\beta}              \newcommand{\del}{\delta}

\def\ie{{\it i.e.}}                    
\def\h#1{\hat{#1}}                    \def\b#1{\overline{#1}}    
                      
\def\oop#1{\vspace{#1}}                 
               
\def\bc{\bar{c}}  
  
\newcommand{\g}{{  \Gamma}}          
   \newcommand{\gh}{\hat{\g}}

\def\LR{({\rm L}\rightarrow{\rm R})}  
  
\newcommand{\smallz}{{\scriptscriptstyle Z}} 
\newcommand{\smalls}{{\scriptscriptstyle S}} 
\newcommand{\smalla}{{\scriptscriptstyle A}} 
\newcommand{\smallw}{{\scriptscriptstyle W}}

\def\dfa#1#2#3{\frac{\delta}{\delta {#1}^{#2}_{#3}}}    
\def\dfu#1#2{\frac{\delta}{\delta {#1}^{#2}}}     
\def\dfd#1#2{\frac{\delta}{\delta {#1}_{#2}}}

\def\fd#1#2#3#4{\dms{\frac{\delta #1}{\delta {#2}^{#3}_{#4}}}}    
\def\fdu#1#2#3{\dms{\frac{\delta #1}{\delta {#2}^{#3}}}}    
\def\fdd#1#2#3{\dms{\frac{\delta #1}{\delta {#2}_{#3}}}}

\def\fdL#1#2#3#4{\dms{\frac{\stackrel{\rightharpoonup}{\delta}\! #1}{\delta {#2}^{#3}_{#4}}}}    
\def\fdR#1#2#3#4{\dms{\frac{#1 \stackrel{\leftharpoonup}{\delta}}{\delta {#2}^{#3}_{#4}}}}    
\def\dfaL#1#2#3{\dms{\frac{\stackrel{\rightharpoonup}{\delta}\! }{\delta {#1}^{#2}_{#3}}}}    
\def\dfaR#1#2#3{\dms{\frac{ \!\stackrel{\leftharpoonup}{\delta}}{\delta {#1}^{#2}_{#3}}}}

\def\fddL#1#2#3{\dms{\frac{\stackrel{\rightharpoonup}{\delta}\! #1}{\delta {#2}_{#3}}}}    
\def\fduR#1#2#3{\dms{\frac{#1 \!\stackrel{\leftharpoonup}{\delta}}{\delta {#2}^{#3}}}}

\def\fdaL#1#2#3#4{\dms{\frac{\stackrel{\rightharpoonup}{\delta}\! #1}{\delta {#2}^{#3}_{#4}}}}    
\def\fdaR#1#2#3#4{\dms{\frac{#1 \!\stackrel{\leftharpoonup}{\delta}}{\delta {#2}^{#3}_{#4}}}}

\begin{document}     
\begin{titlepage}  
\begin{flushright}    
MPI/PhT-98-57 \\  
hep-th/9908188 \\  

\end{flushright}

\vspace{1cm}  
\begin{center}   
  {\LARGE Renormalization of Non-Semisimple Gauge   
    Models \vspace{.3cm} \\ with the Background Field Method}  
   
  \vspace{1cm}  
  
  {\large \bf {Pietro Antonio  
      Grassi\footnote{e-mail:pgrassi@mppmu.mpg.de}}}  
  
  {\it Max-Planck-Institut f\"ur Physik, (Werner-Heisenberg-Institut) \\  
        F\"ohringer Ring 6, 80805, Munich, Germany}  
\end{center}

\vspace{1cm}  
\begin{center}  
  {\bf  Abstract}  
\vspace{.3 cm}  
  
\begin{minipage}{15cm}  
{\small We study the renormalization of non-semisimple gauge models   
quantized in the `t Hooft-background gauge to all orders.    
 We analyze the normalization conditions for masses and couplings compatible    
with the  Slavnov-Taylor and Ward-Takahashi Identities and   
with the IR constraints.   We take into account both the problem   
of renormalization of CKM matrix elements and the problem of CP   
violation and we show that the Background Field Method (BFM) provides proper
normalization conditions for fermion, scalar and  gauge field mixings. 
We discuss the hard and the soft anomalies of the 
Slavnov-Taylor Identities and the conditions under which they are absent.}  
\end{minipage}  
\end{center}  
\end{titlepage}  
  
\vspace{1 cm}  
\section{Introduction}  
\label{sec:intro}  
  
  
The Standard Model (SM) \cite{SM} is the widely accepted Quantum Field Theory (QFT)  
describing the physics of elementary particles up to the TeV scale.  
As any interacting, local QFT in four space-time dimensions it has a  
singular high energy limit leading to UV divergences in perturbative  
calculations. The removal of these divergences through the process of  
renormalization can be achieved in various ways and hence it is important to  
keep control of possible inconsistencies and ambiguities accompanying  
such procedures. The most satisfactory way to do so, is to put up a system of  
``axioms'' of renormalization theory, with the aim to characterize the  
result through physical requirements. This approach to the problem,  
motivated by development of axiomatic QFT, led to a rather satisfactory  
framework \cite{bogo} culminating in the work of Epstein  
and Glaser \cite{Epstein}. However, this unfortunately does not exhaust the 
problem of renormalization for the SM, since it involves the introduction of various  
``unphysical'' fields required for the explicitly local formulation of gauge  
theories, whose decoupling in physical processes has to be guaranteed for the  
renormalized theory. Also this problem found a satisfactory solution in the  
form of certain symmetry relations \cite{brs}, the  
Slavnov-Taylor Identities (STI) expressing the BRST invariance of the theory.  
In order to construct a renormalized perturbative series satisfying these  
identities the most natural strategy is to find an explicit BRST invariant  
regularization and subtraction procedure. Despite many attempts in the past  
no really satisfactory such method is known to the author.  
Hence it offers itself to make recourse again to the ``axiomatic'' approach   
and try to construct a BRST invariant renormalized perturbation series making   
use of the full freedom allowed by the axioms of renormalization theory. This   
strategy, known as ``algebraic'' renormalization (since the result is  
characterized by the symmetry relations pertaining to some Lie algebra),  
advocated mainly by the ``BRS school'' (\cite{brs}, \cite{bec}   
\cite{henri}, \cite{libro}) will be  
adopted in the present work. Although this strategy may seem unduly  
complicated for practical purposes, it is the only one surpassing unresolved  
consistency problems of all the hitherto proposed ``invariant''  
regularizations. 
In fact, combined with efficient  strategies to determine the required non-invariant 
counter terms, it may  
turn out that ``algebraic'' renormalization is a liable method for the SM or  
even its generalizations like its supersymmetric extension.  
  
As a first case, the ``algebraic renormalization'' \cite{zimm} has been  
applied to the renormalization of the Abelian Higgs-Kibble model.   
This models describes some aspects of the SM, namely the spontaneous   
symmetry breaking,  and it   
was the first example of gauge theory handled with the BRST   
formalism \cite{brs}.   
  
The essential results of those papers   
were to establish a formalism independent of the   
regularization and based on the locality and Lorentz invariance to   
proof the renormalizability of the gauge theories and the unitarity of  
the S-matrix. Later, the formalism was generalized to   
non-abelian gauge group in \cite{brs,henri,bbbc_1} and \cite{henn}.    
However, up to now, although the main problems of the renormalization 
of the SM \cite{aoki,Baulieu:1982ux,hollik_2,jegherlener,langacker,krau_ew} 
are solved, some peculiar feature of physical interesting models are  
not yet taken into account, in particular we want to recall  
\begin{enumerate}  
\item    
The IR problems of off-shell Green's functions. They have been recently 
examined in \cite{krau_ew}.    
\item   
The CP violation and the renormalization of the mixing. Although these are well known   
from phenomenological point of view,   
a complete (algebraic renormalization) analysis is still missing.  
\item  
Unstable particles.  
\end{enumerate}  
    
In this paper we use the word {\it minimal}~ in  
order to distinguish the field content of the SM from that of its 
extensions like the Two Higgs Doublets  
Model (2HDM) \cite{2HD}, 
the Minimal Supersymmetric Standard Model (MSSM) \cite{MSSM},
the Grand Unified Theory (GUT) $SU(5)$ models   
\cite{geo-gla} and their supersymmetric versions.   
  
However, in order to easily generalize our results for the SM,    
we shall use a general framework as long as  
possible and we shall specify the equations when the features of a particular  
model are considered. We, therefore, examine     
a non-semisimple gauge model with spontaneous symmetry breaking coupled to   
scalar fields and to chiral fermion fields. This comprises field mixings   
such as the fermion mixing, scalar mixing and the system as the photon and the  
$Z^{0}$ vector boson; furthermore we do not require the $P$ and $C$  
discrete symmetries, but only the hermiticity, locality and covariance   
which imply the $CPT$ invariance. In this general framework     
the mass eigenstates do not coincide with the gauge eigenstates and    
the mass matrices are only semi-definite positive.   
  
The invariance under the local gauge transformations   
essentially guarantees the consistency between a covariant  
formalism for relativistic   
quantum fields and the physical degrees of freedom of vector particles. 
However for quantization purposes, it is necessary     
to choose a gauge which breaks this invariance and introduces   
new unphysical ghost fields. The gauge fixing procedure is   
conveniently   
realized {\it \'a la} BRST \cite{brs} requiring   
the invariance of the theory under the BRST transformations \cite{bec}.   
At the quantum level this invariance ensures the decoupling of unphysical   
states for the physical Fock space, guaranteeing the unitarity   
of the scattering operator and the gauge independence   
of the physical observables (see \cite{brs,bec}).   
  
Regarding the choice of gauge fixing the   
background-`t Hooft type \cite{krau_ew,bkg,msbkg,pg_1,pg_2}
is found to be particularly    
convenient for the SM. In that case, in fact, the $S-matrix$ computed whit the BFM   
is completely defined in terms of (background) gauge invariant Green's functions \cite{ags}.  
Moreover the background gauge invariance, implemented at the quantum level 
by means of Ward-Takahashi Identities (WTI), provides very useful   
constraints on the possible counterterms (CP violating   
counterterms, renormalization of the mixings \cite{qmr,gg_1} and   
wave function renormalizations (w.f.r.)).

As is well known the quantization of gauge models 
can be consistently performed if the tree level BRST symmetry can be implemented at the 
quantum level and it is not anomalous.  
In a general case two kinds of anomaly candidates could be present: 
the {\it hard anomalies} characterized by local operators with the 
highest dimension and the {\it soft anomalies} 
\cite{henri,bbbc_2} with operators of lower dimension.  
In the case of the Adler-Bardeen-Jackiw anomaly  
\cite{abj} --a candidate for hard anomalies-- 
the non-renormalization theorem \cite{bbbc_1} and the   
fermion content of the SM ensures the vanishing of its   
coefficients. The other hard candidates are ruled out by introducing   
new constraints as  will be discussed later and,    
in absence of hard anomalies, the Callan-Symanzik equation guarantees   
the vanishing of soft ones \cite{henri,krau_ew,bbbc_2}. However 
some of the candidates for the  {soft anomalies}  can also be IR dangerous 
({\it IR anomalies}), 
{\it i.e.} the counterterms which can remove them can produce IR divergence. 
This means that their coefficient must vanish and we will check this 
explicitly in the Sec.~\ref{IRanomalies}. It turns out that as a consequence of  
normalization conditions and of the invariances of the model the IR anomalies 
can be excluded in the case of the SM and for extended models.

Besides IR problems the normalization conditions play a prominent r\^ole 
in defining the quantum theory and require a separate discussion. 
Our analysis is divided into three steps. 

First, we solve the symmetry constraints in order to 
single out the free parameters of the models without taking into account 
the IR constraints. This allows us to define the space of free parameters 
independently of the basis chosen for the fields ({\it e.g.} it appears convenient to 
work in the basis of gauge eigenstates because, in that case, the result is 
already available in the literature \cite{henri,henn}). 

Second, we establish a set of normalization conditions 
which is compatible with the symmetry constraints and which ensures the 
particle content of the model (definition of the physical states). Much emphasis is posed on the 
IR normalization conditions which are relevant in order to ensure the 
absence of IR anomalies. Concerning the normalization conditions, 
we study the problem of the renormalization of the mixing angles for fermions (CKM) and 
we propose a scheme to fix the mixing angles among scalars in the case of extended 
models (2HDM, MSSM etc.).
This is based on the study of the cohomology classes  with background fields   
and, subsequently, by imposing the background gauge invariance  
through the WTI.  The latter implements severe constraints on the renormalization   
of mixings. Third, we check the absence of IR anomalies.  

Regarding the normalization conditions and the 
choice of counterterms we have to take into account the CP violation. 
It arises only from the complex Yukawa couplings \cite{CP}, but    
new UV divergent CP-odd Green's functions --as for instance 
the mixed two-point function for the Higgs $H$ and the would-be-Goldstone 
boson $G_0$ \cite{CP-mixings}-- and the   
CP-odd tadpoles \cite{odd-tadpole}) appear at higher orders. 
These new divergences require  
new counterterms and new normalization conditions. In the paper   
we analyze the  CP-odd counterterms and  CP-odd anomalies. The number of   
CP-odd counterterms\footnote{Related to this  
point we have to recall that the algebraic proof of the  
renormalization of non-semisimple gauge model, given in \cite{henri},   
does not rely on any discrete symmetry (up to CPT). The authors, by  
introducing some supplementary  
constraints in order to avoid the CP-odd mixing with the abelian  
ghost field and the rest of the field content and to cancel the  
anomalies beyond the ABJ anomaly,  are able to prove the renormalizability  
of the theory only by means of BRST symmetry.} 
and their normalization conditions are 
considerable reduced   
by using the WTI for the background gauge invariance.  On the other side  
the WTI are  not enough to ensure the absence of CP-odd   
anomalies. The problem is solved by a further  
functional identity, the {\it Abelian Antighost Equation} (AAE) \cite{pg_3}, 
which guarantees the cancellation of the anomaly coefficients order-by-order.   
  
Concerning the system of scalar components of the neutral   
vector fields $Z^0, \gamma$ and the scalar fields $G^0, H$, the   
CP-violation makes the usual definition of the   
Higgs mass, the only physical parameter of the system, meaningless. 
If  the CP symmetry were conserved, (the real part of) the zero of the two-point   
function for the Higgs would correspond to its physical mass and   
the pole structure of the unphysical fields would be fixed by means of the STI.   
However, due to the   
CP violation, which generates a mixing between the Higgs and the  
Goldstone field $G^0$,  the mass of the Higgs can not be identified with 
the zero of the Higgs two-point functions. This must be replaced with 
non-trivial zero of the eigenvalues of the two-point   
function matrix for the system $Z-\gamma-H-G_0$.   
  
Concerning the BFM, we have to recall two relevant results.   
The first is the algebraic proof of the stability under  
radiative corrections  of the splitting of the gauge field into a quantum part and into a classical  
background \cite{pg_1}. The proof also includes the absence of  
anomalies beyond the conventional Adler-Bardeen-Jackiw anomaly  
\cite{abj}, excluded by a suitable content of fermions of  
the model.  The second fact is the extension of this result   
to scalar fields in order to generalize the   
`t\,\,Hooft gauge fixing to a background-`t Hooft gauge fixing.   
This problem was also covered in \cite{krau_ew}    
where the background scalar fields are introduced without imposing   
the (local) background gauge invariance.   
The present work follows the lines of \cite{pg_2,pg_3}   
where to each bosonic field corresponds a background partner   
and the background gauge invariance is implemented.   
  
Besides the Slavnov-Taylor Identities \cite{brs,bec},   
which implement the BRST symmetry at the quantum level,   
the Nakanishi-Lautrup equations \cite{kugo}, the equations of motion   
of the Faddeev-Popov ghost fields, the Abelian Antighost Equation \cite{pg_3} 
and the  Ward-Takahashi Identities for the background gauge invariance are very  
useful to renormalize the SM. However,    
the complexity of the problem requires a method to simplify   
the system of equations by   
reducing the problem to  a restricted functional space (the   
cohomology $H^k({\cal S}|d)$ \cite{henn}, in the space of   
background gauge invariant functionals). We propose the following
hierarchy: {\it i)} we first implement the linear identities, {\it ii)} we use  
the WTI for the background gauge invariance, and finally {\it iii)} 
we study the STI.  As a result we are able to enumerate the   
free parameters which must be tuned to renormalize the theory   
on a finite number of experimental data.   
  
By imposing the WTI, the  number of free parameters is reduced and, consequently,
a particular care has to be devoted  to check if the number of parameters 
is enough to ensure the correct IR  behavior of two- and three-point functions.   
In particular it can not be easily achieved in the   
renormalization of two-point functions for the ghost fields. In fact, in that 
sector,  the matrix of two-point functions is non symmetric    
and the number of the free parameters to fix their IR behavior turns out to be    
larger than the number of parameters needed for the system ($\gamma,Z^0$).   
  
Furthermore it is interesting to clarify the relations among   
the two-point functions with only external background fields and   
the two-point functions with quantum fields. The main point is to   
compare their pole structures and to show that the normalization  
conditions fix consistently both of them.   
  
A possible approach is to avoid finite renormalizations (except those
needed to exclude IR divergences) and to rely only on   
a subtraction scheme such as BPHZL or MS in  
dimensional regularization \cite{zimm,low,maiso}.  Although   
these schemes seem to be very convenient from a theoretical   
and computational point of view, some difficulties arise when the comparison with the  
physical data has to be performed.  In fact the MS or BPHZL parameters  
have to be expressed in terms of the physical input parameters and  
this implies that some algebraic equations are to be solved  
\cite{zimm,sirlin}.    
A different way of proceeding is to impose normalization conditions  
such that the correct parameterization is fixed from the beginning.   
An available scheme of this type is the on-shell scheme   
applied to the SM in \cite{aoki} and rigorously studied in  
\cite{krau_ew}.   
  
The advantages of the on-shell scheme are due to the fact that the  
``renormalized'' fields coincide with the asymptotic fields whose  
normalization conditions are fixed by the LSZ conditions  
\cite{itzy}. However this implies a large number of  
normalization conditions which appear  
unnecessary.  In addition, the success of the on-shell scheme is mostly due to the  
construction of physical amplitudes at the one loop order without the  
computation of LSZ conditions for external particles.    
One price to pay working with   
the on-shell scheme for the SM is the deformation of the STI and of the   
WTI \cite{krau_ew} needed to avoid the IR anomalies.   
One possible alternative consists in requiring a minimal set of normalization conditions   
for the physical parameters and the remaining parameters are  
fixed by a consistent subtraction scheme, like for MS.   

Our choice will be to keep the WTI and the STI in their tree level form. 
This reduces the space of free parameters and therefore we can only establish a 
{\it partially on-shell scheme} where  
a small set of normalization conditions are necessary for the  
physical computations. We divide the complete set into four   
different classes of normalizations:  
\begin{enumerate}  
\item At first in order to have spontaneous symmetry breaking one has to 
  renormalize the tadpoles. This is essential in  
  order to implement the Higgs mechanism at higher orders. Furthermore due to the CP  
  violation of the SM some CP-odd tadpoles could emerge  
  from radiative corrections and these have to be fixed to zero in order to  
  distinguish the physical Higgs boson from the unphysical scalars.  
  
\item The second essential class of normalization conditions are the  
  masses of particles. There are three subclasses of masses which  
  have to be fixed:  
\begin{enumerate}  
\item The masses of physical asymptotic particles. Only the  
  electron, the neutrinos and the photon are asymptotic fields. For  
  them there is no width and their only physical parameter is their mass.  
\item The masses of unstable particles. For unstable particles the  
  problem of correct definition of masses was already taken into  
  account in \cite{sirl,stuart}. Instead of  
  a real pole, a complex pole has to be considered. Clearly   
  only the real part of the pole has to be compared with the measured   
  physical masses.    
\item The masses of unphysical particles. Their masses are obviously  
  un-observable, but their definition is very important to   
  guarantee the correct pole structure of two-point functions ensuring  
  the unitarity of the model.  
\end{enumerate}  
  
\item IR constraints.  
  
  As we will discuss in the next paragraphs, the IR constraints are  
  essential to guarantee the correct pole structure of the massless  
  fields. In fact by ensuring the vanishing of the two-point  
  functions for the massless fields and their mixed two-point  
  functions with massive fields at zero momentum we can avoid any  
  IR divergences caused by regularization and subtraction.  In  
  particular we have to discuss the following three classes of IR  
  normalization conditions:  
  
\begin{enumerate}   
  
\item Transverse component of two-point functions for vector fields.  
  
\item Two-point functions for the longitudinal component of vector fields  
  and scalar fields. Since in the presence of CP violation (as for the  
  SM) there is no discrete quantum number which allows us to  
  disentangle the complete scalar sector into unphysical (would-be-Goldstone)   
bosons and physical ones (Higgs or its partners in the extended versions), we have 
to discuss those   
fields in the same class. We have furthermore to clarify   
how to implement correctly these conditions in order to avoid completely the   
IR problems to all orders.   
  
\item Ghost fields. For the ghost fields which are the unphysical partners of massless   
gauge fields attention has to be paid in order to get rid   
of the non-symmetric two-point functions. In the following we will   
discuss the {\it partially on-shell scheme} comparing   
our results with those given in the paper \cite{krau_ew}.   
\end{enumerate}   
  
\item Couplings.   
  
The most delicate set of normalization conditions for the non-semi-simple   
gauge models are the normalization of the couplings. Generically the problem of fixing the   
gauge couplings and the Yukawa couplings is a hard problem. In fact   
the best situation which can happen is a one-to-one correspondence   
among parameters and measurable physical quantities.   
For couplings it is in general quite difficult to find out a correspondence   
among Green's functions (computed at a specific point and at fixed perturbation order)   
with a single measurable quantity \cite{eff}. Generically it turns out that the most   
commonly used combinations of Green's functions are gauge parameter dependent   
forbidding their comparison with physical quantities.   
Alternative approaches may be applied to by-pass this   
problem \cite{eff}, however at the moment there is no definitive unique   
procedure to fix the gauge couplings at some physical value (for instance the 
Fermi's constant is established by means of the $\mu$-decay amplitude, 
which is satisfactory from practical point of view, but it depends on the specific process 
considered).    
  
For the couplings we have to single out three different classes of normalization   
conditions  
\begin{enumerate}   
\item Gauge Couplings.  
   
In this class, as is well known, due to the   
spontaneous symmetry breaking mechanism some of the   
gauge couplings can be computed in terms of some physical   
masses and they can be fixed by measurable   
quantities (for instance the Weinberg's angle $\theta_\smallw$). 
However not all gauge couplings can be obtained in this way and   
some vertices have to be taken into account (like for the electric charge $e$ and the 
QCD strong coupling $g_s$). 
This corresponds to consider three- or four-point functions which generically depend on   
gauge parameters  in a complicate way and the best   
defined objects, also in the context of pure Yang-Mills theory,   
are the S-matrix elements \cite{alpha_s}.     
  
\item Yukawa couplings.   
  
For the Yukawa couplings we have to take into consideration   
two different types of couplings. Those couplings which   
can be expressed as   
fermion masses and those which are expressed in terms   
of fermion mixing matrices (CKM). In general some Yukawa couplings  
may not even generate any contributions to fermion mass matrices, and   
they have to be considered as genuine Yukawa couplings. A proposal is fixing them by  
computing S-matrix elements involving the scalar and fermion coupling.   
  
\item Higgs potential.   
  
For the Higgs potential a similar considerations as for the above classes   
has to be done. Some of the couplings of the Higgs potential   
are fixed by means of the scalar masses (e.g. for the Higgs mass in the minimal   
SM) and by means of vanishing of the tadpoles.   
Furthermore some couplings have to be defined in terms of S-matrix elements   
and this requires the  inspection of scalar   
self-couplings.  Notice that in extended models like 2HDM and the MSSM the 
Higgs potential is parameterized by the mixing angles among physical Higgs particles. Therefore 
they correspond to physical measurable quantities  like the CKM. We propose a 
scheme to fix them by using the WTI for the background gauge invariance.  
  
\end{enumerate}  
 \end{enumerate}

The paper is organized as follows: in Sec.~2, we briefly review the 
notations, we introduce the gauge fixing and we discuss the 
relevant symmetries (BRST, background gauge invariance and the 
Abelian Antighost Equation). In Sec.~3 we discuss the free parameters 
by solving the complete set of identities. Then we systematically 
analyze the normalization conditions sector by sector. 
The hard and soft anomalies are studied in Sec.~4. In particular the cancellation 
of IR anomalies is explained. 
Finally, in the App.~A we provide the complete set of identities and
their algebra; in App.~B and in  App.~C we discuss the 
renormalization of Nakanishi-Lautrup, the equations of the 
ghost field equations (Faddeev-Popov equation and the Abelian Antighost 
Equation), respectively. 

\section{General Settings}   
\label{notations}  
\subsection{Gauge Group, Representations and Fields}  
\label{sub.sec:quantum}  
  
The field content is specified by the quantized gauge vectors  
$W^{a}_{\mu}$, their background partners $\h{W}^{a}_{\mu}$, the  
scalars $\phi_{i}$, the background scalars $\h{\phi}_{i}$, the  
fermions $\left\{\psi^L_{I}, \psi^R_{I} \right\}$\footnote{The index I for the   
fermion fields is a multi-index for isospin, flavour,  
color (for quarks). }, the Faddeev-Popov  
ghosts $c^{a}, \b{C}^{a}$, the Nakanishi-Lautrup multipliers $b^{a}$,  
the BRST sources $\gamma^{a}_{\mu}, \gamma_i, \eta^{L/R}_I, 
\b{\eta}^{L/R}_I, \zeta^{a}$ for  
quantized fields and the external fields $\Om^{a}_{\mu}, \Om_i$.  The  
component of the fields $W^{a}_{\mu}, c^{a}, \bar{C}^{a}, b^{a}$ are  
identified by the index $a$ labeling a basis of the Lie algebra ${\cal  
  G}$ of the gauge group; the fields $\h{W}^{a_\smalls}_{\mu},  
\gamma^{a_\smalls}_{\mu}, \zeta^{a_\smalls}, \Om^{a_\smalls}_{\mu}$ are restricted to  
the semi-simple factors ${\cal G}_{S}$ of ${\cal G}$ while  
$\phi_{i},\h{\phi}_{i}, \gamma_i,\Om_i$, $ \psi^{R}_I, \eta^{R}_I $ and   
$ \psi^{L}_I, \eta^{L}_I$ belong to the scalar, the left handed  and the   
right handed fermion representation spaces for ${\cal G}$. The fields are also  
characterized by a conserved Faddeev-Popov (ghost) charge which is: $0$ for  
$W^{a}_{\mu}, \h{W}^{a}_{\mu}, \phi_{i}, \h{\phi}_{i}, \psi^{R}_I, \psi^{L}_I, b^{a}$,  
$-1$ for $\bar{C}^{a}, \gamma^{a_\smalls}_{\mu}, \gamma_i, \b{\eta}^{L/R}_I, \eta^{L/R}_I$, 
$-2$ for  $\zeta^{a_\smalls}$ and $+1$ for $c^{a}, \Om^{a_\smalls}_{\mu}, \Om_i$.  
  
In order to describe the general model it is also useful to introduce the  
charge matrices involved in the coupling of the gauge fields  
$W^{a}_{\mu}$. First of all we specify the symmetric, positive  
definite charge matrix $e_{ab}$ on adjoint representation of the  
algebra ${\cal G}$. Clearly $e_{ab}$ has no elements connecting the  
semi-simple factors ${\cal G}_{S}$ to the abelian ones ${\cal G}_{A}$.  
Furthermore the restriction of $e_{ab}$ to each simple component is  
proportional to the Killing form, and, in a basis where the latter is  
diagonal, we have $e_{a_\smalls b_\smalls}= e_\smalls \del_{a_\smalls b_\smalls}$ 
and $ e_\smalls$ is  
identified with the charges of the simple factors. The charge matrix 
$e_{ab}$ of the abelian factors ${\cal G}_{A}$ must be   
symmetric and positive definite.  
  
The gauge group generators $t^{a}, T^{a}_{R}, T^{a}_{L}$ 
in the scalar and fermion representations obey  
\bea\label{setti.1}   
&& (t^{a})^{T} = - t^{a}, ~~~~  
(T^{a})^{\dagger}_{R/L} = - T^{a}_{R/L}, \nonumber \\ && \left[  
t^{a},t^{b} \right] = f^{abc} t^{c} , ~~~~ \left[  
T^{a}_{R/L},T^{b}_{R/L} \right] = f^{abc} T^{c}_{R/L}, ~~~~ \left[  
T^{a}_{R},T^{b}_{L} \right] = 0 \nonumber   
\eea   
where $t^{a}$ are real  
and $f^{abc}$ are the structure constants of ${\cal G}$, with $f^{abc_\smalla}=0$.    
The couplings of the gauge fields $W^{a}_{\mu}$ (for each simple factor
${\cal G}_\smalls$) and the couplings with scalar and fermion fields can now be  
expressed in terms of the tensors   
\beq\label{setti.2}   
(ef)^{a_\smalls b_\smalls c_\smalls} = e_{S} f^{a_\smalls b_\smalls c_\smalls}, ~~~~~
(et)^{a} = e_{a b} t^{b}, ~~~~ (eT)^{a}_{R/L} = e_{a b} T^{b}_{R/L}.   
\eeq  

The tree level action $\g_0$ is a local Lorentz invariant hermitian functional    
\beq\label{setti.5}  
\g_0 = \dms{\int \dx} {\cal L}(W^{a_\smalls}_\mu, \dots, \Om^{a_\smalls}_\mu)(x)  
\eeq  
where ${\cal L}(W^{a_\smalls}_\mu, \dots, \Om^{a_\smalls}_\mu)(x)$ 
is a local function of fields, of sources and their derivatives.   

Although we adopt the BPHZL \cite{zimm,low} as a renormalization scheme in order to   
get rid of divergences, our results 
can be extended to other renormalization schemes \cite{maiso}.  We will use the 
usual symbol $\g$ to denote the effective action which generates the one particle 
irreducible Green's functions \cite{zimm,itzy}.   

According to the BPHZL the fields are characterized by their UV and IR 
degrees so that the power counting renormalizability of the theory
and the IR finiteness of the off-shell Green's functions requires   
\beq\label{setti.5.1}  
d_{UV} \g_0 \leq 4 ~~~{\rm and}~~~ d_{IR} \g_0 \geq 4.  
\eeq   
We also assume the charge neutrality for $\g_0$ with respect to   
QED charge, Faddeev-Popov charge, lepton numbers and baryon number.   
The  tree level action $\g_0$, in this paper, is the bookkeeping of all   
 (invariant and non-invariant) counterterms of the model and   
we refer to the classical terms of the action $\g_0$ as the Lagrangian.    
  
As is well known \cite{brs,bec,libro} the  
renormalization of gauge theories requires 
external sources coupled to non-linear BRST  
transformations. The BRST-sources or, equivalently, the {\it  anti-fields}
enter on the same footing as the quantum fields and they also require their own IR and UV degree.   
If $\gamma_{\phi}$ is the BRST-source for the field $\phi$, Eqs.~(\ref{setti.5.1}) imply  
$$  
d_{UV} \gamma_\phi \leq 4 - d_{UV} \phi~~~{\rm and}~~~  
d_{IR} \gamma_\phi \geq 4 - d_{IR} \phi   
$$  
with the condition $d_{UV} \leq d_{IR}$.   
   
At tree level the gauge eigenstates are defined  by the background 
gauge symmetry, but they have no definite IR power
counting\footnote{In the SM
  this applies to the mixing between the photon field $A$ and the $Z$
  bosons. Moreover also the ghost system $c^A-c^Z$ has no definite IR
  power counting in the gauge eigenstates basis. For the quark fields
  it is reasonable to work in the gauge eigenstates, but this is not very
  practical.}. This is clearly due to non-diagonal non-negative definite mass matrices. A 
meaningful subtraction procedure with the correct IR power counting 
can only be achieved by expressing the fields in terms of combinations corresponding 
to mass eigenstates. Then we also introduce 
the tree level mass eigenstates  
$A^{a'}_{\mu}, Z^{a''}_{\mu}, G_{i'}, H_{i''},  \b{f}_I, f_J,   
c^{a'}_\smalla, c^{a''}_\smallz$ for massless and massive gauge fields\footnote{In the 
paper we will follow the convention given by \cite{aoki} for the SM: 
$Z_\mu = c_\smallw W^3_\mu -s_\smallw W^0_\mu$.}, for 
would-be Goldstone bosons and Higgs fields, for fermions 
and for massless/massive ghost fields 
respectively. $ \Xi'_{\alp},  \h{A}^{a'}_{\mu},  \h{Z}^{a''}_{\mu},  \h{G}^{i'}, \h{H}^{i''}, \Theta'_{\alp},   
\bc'_a, b'_a $ are the corresponding BRST-sources, background fields, the   
BRST variations of the background fields, the antighost fields and their 
BRST variations in the mass eigenstates representation.   
The definition of the mass-eigenstates will be 
extended at the quantum level in the next sections. 


\subsection{Symmetries and Background Gauge Fixing}     
\label{sub.sec:symmetries}  
  
To implement the BRST symmetry \cite{brs,bec,bbbc_1,aoki,krau_ew} 
at the quantum level, the non-linear 
BRST transformations are coupled to the BRST-sources 
(summation over repeated indices is understood)  
\bea  
\hspace{-1cm} {\cal L}^{S.T.} & = &  
\gam_{a_{S}}^{\mu} \left(\de_{\mu} c_{a_{S}} - (ef)^{a_\smalls b_\smalls c_\smalls}  
W_{\mu}^{b_{S}} c_{c_\smalls} \right) + \gam^{i} \left(c_a (et)^a_{ij}  
(\phi_{j} + v_j)\right) + \nonumber \\   
& + & \zeta^{a_{S}} \left( 
-\frac{1}{2} (ef)^{a_\smalls b_\smalls c_\smalls} c_{b_{S}} c_{c_\smalls} \right) +  
\b{\eta}^{R}_{I} c^{a}(eT)^{a}_{R,IJ} \psi^{R}_{J} + \b{\psi}^{R}_{I}  
c^{a}(eT)^{a}_{R,IJ} \eta^{R}_{J} + {\rm h.c. }. \label{sou} \eea   
The invariance of the vertex functional $\g$ 
is expressed by the Slavnov-Taylor Identities  ($L/R$ is suppressed, but
the summation over the two types of fermion is understood)
\bea\label{st}\ber{l}  
  \oop{.3cm} {\cal S}(\g) = \dms{\int} \dx \left[  
  \dms{\fd{\g}{\gamma}{a_\smalls}{\mu}} \dms{\fd{\g}{W}{a_\smalls}{\mu}} +  
  \de_{\mu}c^{a_\smalla} \dms{\fd{\g}{W}{a_\smalla}{\mu}} +  
  \dms{\fdd{\g}{\zeta}{a_\smalls}} \dms{\fdu{\g}{c}{a_\smalls}} +  
  \dms{\fdd{\g}{\gamma}{i}} \dms{\fdu{\g}{\phi}{i}} +  
  \dms{\fddL{\g}{\b{\eta}}{I}} \dms{\fduR{\g}{\psi}{I}} + \right. \\   
  \left. \hspace{2.5cm} + \dms{\fddL{\g}{\b{\psi}}{I}}  
    \dms{\fduR{\g}{\eta}{I}} + b_{a} {G}^{ab}  
    \dms{\fdd{\g}{\bar{C}}{b}} + \Om^{a_\smalls}_{\mu}  
    \dms{\fd{\g}{\h{W}}{a_\smalls}{\mu}} + \Om_{i}  
    \dms{\fdu{\g}{\h{\phi}}{i}} \right] = 0.    
\eer \eea   
The  general symmetric and invertible matrix ${G}^{ab}$ was introduced  
in the papers by Becchi {\it et al.} \cite{bec,henri}. It  
can be used to get rid of the possible IR problems in the on-shell  
renormalization scheme, proposed by Aoki {\it et al.} in  
\cite{aoki}, for the SM as pointed out by E.Kraus in \cite{krau_ew}. 
However we can introduce the ``rotated'' anti-ghost  
fields $\bc^{a} = ({G}^{-1})^{ab}\bar{C}_{b}$ in order to simplify  
our derivations and we will use this matrix discussing the  
normalization conditions for the ghost fields in Sec.~3 and in App.~C.
  
The nilpotency of the BRST transformations is a crucial ingredient in the analysis of the 
quantum theory; however, in the case of non-semisimple  
gauge models, this  is not sufficient to guarantee that 
the gauge group of the renormalized theory and, 
in particular, its scalar and fermion representations coincide with 
the tree level ones \cite{henri}.  
This is equivalently expressed by saying that the 
couplings among the matter fields and  abelian gauge bosons with 
scalars and fermions are not protected against the mixing 
with other conserved abelian currents \cite{acci}. 
For that reason, 
 we impose the Abelian Antighost Equation  (AAE)
\bea\label{aa} {\cal G}_{a_\smalla}(\g) &  
  \equiv & \fdu{\g}{c}{{a_{A}}}- (\h{\phi}+v)_{k} (et)^{a_{A}}_{kj}
  \fdu{\g}{\Om}{j} = \nonumber \\ && =  
\partial^{2} \bc^{a_{A}} + \gam_{j}(et)^{a_{A}}_{jk}(\phi+v)_{k}  
+ \b{\eta}^{R}_{I} (eT)^{a_\smalla}_{R,IJ} \psi^{R}_{J} + h.c. \equiv  
\Delta^{Cl}_{c_{a_\smalla}} \eea 
as supplementary constraints on $\g$ 
(see \cite{henri,pg_3} for further details).  In App.~A (Eq.~(\ref{wt})
it will be shown that the commutation relation between the STI (\ref{st}) and the 
AAE coincides with the WTI for the background gauge invariance with 
respect to abelian factors. Those WTI can be used to fix the abelian couplings 
in the same way as the Eq.~(\ref{aa}) as in \cite{krau_ew}. 

The linear terms in Eq.~(\ref{st}), proportional to the external fields $\Om^{a_\smalls}_{\mu},  
\Om_i$, which implement the non-trivial BRST transformation of the 
background fields $\h{W}^{a_\smalls}_{\mu}, \phi_i$, are essential for proving the 
renormalizability \cite{krau_ew,pg_1}.   
In fact the introduction of background fields $\h{W}^{a_\smalls}_{\mu}, \phi_i$   
triggers new couplings with dimension less   
or equal to four among background and quantum fields: 
the latter are absent in other formalisms and they require a proper renormalization.   
The extension of the BRST symmetry to the background fields is a necessary condition   
to ensure the renormalizability of the model, but it is not sufficient to guarantee     
the equivalence between the conventional approach (that is without   
background fields) and the BFM. The equivalence can be only achieved requiring  
the background gauge invariance of the simple factors ${\cal G}_\smalls$
 \bea\label{wt_S} && \hspace{-1.5cm}  
    \mbox{\large \bf W}_{a_\smalls}(\g) \equiv - \nabla^{a_{S}b_{S}}_{\mu}  
    \fd{\g}{W} {{b_{S}}}{\mu} - \nabh^{a_{S}b_{S}}_{\mu}  
    \fd{\g}{\h{W}}{{b_{S}}}{\mu} + \nonumber  
\\ &&   
+ (et)^{a_\smalls}_{ij} \left[ (\phi+v)_{j} \fdd{\g}{\phi}{i} +  
(\h{\phi}+v)_{j} \fdd{\g}{\h{\phi}}{i} + \Om_{j} \fdd{\g}{\Om}{i}+  
\gam_{j} \fdd{\g}{\gam}{i} \right] +  
\\ &&    
+ (et)^{a_\smalls}_{R,IJ} \left[ 
{\b{\eta}}^{R}_{I}  \dms{\fdaL{\g}{\b{\eta}}{R}{J}} + 
\dms{\fdaR{\g}{\psi}{R}{I}} {{\psi}}^{R}_{J}  +
\dms{\fdaR{\g}{\eta}{R}{I}}  {\eta}^{R}_{J}  +  
{\b{\psi}}^{R}_{I} \dms{\fdaL{\g}{\b{\psi}}{R}{J}} \right] + \LR  
+ \nonumber  
\\ &&     
+ (ef)^{a_\smalls}_{bc} \left[ \Om^{b}_{\mu} \fd{\g}{\Om}{c}{\mu} +  
\gam^{b}_{\mu} \fd{\g}{\gam}{c}{\mu} + \bc^{b} \fdu{\g}{\bc}{c} +  b^{b} \fdu{\g}{b}{c}
+ c^{b} \fdu{\g}{c}{c} + \zeta^{b} \fdu{\g}{\zeta}{c} \right] = 0  
\nonumber \eea   
where $\nabla^{a_{S}b_{S}}_{\mu}, \nabh^{a_{S}b_{S}}_{\mu}$ are the covariant 
derivatives with respect to the gauge fields $W^{a_\smalls}_\mu$ and to the 
backgrounds $\hat{W}^{a_\smalls}_\mu$. 
The invariance  under background gauge transformations of the action 
implies also the invariance under rigid transformation of the group  ${\cal G}$. 

The quantization of the classical gauge invariant action   
$\g_0^{Inv}[W,\phi, \psi]$  requires a  
{\it gauge fixing} which breaks the local gauge symmetry. In particular for the BFM the   
choice\footnote{ For the computations of radiative  corrections 
                \cite{msbkg}, it is convenient to introduce the  
                background gauge fields for the abelian gauge bosons  
                ${A}^{a_\smalla}_{\mu}$.  However those background fields  
                are unessential from a theoretical point of view, since   
                their equations of motion are   
                $$\fd{\g}{\hat{W}}{a_\smalla}{\mu} = \de_{\mu} b^{a_\smalla}$$ and they are   
                free after removing the $b_a$ fields.   
                This equation has to  
                be compared with the analogous equations for the non-abelian  
                background gauge fields \cite{pg_1} and for the scalar fields.  
                They are non-trivial and require the sources $\Om^{a_\smalls}_{\mu}, \Om_{i}$.}      
\beq  
{\cal L}^{g.f.} =  
b^{b} \left[\del^{ba_{S}} \nabh ^{a_{S}b_{S}}_{\mu}  
(W-\h{W})^{\mu}_{b_{S}} + \del^{ba_{A}} \partial_{\mu}  
(W-\h{W})^{\mu}_{a_{A}} + \rho^{bc} (\h{\phi}+v)_{i} (et)^{c}_{ij}  
(\phi+v)_{j} + \frac{\dms{\Lambda^{bc}}}{2} b_{c} \right]  
\label{gaugefixing}  
\eeq  
is invariant under background gauge transformations (\ref{wt_S}) if    
the 't Hooft parameters $\rho^{ab}$ and the gauge   
parameters $ \Lambda^{ab}$ satisfy    
\beq\label{gau_par} 
\Lambda^{ab} = \xi_\smalls \del^{a a_\smalls} \delta^{b b_\smalls} \delta^{a_\smalls b_\smalls} +  
\xi^{a_\smalla b_\smalla} \del^{a a_\smalla} \del^{b b_\smalla},~~~ 
\rho^{ab} = \rho_\smalls \del^{a a_\smalls}   
\del^{b b_\smalls} \delta^{a_\smalls b_\smalls} +   \rho^{a_\smalla b_\smalla}
\del^{a a_\smalla} \del^{b b_\smalla} .  
\eeq   
Here a single gauge parameter $\xi_\smalls$ and a single `t Hooft parameter  
$\rho_\smalls$ is allowed for each simple factor ${\cal G}_\smalls$. The symmetric and positive   
definite\footnote{The 't Hooft parameters, in the  
        tree level approximation, are proportional to ghost masses. If some  
        masses vanish, as, for instance, for the ghost $c_\smalla$ for the photon,  
        the `t Hooft parameters $\rho^{ab}$ are constrained to ensure this  
        feature to all orders.} 
matrices $ \xi_{a_\smalla b_\smalla}, \rho_{a_\smalla b_\smalla}$ fix the gauge for the abelian
factors ${\cal G}_\smalla$.    
  
We assume that $\g_0^{Inv}[W, \phi, \psi]$ does not explicitly depend on the background   
fields $\h{W}^{a_\smalls}_{\mu}, \phi_i$. The latter enter only in the 
gauge fixing terms (\ref{gaugefixing}).   
This, together with the WTI (\ref{wt_S}), 
ensures the equivalence between the physical observables computed   
in the conventional approach (namely with the `t Hooft gauge fixing) 
and those computed by the BFM 
(see for example \cite{ags} for further details).     
Finally, the variation of the gauge fixing with respect to the BRST 
provides the following Faddeev-Popov terms
\begin{eqnarray}
   {\cal L}^{\fp} &=& - \bc_{a} \left[\del^{aa_{S}} 
    \nabh^{a_{S}b_{S}}_{\mu} \nabla_{\mu}^{b_{S}c_{S}} c_{c_{S}} + 
    \del^{aa_{A}} \partial^{2} c_{a_{A}}  + 
    \rho^{ab} (\h{\phi}+v)_{i} (et)^{b}_{ij} (et)^{c}_{jk}(\phi+v)_{k} c_{c}  +  
  \right. \nonumber \\ 
&&\left.
    + \nabla_{\mu}^{a_{S}b_{S}} \Om^{b_{S}}_{\mu} + 
    \rho^{ab} \Om_{i} (et)^{b}_{ij} (\phi+v)_{j}  \right] \label{gua_fi_2} 
\end{eqnarray}
Non-semisimple gauge models, quantized in the `t Hooft-background gauge,  
are completely determined by the 
complete set of identities, derived (see App.~A) by computing 
the commutators among STI (\ref{st}), the AAE (\ref{aa}) and the 
WTI (\ref{wt_S}) for simple factors, and by normalization conditions which will be discussed in 
the next section. 
 
\section{Normalization Conditions and Parameterization}
\label{normalizations}
\subsection*{Renormalization Scheme}

In the next sections we study the free parameters and the 
normalization conditions according to the following scheme. 
\begin{enumerate}
\item We derive the solution to the functional equations 
  (\ref{st}),~(\ref{wt_S}),~(\ref{aa}),~(\ref{wt}),~(\ref{nl}) 
  in the space of local integrated functionals.
  We assume that the identities are not spoiled by anomalies and this hypothesis will be verified
  in the Sec.~\ref{ren_iden}. Furthermore we assume that these identities 
  maintain their tree level form \cite{brs,bec,henri}. This implies severe constraints on the 
  renormalizations of the field mixings and on the renormalization of the background fields. 
  As a result we are able to classify the free parameters of the model
  and to derive the classical action $\g_{Cl}$ including 
  all counterterms. 
\item In order to separate the normalization conditions for the unphysical 
  states like ghost fields
  from the physical ones (masses and
  couplings of physical particles), it is convenient to start considering 
  the ghost sector. There, we discuss the normalization conditions 
  which prevent IR problems and provide a reasonable 
  definition of the Z-ghost mass. As a consequence we find that 
  the free parameters compatible
  with the complete set of functional identities are not sufficient to
  fix the proper normalization conditions. Nevertheless we show that ghost  
  equations (\ref{fp})-(\ref{aa}) must be slightly deformed in order to be consistent
  with the normalization conditions. Therefore the equations of motion for the $b$ fields 
  (NL Eqs.~(\ref{nl})) 
  are accordingly modified to higher orders. These deformations 
  do not influence either the STI and the WTI nor the physical amplitudes. 
\item By using the renormalized ghost equations (\ref{fp}) and 
  the NL Eqs.~(\ref{nl}) we can simplify the STI for the longitudinal 
  (scalar) gauge bosons and the mixing with scalar fields 
  introducing the reduced functional $\hat{\g}$. 
  Within this sector we discuss the renormalization of
  tadpoles, the IR problems (related to the IR anomalies studied in Sec.~\ref{IRanomalies}) 
  and we discuss the renormalization of the would-be Goldstone masses. As a consequence of 
  the STI we are also able to identify the masses of the Higgs fields and 
  to provide a gauge parameter independent definition for them. In the sector of scalar fields 
  we have to consider the possible mixing and the CP-odd counterterms and renormalizations.   
\item Having fixed the unphysical degrees of freedom and the Higgs
  masses, we consider the gauge bosons. By means of STI for two-point
  Green's functions, we discuss the renormalization of massless and massive
  gauge bosons and their mixings. We show that by using a minimal 
  set of free parameters compatible with the WTI we are able to 
  avoid completely the IR problems. Furthermore, independently of the mixings 
  and in the case of CP violation we are also able to fix the proper, gauge parameter independent 
  mass renormalizations. 
\item For fermion fields we have to select the normalization conditions 
  which must be fulfilled in order to compute gauge invariant amplitudes: the
  renormalization of the mass and of the CKM mixing matrices.  
  We provide a gauge independent definition of the mass renormalization
  which is also appropriate in the presence of mixings. Moreover, we define the 
  mass eigenstates and in terms of them we consider the WTI. Finally by 
  using the WTI expressed in the mass eigenstates, we are able to fix the CKM
  renormalizations. 
\item Finally we show that the renormalization 
Green's functions with external background fields 
is related to the renormalization of Green's
functions with external quantum fields. In particular we show that the zeros of 
two-point functions with external background fields coincide with zeros of two-point functions 
with external quantum partners. 
We also show that the constraints coming form the WTI are not sufficient
to fix the IR problems for quantum Green's functions and proper 
normalization conditions must be used (see Item~(4) above). 
\item At the end we discuss the gauge coupling renormalization. 
\end{enumerate}  

\subsection{General Solution of the Symmetry Constraints at Tree Level}  
\label{Generalsolution}  
 
The unknown quantity of the problem is a Lorentz-invariant CPT-even\footnote{In the  
  case of the SM we cannot assume the P and the CP invariance because  
  they are broken by the presence of explicit chiral vertices and by the  
  CP-violating phase (and the non-degeneracy of the quark masses) of the  
  CKM matrix.} local   
functional\footnote{This functional must not be confused with 
$\g_0$. This new quantity contains all the counterterms needed to impose
the normalization conditions \cite{libro,zimm}.} 
$\g_{Cl} = \int \dx {\cal L}_{Cl}$ with zero Faddeev-Popov charge.  Furthermore the power  
counting renormalizability  requires  
$d_{UV} \g_{Cl} \leq 4$ and $ d_{IR} \g_{Cl} \geq 4 $  
and, since we perform the present calculation in the symmetric  
variables, we derive the most general solution which respect to the UV power
counting. In the next sections we show that also the IR constraint can be satisfied 
by a suitable choice of normalization conditions. 
  
Due to the complexity of the problem it is convenient to establish 
a precise hierarchy among functional identities, solving the linear  
constraints at the first step, then solving the WTI and, finally, the
STI.  This hierarchy shows how the abelian anti-ghost equation (\ref{aa})   
fixes the abelian couplings among the quantum fields $\phi^i,  
\psi^{R}_{I}, \psi^{L}_{I}$ and their BRST sources   
$\gamma_{i}, \eta^{R}_{I}, \eta^{L}_{I}$ and the   
renormalization of the abelian ghost fields $c^{a_\smalla}$.   
  
The solution can be expressed in terms of the following separated terms
\beq\label{sol_1}
\g_{Cl} = \g^Q_{Cl} + \g^{S.T.}_{Cl}  + \g^{\Om}_{Cl}  +  \g^{g.f.}_{Cl}
\eeq
where $\g^Q_{Cl}$ depends only on quantum fields, 
$\g^{S.T.}_{Cl}$ contains the BRST-sources couplings and the w.f.r. of the quantum fields, 
$\g^{\Om}_{Cl} $ contains the couplings among the BRST-sources and the 
BRST variation of the background fields $\Om^{a_\smalls}_\mu,\Om_i$ and $\g^{g.f.}_{Cl}$ is
\bea\label{re_1}   
\g^{g.f.}_{Cl}  &{=}&    
\dms{\int \dx}   
\left\{ b_c   
  \left[   
    \del^{c a_{S}}  \nabh^{a_{S}b_{S}}_{\mu} (W-\h{W})^{\mu}_{b_{S}} +   
    \del^{c a_{A}} \partial_{\mu} (W - \h{W})^{\mu}_{a_{A}}    
  \right.   
\right. 
\nonumber \\
&+&     
\left.   
  \left.   
    \rho^{c b} (\h{\phi} + v)^{i} (et)^{b}_{ij} (\phi + v)^{j}   
  \right]  +  
    \dms{\frac{\Lambda^{c a}}{2}} b_{c} b_a  - \bar{c}_{a_\smalla} \de^2 c_{a_\smalla}
  \right\}  
\eea  
contains the gauge fixing terms. Furthermore by the Eqs.~(\ref{fp}) and Eq.~(\ref{aa}) the 
reduced functional $\hat{\g} = \g -\g^{g.f.}_{Cl}$ depends on the new variables
\bea\label{re_2}   
\hg^{a_{S}}_{\mu}  =  \gam^{a_{S}}_{\mu} + \nabh_{\mu}^{a_{S}b_{S}}
\bc^{b_{S}}, ~~~~   
\hg_{i} =  \gam_{i} + \bc^{a}  \rho_{ab} (et)^{b}_{ij}   (\h{\phi} +
v)^{j}, ~~~~ 
\h{\Om}_{i} = \Om_{i} + (et)^{a_\smalla}_{ij}  (\h{\phi}+v)_{j} c^{a_\smalla}.  
\nonumber \eea
As a consequence the complete dependence on the abelian ghost is re-absorbed
by the new variable $\hat{\Om}$. This is due to the AAE (\ref{aa}).   
Because of the negative Faddeev-Popov charge of the BRST-sources $\hg^{a_\smalls}_{\mu},  
\hg_{i}$, their UV dimensions and Lorentz invariance 
$\g^{\Om}_{Cl}$ can be parameterized by   
\beq\label{tree.2}   
\g^{\Om}_{Cl}   
= \dms{\int \dx} \left(\hat{\gamma}^{a_\smalls}_{\mu} X_{a_\smalls b_\smalls}  
\Om^{b_\smalls}_{\mu} + \hat{\gamma}^{i} X_{i j} \hat{\Om}^{j} \right)  
\eeq  
where $X_{a_\smalls b_\smalls},  X_{i j}$ are arbitrary non singular and,  
respectively, real and complex matrices.  These are constrained by   
the WTI which imply   
(assuming the irreducibility of the scalar field representation)  
\bea\label{tree.3}   
X_{a_\smalls b_\smalls} =  X_{S} \delta^{a_\smalls b_\smalls}, ~~~~ 
X_{ij} =  X_{0} \delta_{ij } + \sum_{a_\smalla} X_{a_\smalla} t^{a_\smalla}_{ij}   
\eea   
where $X_{S}$ are single free real parameters for each simple factor  
${\cal  G}_\smalls$ of the group and $X_{0}, X_{a_\smalla}$ for the scalar representation.   
For a complex representation (as for the Higgs minimal sector of the SM) 
$X_{0}, X_{a_\smalla}$ are complex.  For this reason there is no counterterms 
for the mixing between the would-be-Goldstone boson and the Higgs field in the 
case of the minimal SM where only a restricted Higgs potential is allowed.  
In the case of reducible representations for the scalar fields (such as  
in the 2HDM and in the Higgs sector of the MSSM) we have   
a set of free complex constants $\{ X_{\alp,0}, X_{\alp,a_\smalla} \}$ for each   
irreducible representation $\alp$.  It is convenient to rewrite the $X_{ij}$ 
in terms of a product of a  multiplicative factor times a  
rotation $X_{ij} = X_{0} {\cal R}^{X}_{ij}$.  The rotation is  
generated by the abelian generators $t^{a_\smalla}_{ij}$ in the  
scalar representation.  Finally by imposing the STI and the WTI 
 we immediately get  
\bea\label{tree.6}   
\hspace{.5cm}   
\tilde{W}^{a_\smalls}_{\mu}  =  Z^{W, a_\smalls b_\smalls} \left({W}^{b_\smalls}_{\mu} + 
X_\smalls \hat{W}^{b_\smalls}_{\mu} \right), ~~~~~  
\tilde{\phi}^{i}  = Z^{\phi, ij} \left(\phi^{j} + X_{0} 
{\cal R}^{X, jk} \hat{\phi}^{k} \right)   
\eea   
where $Z^{W, a_\smalls b_\smalls}, Z^{\phi, ij}$ are arbitrary non singular matrices
and the $X$ are related to them by the equations 
\bea\label{tree.7}   
\hspace{.5cm}   
&&Z^{W}_{a_\smalls b_\smalls} =  Z^{W}_\smalls \delta^{a_\smalls b_\smalls}, ~~~~~
\hspace{.5cm}   
Z^{\phi}_{ij}  =  Z^{\phi}_0 \delta_{ij } + \sum_{a_\smalla} Z^{\phi}_{a_\smalla}  
t^{a_\smalla}_{ij} \equiv Z^{\phi}_0 {\cal R}_{ij}   
\\   
&&X_\smalls  =  Z_{S}^{{W},-1} - 1, ~~~~~ 
X_0 {\cal R}^{X}_{ij}  =  Z_{0}^{{\phi},-1} {\cal R}^{-1}_{ij} -
\delta_{ij}.    
\eea   
As a consequence we derive the renormalization of the 
gauge fields and the splitting among the quantum parts from the 
background ones   
\bea\label{tree.7.1}  
\tilde{W}^{a_\smalls}_{\mu} = Z^{W, a_\smalls b_\smalls} \left({W}^{b_\smalls}_{\mu} -  
\hat{W}^{b_\smalls}_{\mu} \right) + \hat{W}^{a_\smalls}_{\mu}, ~~~~~
\tilde{\phi}^{i} = Z^{\phi, ij} \left(\phi^{j} - \hat{\phi}^{j} \right)  
+ \hat{\phi}^{i}.  \nonumber   
\eea   
This implies that the background field part of  
the gauge fields and of scalar fields $\tilde{W}^{a_\smalls}_{\mu},  
\tilde{\phi}^{i}$ do not get independent radiative corrections and only the  
quantum part of these fields is renormalized 
by a wave function  $Z_\smalls, Z_0 {\cal R}_{ij}$.  By rescaling the gauge fields  
${W}^{a_\smalls}_{\mu} \rightarrow e_s {W}^{a_\smalls}_{\mu}$ and their background 
partners $\tilde{W}^{a_\smalls}_{\mu} \rightarrow e_s \tilde{W}^{a_\smalls}_{\mu}$  and 
introducing the wave function renormalization for the background gauge fields 
$\h{Z}^{W}_{S}\delta^{a_\smalls b_\smalls}$, the usual relation \cite{bkg} 
between the charge renormalization and the two-point functions with external 
background gauge fields  is easily recovered.  
   
We introduce the redefined sources   
$\tilde{\hg}^{a_\smalls}_{\mu} = Z^{W,-1}_\smalls  
\hg^{a_\smalls}_{\mu},  \tilde{\hg}_{i} = Z^{\phi,-1}_{ij}  \hg_{j}$, 
the BRST-sources part  $ \g^{S.T.}_{Cl}$ 
is given by
\bea\label{tree.10}   
\g^{S.T.}_{CL} & = &   
\dms{\int \dx}    
\left[  
  \sum_{S} \widetilde{\hat{\gamma}}^{a_\smalls}_{\mu}   
  \widetilde{\nabla}_{a_\smalls b_\smalls}^{\mu} 
\left(Z^{C}_{b_\smalls d_\smalls} c^{d_\smalls}\right) +   
  \widetilde{\hat{\gamma}}^{i} Z^{C}_{a_\smalls b_\smalls} c^{b_\smalls}  
  (et)^{a_\smalls}_{ji} \left(\widetilde{\phi} + v \right)_{l}    
\right.  
\nonumber \\   
&&   
\left.   
  + \, \sum_{S}  \zeta_{a_\smalls} 
Z^{C,-1}_{a_\smalls a'_\smalls}(ef)^{a_\smalls b_\smalls c_\smalls}   
  Z^{C}_{b_\smalls b'_\smalls} c^{b'_\smalls}  Z^{C}_{c_\smalls c'_\smalls} c^{c'_\smalls}   
\right.  
\nonumber \\   
&&   
\left.   
  + \, \b{\eta}^{R,I} (Z^{\psi,R})^{-1}_{IJ}  
Z^{C}_{a_\smalls b_\smalls} c^{b_\smalls} (eT)^{a_\smalls}_{R,K}    
  Z^{\psi,R}_{KM} {\psi}^{R,M}     
\right.   
\nonumber \\   
&&   
\left.  
  +  \, \b{\psi}^{R,I} (\overline{Z}^{\psi,R})^{-1}_{IJ} Z^{C}_{a_\smalls b_\smalls} c^{b_\smalls}  
  (eT)^{a_\smalls}_{R,JK} \overline{Z}^{\psi,R}_{K M} {\eta}_{R, M} + ({\rm R} 
\rightarrow {\rm L})  \right]   
\eea   
where   
\beq\label{tree.10.1}  
 \widetilde{\nabla}_{a_\smalls b_\smalls}^{\mu} c^{b_\smalls} =   
\de_{\mu}  c_{a_\smalls} - (ef)_{a_\smalls b_\smalls d_\smalls} 
\left( Z^{W}_{a_\smalls b_\smalls} ({W} -  
\hat{W} ) ^{b_\smalls}_{\mu}  + \hat{W}^{a_\smalls}_{\mu} \right)c^{d_\smalls}  
\eeq  
where   
$Z^{C}_{a_\smalls b_\smalls} $   
are the w.f.r.s  
for ghost fields. The invariance under the WTI implies   
$Z^{C}_{a_\smalls b_\smalls} = Z^{C}_{S} \delta_{a_\smalls b_\smalls}$ with an arbitrary real  
constant $Z^{C}_{S}$ for each simple factor.    
  
For the fermions we introduce   
$Z^{\psi,R}_{IJ},Z^{\psi,L}_{IJ},  
\overline{Z}^{\psi,R}_{IJ}, \overline{Z}^{\psi,L}_{IJ}$   
and applying the WTI to the fermionic terms we have  
\bea\label{tree.10.2}  
&&   
Z^{\psi,R}_{IJ} = Z^{\psi,R}_{I,0} \delta_{IJ} + \sum_{a_\smalla}
Z^{R,\psi}_{I,a_\smalla} T^{R, a_\smalla}_{IJ},~~~~   
Z^{\psi,L}_{IJ} = Z^{\psi,L}_{I,0} \delta_{IJ} + \sum_{a_\smalla} Z^{L,\psi}_{I,a_\smalla} 
T^{L,a_\smalla}_{IJ}   
\\ &&  
\overline{Z}^{\psi,R}_{IJ} = \overline{Z}^{\psi,R}_{I,0} \delta_{IJ} + \sum_{a_\smalla} 
\overline{Z}^{R,\psi}_{I,a_\smalla} T^{R,a_\smalla}_{IJ},~~~~   
\overline{Z}^{\psi,L}_{IJ} = \overline{Z}^{\psi,L}_{I,0} \delta_{IJ} + \sum_{a_\smalla} 
\overline{Z}^{L,\psi}_{I,a_\smalla} T^{L,a_\smalla}_{IJ}   
\nonumber   
\eea  
for each independent multiplet of fermions.   

In order to compare these free parameters
with the conventional formalism of the   
SM, we translate the compact notation $\psi_I, \bar{\psi}_I$   
for fermions used here into physical fields  
\bea\label{tree.14}  
\psi^{L}_{I}  =  \left\{ Q^{L}_{\alp, a, i},   L^{L}_{\alp, i}  
\right\}, ~~~~~   
\psi^{R}_{I} =  \left\{ u^{R}_{\alp, a},   d^{R}_{\alp, a},  
e^{R}_{\alp}   
\right\}  
\eea  
where $\alp$ is the flavour index, $a$ is the color index and   
$i$ is the $SU(2)$ isospin.
In this formalism, we obtain  
\bea\label{tree.10.2.1}    
Z^{\psi,L}_{IJ} = Z^{0,L}_{\alp,\bet} \delta_{ij} \delta_{a b} +
 \sum_{a_\smalla}
Z^{L,a_\smalla}_{\alp,\bet} T^{L, a_\smalla}_{i j} \delta_{a b},~~~~   
\eea
for $Z^{\psi,L}_{IJ}$ and analogous equations hold for
$Z^{\psi,R/}_{IJ},\bar{Z}^{\psi,R/L}_{IJ}$. In the above equation
the sum runs over the five abelian conserved currents (see App. A).
Notice that without the BFM, and therefore, without imposing the
WTI, we can add a further set of parameters
$Z^{3,L}_{\alp,\bet} \sigma^3_{ij}$ where $\sigma^3$
is the third component of the $SU(2)$ factor of the gauge group. 

Finally the terms $\g^{Q}_{Cl}$ involving the quantum fields only are   
(compare with \cite{henri} and \cite{henn}):   
\bea\label{tree.11}   
\g^{Q}_{Cl} & = &   
\dms{\int \dx} \left[ - \frac{1}{4} \sum_{S}  
{\cal F}^{a_{S}}_{\mu \nu } {\cal F}_{a_{S}}^{\mu \nu } -  
\frac{1}{4}   
{\cal F}^{a_{A}}_{\mu \nu } {\cal F}_{a_{A}}^{\mu \nu } +  
\widetilde{\nabla}_{ij}^{\mu} \left(\widetilde{\phi} + v \right)^{j}  
\widetilde{\nabla}^{ik}_{\mu} \left(\widetilde{\phi} + v \right)_{k}   
\nonumber \right. \\   
&& \left.   
+ \, \mu_{ji} \left(\widetilde{\phi} + v \right)^{i} \left(\widetilde{\phi} +
v \right)^{j} +   
\lambda_{ijkl}  
\left(\widetilde{\phi} + v \right)^{i} \left(\widetilde{\phi} + v  
\right)^{j} \left(\widetilde{\phi} + v \right)^{k} \left(\widetilde{\phi} +  
v \right)^{l} + \nonumber \right. \\   
&& \left.    
+ \, i \widetilde{\b{\psi}}^{R}_{I}   
\gamma^{\mu} \widetilde{\nabla}^{R,IJ}_{\mu} \widetilde{{\psi}}^{R}_{J} + i \widetilde{\b{\psi}}^{L}_{I}  
\gamma^{\mu} \widetilde{\nabla}^{L,IJ}_{\mu} \widetilde{\psi}^{L}_{J} + \widetilde{\b{\psi}}^{R}_{I}  
Y^{ IJ}_{i} \widetilde{\psi}^{L}_{J} \left(\widetilde{\phi} + v \right)^{i}   
+ {\rm h.c.}   
\right]   
\eea  
where ${\cal F}^{a_{S}}_{\mu \nu }$ is the field strength tensor  
for the non abelian gauge fields   
$\widetilde{W}^{a_\smalls}_{\mu}$ and ${\cal F}^{a_{A}}_{\mu \nu } $ for the  
abelian ones ${W}^{a_\smalla}_{\mu}$. $ \widetilde{\psi}^{L}_{J}$ are 
the rescaled fermions $\widetilde{\psi}^{L}_{J} = Z^{L}_{JK} {\psi}^{L}_{J}$.  
The covariant derivatives are given by   
\bea\label{tree.11.1}  
&& \widetilde{\nabla}_{ij}^{\mu} {\phi}^{j} = \de^\mu  {\phi}_{i} +   
\left[ (e t)^{a_\smalls}_{ij} \left(Z^{W}_\smalls ({W} -  
\hat{W} )_{a_\smalls}^{\mu}  + \hat{W}_{a_\smalls}^{\mu} \right) + 
(e t)^{a_\smalla}_{ij}  {W}_{a_\smalla}^{\mu}  \right]  {\phi}^{j}  
\nonumber \\   
&&  
\widetilde{\nabla}^{R, IJ}_{\mu} \widetilde{\psi}_{J} =   
Z^{\psi,R}_{I K} \de_\mu  {\psi}^{R}_{K}  +   
\left[ (eT)^{R,a_\smalls}_{IJ} \left(Z^{W}_\smalls ({W} -  
\hat{W} ) ^{a_\smalls}_{\mu}  + \hat{W}^{a_\smalls}_{\mu} \right) + (eT)^{R,a_\smalla}_{IJ}  
{W}^{a_\smalla}_{\mu}  \right]  Z^{\psi,R}_{JK} {\psi}^{R}_{K}   
\nonumber  
\eea  
and analogously for the left-handed fermions.   
  
The remaining free parameters (in the gauge sector) are  the gauge couplings $e_{S}$ for   
each simple factor and the gauge couplings $e_{a_\smalla b_\smalla}$ for the abelian  factors.   
Notice that for several abelian gauge fields the  
kinetic terms of the action could mix the gauge fields. For the SM, for   
a single $U(1)$ factor, this is obviously excluded. A complete   
discussion on the free abelian gauge fields and abelian factors is   
given in paper \cite{henn}.   
  
The free invariant parameters $\mu_{ij}, \lambda_{ijkl}$ are the  
quadratic and the quartic couplings for the scalar fields, respectively.   
They satisfy the following algebraic relations \cite{bbbc_1}  
\bea\label{tree.12}  
&& (et)^{a}_{ik} \mu_{kj} + \mu_{ik} (et)^{a}_{kj} = 0 \\  
&& (et)^{a}_{im} \lambda_{mjkl} + (et)^{a}_{jm} \lambda_{imkl} +  
(et)^{a}_{km} \lambda_{ijml}  + (et)^{a}_{lm} \lambda_{ijkm}  
  = 0.   
\eea  
The number of these free parameters depends on the number of scalar multiplets.  
For the minimal SM the number of free parameters are just one   
quadratic term ($\mu$) and one quartic coupling ($\lambda$), however for the non-minimal  
SM such as the 2HDM the number of invariant  
scalar terms increases. 

Concerning the CP-violating counterterms we find the following result: without the 
background gauge invariance the CP-violation permits the existence of 
non-vanishing CP-odd Green's 
functions --for instance the mixing between the would-be-Goldstone boson and the Higgs-- and 
the STI allows for CP-violating counterterms which fix these new divergences. By using the BFM 
in the case of CP-violation induced only by fermion couplings, 
these CP-odd Green's functions are automatically fixed by the background gauge invariance. 
For extended models, where the Higgs potential violates the CP symmetry, the background gauge 
invariance is not sufficient to fix all these CP-odd couplings and 
new symmetries must be invoked.   
  
In order to respect the symmetry constraints 
the Yukawa coupling  $Y^{IJ}_{i}$ have to satisfy the algebraic constraints  \cite{bbbc_1} 
\bea\label{tree.13}  
(eT)^{R,a}_{IK} Y^{KJ}_{i} + Y^{IK}_{i}  (eT)^{L,a}_{KJ} +  (et)^{a}_{ij}  
Y^{IK}_{j} = 0  
\eea  
and this restricts further the number of free parameters. 
 
Notice that the covariant   
kinetic terms for fermions and scalars have no  free parameter,   
beyond the w.f.r., in contrast to   
the kinetic terms for the gauge fields or mass terms for the scalar fields. This is due   
to the linear dependence among these terms and the   
invariant counterterms of the form  
\beq\label{tree.13b}  
{\cal S}_0 \dms{\int \dx} \gamma_i  \left(\widetilde{\phi} + v  
\right)_{j}, ~~ {\cal S}_0 \dms{\int \dx}   
\left(\bar{\eta}_I \psi_J +  \bar{\psi}_I \eta_J \right)   
\eeq  
The analysis of this linear dependence it is essential to separate the 
unphysical parameters such as the w.f.r. and the physical CKM matrix elements.   
   
The Yukawa matrices for the SM   
in the basis (\ref{tree.14})
are given by (the indices of the representation of   
$SU(2)$ for scalar fields and fermions are the same because   
they live in the same representation)   
\bea\label{tree.15}  
Y^{JK}_{i} = \left(\ber{lll}   
\vspace{.5cm} Y^{Qu, i}_{(\alp, a, j | \bet,b)} & Y^{Qd, i}_{(\alp, a, j | \bet,b)} &  
Y^{Qe, i}_{(\alp, a, j | \bet)} \\  
Y^{Lu, i}_{(\alp, j | \bet,b)} & Y^{Ld, i}_{(\alp, j | \bet, b)} &  
Y^{Qe, i}_{(\alp, j | \bet)}   
\eer \right)   
\eea   
and by the covariance with respect to $SU(3)$ gauge symmetry   
we have $Y^{Qe, i}_{(\alp, a, j | \bet)} = Y^{Lu, i}_{(\alp, j | \bet,b)}  
= Y^{Ld, i}_{(\alp, j | \bet, b)} = 0$ and   
$Y^{Qu, i}_{(\alp, a, j | \bet,b)} =   
Y^{Qu, i}_{(\alp, j | \bet)} \delta_{ab},  Y^{Qd, i}_{(\alp, a, j | \bet,b)} =   
Y^{Qd, i}_{(\alp, j | \bet)} \delta_{ab}$. The covariance with respect  
to the $SU(2)$ is more delicate and   
the instabilities of the fermion representation of the abelian factor  
$U(1)$ have to be taken into account.     
    
\subsection{Ghost Sector}    
\label{sub.sec:ghost_BFM}    
    
This section deals with the normalization conditions for the ghost fields  
and it is divided into two parts. In the first part we discuss the    
normalization conditions on two-point functions with a special emphasis on    
the renormalization of the mixings $\bc^{A}-c^Z, \bc^{Z}-c^A$. In the second part 
we analyze the Faddeev-Popov equations (\ref{fp}) and their deformations. 
In particular we show that the number of free parameters 
($Z^c_\smalls, Z^c_{a_\smalla b_\smalla}$),
constrained by Faddeev-Popov equations (\ref{fp}) and by the AAE (\ref{aa}) is 
not enough to fix the two-point functions in order to avoid the IR
divergences. We discuss the minimal extension of the set of parameters, namely the 
introduction of the 
w.f.r. for the antighost fields $\bar{Z}^c_\smalls, \bar{Z}^c_{a_\smalla b_\smalla}$ (related to the 
matrix $G_{ab}$ discussed in Sec.~\ref{sub.sec:symmetries}), 
which respects the STI and the WTI and the corresponding normalization conditions.

The ghost masses are independent parameters of the theory   
and their renormalization can   
be achieved by adjusting the independent `t Hooft parameters  
$\rho^{S}, \rho^{a_\smalla b_\smalla}$.     
For practical calculations, in order to avoid the double poles, it is advantageous to     
set the ghost masses $m^{Gh}$ equal to the masses of the would-be-Goldstone bosons    
(restricted `t Hooft gauge fixing). That corresponds, at tree level, to   
setting     
all the `t Hooft parameters $\rho^{S},     
\rho^{a_\smalla b_\smalla}$ equal to the gauge parameters $\xi^{a_\smalls},     
\xi^{a_\smalla b_\smalla}$. However, at higher orders, this degeneracy cannot be  
maintained since,    
to exclude IR  divergences, the `t Hooft parameters must be used to  
enforce   
\beq\label{nor.2}    
\left. \g^{(n)}_{\bc^{a'}_\smalla c^{b'}_\smalla}(p)\right|_{p^2=0} = 0 \hspace{2cm}     
\left. \g^{(n)}_{\bc^{a'}_\smalla c^{b''}_\smallz}(p)\right|_{p^2=0} = 0 \hspace{2cm}     
\left. \g^{(n)}_{\bc^{a''}_\smallz c^{b'}_\smalla}(p)\right|_{p^2=0} = 0       
\eeq      
as normalization conditions. Notice that the ghosts   
($c^{a'}_\smalla$ are the massless ghosts  and $c^{a''}_\smallz $ are the  massive ones)     
and the antighost fields  ($\bc^{a'}_\smalla$ and $\bc^{a''}_\smallz$)    
have independent degrees of freedom and the two-point function matrix is     
non-symmetric. As a consequence, unlike the gauge bosons, we must  
nullify both the mixed two-point functions. To this end, 
we firstly check that we can impose such
normalization conditions by tuning the free parameters 
$\rho_{a_\smalla,b_\smalla},\rho_\smalls,{Z}^c_{S}$  
and the w.f.r. for antighost fields $\bar{Z}^c_{a_\smalla b_\smalla},  
\bar{Z}^c_{S}$. 

We consider the two-point functions matrix     
 \bea\label{two_point_ghosts}    
\left(\ber{cc}     
\g_{\bc^{a_\smalla} c^{b_\smalla}}(p) & \g_{\bc^{a_\smalla} c^{b_\smalls}}(p) \\    
\g_{\bc^{a_\smalls} c^{b_\smalla}}(p) & \g_{\bc^{a_\smalls} c^{b_\smalls}}(p)     
\eer \right)     
\eea     
where $\g_{\bc^{a} c^{b}}(p) \neq \g_{\bc^{b} c^{a}}(p)$ and we   
rotate the fields by two independent transformations     
$\bar{\cal R}(p)$,  ${\cal R}(p)$ obtaining, at fixed momentum $p$,      
\bea\label{rot_gho}    
\hspace{-.5cm} \left(\ber{cc}   
\g_{\bc^{a'}_\smalla c^{b'}_\smalla}(p) & \g_{\bc^{a''}_\smallz c^{b'}_\smalla}(p) \\    
\g_{\bc^{a'}_\smalla c^{b''}_\smallz}(p) & \g_{\bc^{a''}_\smallz c^{b''}_\smallz}(p)     
\eer \right) \hspace{-.1cm} =
\hspace{-.1cm}   
\left(\ber{cc}    
\bar{\cal R}_{a' a_\smalla} & \bar{\cal R}_{a' a_\smalls} \\    
\bar{\cal R}_{a'' a_\smalla} & \bar{\cal R}_{a'' a_\smalls}    
\eer \right)\!   
\left(\ber{cc}   
\g_{\bc^{a_\smalla} c^{b_\smalla}}(p) & \g_{\bc^{a_\smalla} c^{b_\smalls}}(p) \\    
\g_{\bc^{a_\smalls} c^{b_\smalla}}(p) & \g_{\bc^{a_\smalls} c^{b_\smalls}}(p)     
\eer \right)  \!  
\left(\ber{cc}    
{\cal R}_{a_\smalla b'} & {\cal R}_{a_\smalls b'} \\    
{\cal R}_{a_\smalla b''} & {\cal R}_{ a_\smalls b''}    
\eer \right) \nonumber  
\eea    
Furthermore,  we consider    
tree level parameters appearing in the action $\g^{(0)}$, namely   
$\bar{Z}^c_\smalls, \bar{Z}^c_{a_\smalla b_\smalla}$, ${Z}^c_\smalls, {Z}^c_{a_\smalla b_\smalla}$
and setting   $m^{Gh}_{a_\smalls b_\smalls} \equiv \rho_{S}\, e^2_{S}   
v_i t_{a_\smalls}^{ij} \, t_{b_\smalls}^{jk} v_k$, and analogously for 
$m^{Gh}_{a_\smalls b_\smalla}$,   
$m^{Gh}_{a_\smalla b_\smalls}$, $m^{Gh}_{a_\smalla b_\smalla}$,   
the first equation of (\ref{nor.2})   
is rewritten in the form  
\bea\label{gh.2}   
&& 
\left[ \bar{\cal R}_{a_\smalls a'} 
\left(\bar{Z}^c_\smalls m^{Gh}_{a_\smalls b_\smalls} {Z}^c_\smalls \right)   
{\cal R}_{b_\smalls b'} +   
\bar{\cal R}_{a_\smalla a'} \left(\bar{Z}^c_{a_\smalla c_\smalla} m^{Gh}_{c_\smalla b_\smalls} 
{Z}^c_\smalls \right)   
{\cal R}_{b_\smalls b'} \right.  \\   
&& \left.  + \bar{\cal R}_{a_\smalls a'} \left(\bar{Z}^c_\smalls m^{Gh}_{a_\smalls c_\smalla} 
{Z}^c_{c_\smalla b_\smalla}  \right)   
{\cal R}_{b_\smalla b'} +   
\bar{\cal R}_{a_\smalla a'} \left(\bar{Z}^c_{a_\smalla c_\smalla}   
m^{Gh}_{c_\smalla d_\smalla} {Z}^c_{d_\smalla b_\smalla}  \right)   
{\cal R}_{b_\smalla b'} \right] +   
\Sigma^{(n)}_{\bc^A_{a'} c^A_{b'}}(0) = 0 \nonumber  
\eea   
where the ghost fields corresponding to the gauge 
eigenstates are rescaled.   
Equivalent equations are derived for $\g_{\bc^A_{a'} c^Z_{b''}}(0)$,
$\g_{\bc^Z_{a''} c^A_{b'}}(0)$ substituting the matrices $\bar{\cal R}_{a_\smalls a'},   
\bar{\cal R}_{a_\smalla a'} $ with $\bar{\cal R}_{a_\smalls a''},   
\bar{\cal R}_{a_\smalla a''}$ (where the index $a''$ runs over the set of massive   
ghost fields) and   
${\cal R}_{b_\smalls b'},   
{\cal R}_{b_\smalla b'} $ with $\bar{\cal R}_{b_\smalls b''},   
\bar{\cal R}_{b_\smalla b''}$. In Eq.~(\ref{gh.2}), $\Sigma^{(n)}_{\bc^A_{a'} c^A_{b'}}(0)$   
is the $n$-loop contribution to the two-point function   
$\g_{\bc^A_{a'} c^A_{b'}}(0)$ computed in the BPHZL subtraction
scheme\footnote{In the present section we always use the notation
  $\g^{(0)} \equiv \g_0$ to indicate the tree level action;
  $\Sigma^{(n)}$ to denote the $n$-loop correction which 
satisfy the normalization conditions up to $n-1$ order. The field $\widetilde{\phi}(p)$ indicates 
the Fourier transform of the field with in-coming momentum $p$ and the momentum 
conservation is understood.}.   
   
 The determinant of the system which corresponds to the normalization conditions   (\ref{nor.2})   
vanishes. This is due to the presence of massless particles and it is ensured 
by the STI.   
Therefore some equations are redundant   
and, for instance,    
$\g_{\bc^A_{a'} c^A_{b'}}(0)$ can be discarded.   
Nevertheless   
the system is still solvable in terms of w.f.r. 
$\bar{Z}^c_\smalls, \bar{Z}^c_{a_\smalla b_\smalla},{Z}^c_\smalls$ and in terms of 
the 't Hooft parameters. 
On the other side  
$Z^c_{a_\smalla b_\smalla}$ is fixed by the WTI and the AAE to the charge
renormalization $e_{a_\smalla b_\smalla}$. 
As an example we consider the  
system  $c^A - c^Z$ of the SM, and we impose that $\bar{\cal R}$ and  
${\cal R}$ coincide with the tree level rotations, namely the Weinberg  
angle $\theta_\smallw$.  The latter follows from our request of preserving 
the STI and the WTI in their tree level form in the non-minimal sector.   
In fact,  we use the common identifications\footnote{The symbols   
$s_\smallw, c_\smallw$ denote the sine and cosine of the Weinberg's angle.}   
\bea\label{gh.3}   
&& \bar{\cal R}_{3 \, A} = {\cal R}_{3 \, A} = s_\smallw, \ \ ~~     
\bar{\cal R}_{3 \, Z} = {\cal R}_{3 \, Z} = c_\smallw, \\     
&& \bar{\cal R}_{0 \, A} = {\cal R}_{0 \, A} = c_\smallw, \ \   ~~  
\bar{\cal R}_{0 \, Z} = {\cal R}_{0 \, Z} = - s_\smallw, \nonumber   
\eea  
for the rotations and   
\bea\label{gh.4}   
& \tilde{m}^{33}_{Gh} = \bar{Z}^c_{2} {m}^{33}_{Gh} {Z}^c_{2} =   
\bar{Z}^c_{2} {Z}^c_{2} \rho_2 g^2_2 \, v^2 ~~~~ 
& \tilde{m}^{30}_{Gh} = \bar{Z}^c_{2} {m}^{30}_{Gh} {Z}^c_{1} =   
\bar{Z}^c_{2} {Z}^c_{2} \rho_2 g_2 g_1 \, v^2 \\   
& \tilde{m}^{03}_{Gh} = \bar{Z}^c_{1} {m}^{03}_{Gh} {Z}^c_{2} =   
\bar{Z}^c_{1} {Z}^c_{2} \rho_1 g_2 g_1 \, v^2 ~~~~   
& \tilde{m}^{00}_{Gh} = \bar{Z}^c_{1} {m}^{00}_{Gh} {Z}^c_{1} =   
\bar{Z}^c_{1} {Z}^c_{1} \rho_1 g^2_1 v^2 \nonumber   
\eea   
for the renormalized mass parameters. There we put    
$\bar{Z}^c_2 \equiv \bar{Z}^c_{SU(2)}, \bar{Z}^c_1 \equiv \bar{Z}^c_{U(1)}$  
for the antighost and   
${Z}^c_2 \equiv {Z}^c_{SU(2)}, {Z}^c_1 \equiv {Z}^c_{U(1)}$  
for the ghosts; we have also used the parameters  
$\rho_2 \equiv \rho_{SU(2)},   
\rho_1 \equiv \rho_{U(1)}$ and the couplings $g_2 \equiv g_{SU(2)},   
g_1 \equiv g_{U(1)}$. Therefore in the neutral ghost sector of the SM, we have five real   
parameters. Indeed, $Z_1$ is fixed by the   
renormalization of the abelian gauge fields and, by means of the AAE (\ref{aa}),   
coincides with the renormalization of the $U(1)$ gauge coupling.    
   
Then introducing this definition into the equation (\ref{gh.2})   
we finally obtain the system    
\bea\label{gh.5}   
\tilde{m}^{03}_{Gh} &=& - s_\smallw^2 \Sigma_{\bc_\smalla c_\smallz}(0) +   
c_\smallw^2 \Sigma_{\bc_\smallz c_\smalla}(0) - s_\smallw c_\smallw   
\left(\Sigma_{\bc_\smallz c_\smallz}(0)  - \Sigma_{\bc_\smalla c_\smalla}(0) \right) \\   
\tilde{m}^{30}_{Gh} &=& s_\smallw^2 \Sigma_{\bc_\smallz c_\smalla}(0)    
- c_\smallw^2 \Sigma_{\bc_\smalla c_\smallz}(0) - s_\smallw c_\smallw   
\left(\Sigma_{\bc_\smallz c_\smallz}(0)  - \Sigma_{\bc_\smalla c_\smalla}(0) \right) 
\nonumber 
\eea   
which can be solved in terms of two parameters.  
One of the remaining parameters is used to impose   
\bea\label{gh.6}   
\lim_{p\rightarrow 0} \frac{- i \, p_\mu}{p^2} \g_{\gamma_3^\mu c_\smalla}(p) = s_\smallw   
\eea   
where $\gamma_3^\mu$ is the BRST source for the third component of $SU(2)$   
triplet of gauge bosons, which is essential to guarantee the absence of   
IR anomalies  (see Sec.~\ref{IRanomalies}). The remaining parameters are needed to   
have a consistent renormalization of the Eqs.~(\ref{fp})  as we will  be  
shown below, and to fix the mass of the $Z$-ghost. Therefore there is no free parameter   
to fix also the mass of the $W$-ghost  
as also discussed in \cite{krau_ew}.   
To fix those masses independently the   
WTI must be deformed. On the other hand, the degeneracy among the   
ghost masses and masses of the system of Goldstones and scalar components of 
gauge bosons is ensured  by the STI.  
  
In conclusion, this analysis shows that for the SM the free   
parameters, constrained by the equation of motion for the ghost fields 
(Faddeev-Popov equations and AAE), 
are not enough to avoid the IR problems and to fix the   
$Z$-ghost mass. If a new abelian factor is added and the new gauge boson   
has to be massless, the WTI and the STI are necessarily 
modified\footnote{In that case new w.f.r. $Z^c$ for ghost fields are
  needed to impose the normalization conditions.}.   
    
At this point we are able to study the Faddeev-Popov equations   
and the renormalization of the Green's functions which contain   
the BRST sources. In particular, by the Faddeev-Popov   
equations we have the following relations among two-point   
functions   
\bea\label{gh.7}  
&& \g_{\bc^{a_\smalla} c^{b_\smalla}}(p) = - \del^{a_\smalla b_\smalla} p^2 +     
\rho^{a_\smalla c_\smalla} v_{i} (et)^{c_\smalla}_{ij} {\g}_{\gam_j c^{b_\smalla}}(p^2) \nonumber \\    
&& \g_{\bc^{a_\smalla} c^{b_\smalls}}(p) =      
\rho^{a_\smalla c_\smalla} v_{i} (et)^{c_\smalla}_{ij} {\g}_{\gam_j c^{b_\smalls}}(p^2) \nonumber \\    
&& \g_{\bc^{a_\smalls} c^{b_\smalla}}(p) = - i p_{\mu} {\g}_{\gam^{a_\smalls}_{\mu} 
c^{b_\smalla}}(p) +    
\rho_{S} v_{i} (et)^{a_\smalls}_{ij} {\g}_{\gam_j c^{b_\smalla}}(p^2) \\    
&& \g_{\bc^{a_\smalls} c^{b_\smalls}}(p) = - i p_{\mu} {\g}_{\gam^{a_\smalls}_{\mu} 
c^{b_\smalls}}(p) +    
\rho_{S} v_{i} (et)^{a_\smalls}_{ij} {\g}_{\gam_j c^{b_\smalls}}(p^2). \nonumber     
\eea     
   
The UV dimensions     
of  BRST-sources $\gamma^{a_\smalls}_{\mu}, \gamma_i$ and the Lorentz     
invariance imply the superficial divergence of two-point functions   
$ {\g}_{\gam_{\mu}^{a_\smalls} c^{b_\smalla}},    
{\g}_{\gam^{a_\smalls}_{\mu} c^{b_\smalls}}$ 
${\g}_{\gam_j c^{b_\smalla}}, {\g}_{\gam_i c^{b_\smalls}}$.    
They are renormalized by the w.f.r.s $\bar{Z}^{c}_{a_\smalla b_\smalla},     
\bar{Z}^{c}_{S}, {Z}^{c}_{S}$  and the mass renormalization of ghost fields.     
Furthermore,  in order to clarify the relation among the kinetic terms     
of ghost two-point functions and the ${\g}_{\gam^{a_\smalls}_{\mu} c^{b_\smalls}}$, we     
differentiate the forth equation in the system (\ref{gh.7}) with     
respect to momentum $p$ and we consider the limit $p\rightarrow \infty$:  
\bea\label{limit}    
\lim_{p \rightarrow \infty}     
\de^2_{p}  \g_{\bc^{a_\smalls} c^{b_\smalls}}(p) & = & \lim_{p \rightarrow \infty} \left[      
i \de^2_{p} \left( p_{\mu}{\g}_{\gam^{a_\smalls}_{\mu} c^{b_\smalls}}(p) \right) +    
\rho_{S} v_{i} (et)^{a_\smalls}_{ij} \de^2_{p} {\g}_{\gam_j c^{b_\smalls}}(p^2) \right] = 
\nonumber \\ & = &     
 \lim_{p \rightarrow \infty} \left(
2 i \de_{p_{\mu}} + p^\mu \de_{p^2} \right)  {\g}_{\gam^{a_\smalls}_{\mu}  
   c^{b_\smalls}}(p).      
\eea    
The Weinberg's theorem \cite{wein} is used to discard the convergent   
two-point functions.     
On the other hand, because of the AAE Eq.~(\ref{aa}), the Green's functions     
${\g}_{\gam_{\mu}^{a_\smalls} c^{b_\smalla}}(p)$ are superficially     
finite to all orders and no normalization is required. This can be    
easily shown by      
differentiating Eq.~(\ref{aa}) with respect to the sources ${\gam_{\mu}^{a_\smalls}}(p)$     
\beq\label{fini_gaa}    
\de_{p_\mu} \g_{\gam_{\mu}^{a_\smalls} c^{b_\smalla}}(p) +     
v_{i} (et)^{b_\smalla}_{ij} \de_{p_\mu} {\g}_{\gam_{\mu}^{a_\smalls} \Om_{j}}(p^2) = 0     
\eeq    
and observing that $\de_{p_\mu} {\g}_{\gam_{\mu}^{a_\smalls} \Om_{j}}(p^2)$     
is superficially convergent by power counting. We conclude      
that also ${\g}_{\gam_{\mu}^{a_\smalls} c^{b_\smalla}}(p)$ are 
superficially convergent to all orders.     
As a consequence,    
the w.f.r. of abelian ghost fields    
${Z}^{c}_{a_\smalla b_\smalla}$ is not independent of the w.f.r. 
of the abelian gauge fields.      
  
Finally to check the consistency among the Eq.~(\ref{gh.7}) and the  
normalization conditions (\ref{nor.2}), we insert the expressions  
(\ref{gh.7}), computed at zero momentum, in (\ref{gh.2}).   
However, since the system has sensible solutions only for specific  
models, we concentrate on the $c^A,c^Z$ system of the SM.  From identities  
(\ref{gh.7}), by using the identifications (\ref{gh.3}-\ref{gh.4}) we  
obtain the following system   
\bea\label{gh.7.1}   
\g_{\bc^A c^A}(p) &=& - p^2 \left(c^2_\smallw + s_\smallw \g_{\gamma_3 c_\smalla} \right) +   
(s_\smallw \bar{Z}^c_2 g_2   
\rho_2   + c_\smallw \bar{Z}^c_1 g_1 \rho_1) \g_{\gamma_0 c^A} \nonumber \\   
\g_{\bc^A c^Z}(p) &=& - p^2 \left(c_\smallw \,s_\smallw - s_\smallw 
\g_{\gamma_3 c_\smallz} \right) +   
(s_\smallw \bar{Z}^c_2 g_2   
\rho_2   + c_\smallw \bar{Z}^c_1 g_1 \rho_1) \g_{\gamma_0 c^Z}  \\   
\g_{\bc^Z c^A}(p) &=& - p^2 \left(s_\smallw c_\smallw - c_\smallw 
\g_{\gamma_3 c_\smalla} \right) -   
(c_\smallw \bar{Z}^c_1 g_1 \rho_1 - s_\smallw  \bar{Z}^c_2 g_2   
\rho_2 ) \g_{\gamma_0 c^A} \nonumber   
\eea   
where $\gamma_0$ is the BRST source coupled to the BRST variation of the neutral   
Goldstone boson $G_0$.  The vanishing of the   
terms on the l.h.s. at zero momentum implies $\g_{\gamma_3 c_\smalla}(0) =0$ and   
$(c_\smallw  \bar{Z}^c_2 g_2 \rho_2 + s_\smallw  \bar{Z}^c_1 g_1 \rho_1 ) =0$.   
The latter imposes  a relation among the   
't Hooft parameters and the w.f.r. which reduces the number of free  
parameters. Furthermore those relations imply that also the first  
Eq.~(\ref{gh.7.1})  is automatically satisfied showing that  
$\g_{\bar{c}_\smalla c_\smalla}(0)= 0$. This is a consequence of the   
STI (which can also be checked from (\ref{gh.7}))   
and the normalization conditions (\ref{nor.2}) are   
compatible with the Faddeev-Popov equations.   
The resulting condition $\g_{\gamma_0 c_\smalla}(0) =0$ turns out to be    
useful to exclude the IR anomalies.   
 
Clearly the Faddeev-Popov Eqs.~(\ref{fp}) are deformed by the mass  
renormalization of the mass of $c_\smallz$ and by the conditions  
(\ref{nor.2}). This means that the parameters $\rho_1, \rho_2$   
in Eqs.~(\ref{gh.7.1}) do not coincide with to tree level parameters.   
Moreover, the constraints coming from the STI  
and the WTI are not affected by these deformations.   
  
\subsection{Scalar Sector}    
\label{sub.sec:longi_gauge_BFM}  
  
The unitarity of the S-matrix relies on the {\it quartet mechanism}    
\cite{bec,kugo}; that is, the scalar degrees of    
freedom of gauge bosons and their Goldstone partners must be degenerate    
with the system of ghosts and anti-ghosts. Therefore, it     
must be implemented at higher orders, and this requires a proper   
renormalization of tadpoles, of the Nakanishi-Lautrup Eqs.~(\ref{nl}) and, finally, 
of the STI.      
    
First, we recall that a generic subtraction scheme (for    
instance the $\overline{MS}$ scheme for the Dimensional Regularization)     
shifts the tadpole contribution to Green's functions. Consequently,     
by adjusting the vacuum expectation value $v_i$ and the free parameters     
$\mu_{ij}, \lambda_{ijkl}$, defined in (\ref{tree.12}),  the tadpole    
normalization conditions    
\beq\label{nor.20}    
\left. \frac{\del \g^{(n)}}{\del \phi_i} \right|_{\phi=0}= \left(   
  \mu_{ij} v^j + \frac{1}{6} \lambda_{ijkl} v^j v^k v^l \right)^{(n)} +    
\Sigma^{(n)}_{\phi_i}(0) = 0     
\eeq     
must be imposed to all orders. On the r.h.s.,  $\Sigma^{(n)}_{\phi_i}(0)$ denotes     
n-loop contribution to tadpoles.    
    
Notice that the normalization condition (\ref{nor.20})    
involves both the tadpoles of physical Higgs fields ($H_{i'}$ where    
$i'=1\dots N_{Higgs}$)  and the tadpoles of unphysical Goldstone fields     
($G_{i''}$ with $i''=1\dots N_{Golstone}$) \cite{odd-tadpole}. In fact, although the    
latter are absent at tree level, due to the CP invariance of the    
scalar terms in the action (\ref{tree.11}), they are shifted by CP-odd radiative    
corrections involving the CKM couplings.    
Furthermore notice that for extended models there is a flat direction of the 
Higgs potential ensured by the following STI\footnote{Due to the IR power counting and 
the zero momentum configuration this STI cannot be spoiled by breaking terms due to the 
regularization scheme.}
\bea\label{flat}
\g_{ c^{a''}_\smallz \gamma_i}(0) \g_{\phi_i}(0) = 0
\eea
where $c^{a''}_\smallz$ are the neutral massive ghost fields. For the SM, neglecting  
CP violation, the above equation implies that $\g_{G_0}(0)=0$ where $G_0$ 
is the neutral would-be-Goldstone boson. For the SM, considering the CP violation, 
we have that $\g_H(0)=0$  implies  $\g_{G_0}(0)=0$. And finally, for extended model, 
by a suitable rotation among the scalar fields, one tadpole is automatically nullified. 
    
For two-point functions we have to consider the complete matrix  
\bea\label{new_form_matr}    
{\cal M}^{L}_{\alp \bet}(p) =     
\left(\ber{lll} \g^{L}_{W^{a_\smalla} W^{b_\smalla}}(p) & 
\g^{L}_{W^{a_\smalla} W^{b_\smalls}}(p) &     
\g_{W^{a_\smalla} \phi_j}(p) \\     
\g^{L}_{W^{a_\smalls} W^{b_\smalla}}(p) & \g^{L}_{W^{a_\smalls} W^{b_\smalls}}(p) &     
\g_{W^{a_\smalls} \phi_j}(p) \\    
\g_{\phi_i W^{b_\smalla}}(p) & \g_{\phi_i W^{b_\smalls}}(p) &     
\g_{\phi_i \phi_j}(p)    
\eer \right)     
\eea    
where $\g^{L}_{W^a W^b}(p)$ are     
the longitudinal part of two-point functions      
for the gauge bosons,    
$$ \g_{W^a_\mu W^b_{\nu}}(p) = \left(g_{\mu\nu} -\frac{p_{\mu} p_{\nu}} {p^2}     
\right) \g^{T}_{W^a W^b}(p) +     
 \frac{p_{\mu} p_{\nu}} {p^2}  \g^{L}_{W^a W^b}(p);$$    
in the same way 
$$\g_{W^a \phi_i}(p)= - i p_{\mu} \g_{W^a_{\mu} \phi_i}(p)$$ are the mixed two-point    
functions for gauge bosons and     
scalars and $ \g_{ \phi_i \phi_j}(p)$ are   
the two-point functions matrix for scalars.     
    
In Sec.~\ref{notations} the gauge fixing was given in terms of the  
$b_a$ fields. However, for practical computation these fields are commonly  
integrated out and the resulting tree level contribution to 
mixed gauge -scalar Green's functions and for scalar two-point functions
of (\ref{new_form_matr})  
are   
\bea\label{gau_fix_term}  
&& \g^{(0)}_{W^{a_\smalls} \phi_j}(p)      =      
\left(1 -  \frac{\rho_\smalls}{\xi_\smalls}\right) v_k t^{a_\smalls}_{k i}, ~~~~  
\g^{(0)}_{W^{a_\smalla} \phi_j}(p)      =   
\left(1 -  \frac{1}{\xi_{a_\smalla b_\smalla}} \rho_{b_\smalla c_\smalla} \right) 
v_k t^{c_\smalla}_{k i}    \\  
&& \g^{(0)}_{\phi_i \phi_j}(0)            =    
\mu_{ij} + \frac{1}{2} \lambda_{i j k l} v_k v_l + \left(  
\frac{\rho^2}{\xi_{s}} v_k t^{a_\smalls}_{ki} t^{a_\smalls}_{j l} v_l +   
\rho_{a_\smalla b_\smalla} \frac{1}{\xi_{b_\smalla c_\smalla}} 
\rho_{c_\smalla d_\smalla}v_k t^{a_\smalla}_{k i} t^{c_\smalla}_{j l}   
v_l \right) \,. \nonumber 
\eea  
The masses of the Goldstone $m^{i j}_{G}$   
correspond to the expression in the brackets of $\g_{\phi_i \phi_j}(0)$.   
Notice that, since the relations among the Goldstone masses  
and gauge parameters are invertible, the Goldstone masses can be read as  
gauge parameters. The restricted `t Hooft gauge fixing   
corresponds to setting the masses of the Goldstones equal to the masses  
of the massive ghosts. Clearly this is not true for the gauge  
parameters of massless gauge fields. 

Another very convenient choice is  
the `t Hooft-Feynman gauge fixing ($R_\xi$ gauge) which corresponds to identify the  
masses of Goldstone scalars with the masses of ghosts and with the masses of the  
gauge bosons. This feature can not be achieved in the BFM because the  
background gauge invariance implies the degeneracy among the gauge  
parameters: only a single gauge parameter $\xi_\smalls, \rho_\smalls$  can be  
introduced for each simple factor.   

In conclusion the Nakanishi-Lautrup Eqs.~(\ref{nl}) cannot be maintained in  
their tree level form and the parameters which appear in those equations  
are fixed on the masses of the Goldstone (\ref{nor.112}) and by the normalization 
conditions (\ref{nor.2}) for the ghost fields.  
   
Besides the normalization of the Goldstone masses, we have 
to verify that the equations    
\beq\label{nor.111}  
\lim_{p\rightarrow 0} \frac{-i p_{\mu}}{p^2} \g_{A^{a'}_{\mu} \phi_i }(p) = 
 \g_{A^{a'} \phi_i }(0) = 0,   
\eeq  
where $a'$ runs over the set of massless gauge fields and $i$ over the  
set of scalar fields, are satisfied. This conditions are valid at tree level, ensuring  
that the massless gauge fields remain massless after fixing the gauge     
and spontaneous symmetry breaking. However they must also be maintained at higher orders 
to  guarantee the absence of IR anomalies for the STI. 

In particular for the minimal SM the equation    
\beq\label{nor.112}  
\lim_{p\rightarrow 0} \frac{-i p_{\mu}}{p^2} \g_{A_{\mu} G_0 }(p) \equiv 
 \g_{A G_0 }(0) =  0 
\eeq
expresses that the spontaneous symmetry breaking mechanism and the gauge fixing 
does not generate a mass term for the photon field. However at the quantum level this 
conditions must be ensured by the choice of the gauge fixing, by the STI and 
eventually by using the WTI of the BFM. In particular we are able to prove it as a non-trivial 
consequence of  WTI, of STI and of the normalization conditions for transverse component of the 
two-point functions for the gauge bosons. In the case of CP violation, this is also related to 
the mixing between the Goldstone field and the Higgs as it will be shown below. 

On the other hand, the equation 
\beq\label{nor.112.1} 
\lim_{p\rightarrow 0} \frac{-i p_{\mu}}{p^2} \g_{A_{\mu} H }(p) 
\equiv \g_{A H}(0) = 0 \,,   
\eeq  
where $H$ is the Higgs of the minimal SM, is not relevant for IR anomaly cancellations, 
but we are able to prove it as a consequence of normalization conditions, WTI and STI. 
Notice that, at tree level there is no bilinear coupling  between the Higgs field and the longitudinal 
degree of freedom of the photon 
because of the CP invariance of the Lagrangian of the scalar and gauge sector. However, 
at the quantum level, the CP violating radiative corrections might generate that 
Green's function. 
  
Finally,  we can study the definition of the Higgs masses. These are the  
only physical parameters entering in the present sector. In absence of  
CP-violation the two sectors, namely the Higgs fields and the unphysical  
scalar sector can be analyzed separately. However, the CP-violation  
introduces CP-odd  radiative corrections which mix the scalar fields.  
For instance for the {\it minimal} SM we must consider the  
complete matrix  
\bea\label{mat_scal} 
\left(\ber{cc} \g_{G_0 G_0}(p) & \g_{G_0 H}(p) \\ 
\g_{H G_0}(p) & \g_{H H}(p) \eer \right) 
\eea 
where the mixings $\g_{G_0 H}$ are different from zero only  
starting from three loops.  
  
The aim of this paragraph is to show that, despite several mixings  
among the unphysical fields and the Higgses, it is always possible to  define 
the Higgs masses. To this purpose we use the STI\footnote{The IR anomalies of this 
equation will be discussed in the Sec.~\ref{IRanomalies}.}   
\bea\label{new_set_sti}    
&& \gh^{L}_{V^{a_\smalla} V^{b_\smalla}} + \gh_{c_{a_\smalla} \hg_{c_\smalls}} \gh^{L}_{V^{c_\smalls} V^{b_\smalla}} +     
\gh_{c_{a_\smalla} \hg_{k}} \gh_{\phi_k V^{b_\smalla}} = 0  \\    
&& \gh^{L}_{V^{a_\smalla} V^{b_\smalls}} + \gh_{c_{a_\smalla} \hg_{c_\smalls}} 
\gh^{L}_{V^{c_\smalls} V^{b_\smalls}} +     
\gh_{c_{a_\smalla} \hg_{k}} \gh_{\phi_k V^{b_\smalls}} = 0  \\     
&& - p^2 
\left( \gh_{V^{a_\smalla} \phi_j} + \gh_{c_{a_\smalla} \hg_{c_\smalls}} 
\gh_{V^{c_\smalls} \phi_j} \right) +     
\gh_{c_{a_\smalla} \hg_{k}} \gh_{\phi_k \phi_j} = 0  \\     
&& \gh_{c_{a_\smalls} \hg_{c_\smalls}} \gh^{L}_{V^{c_\smalls} V^{b_\smalla}} +     
\gh_{c_{a_\smalls} \hg_{k}} \gh_{\phi_k V^{b_\smalla}} = 0 \label{1.1} \\    
&& \gh_{c_{a_\smalls} \hg_{c_\smalls}} \gh^{L}_{V^{c_\smalls} V^{b_\smalls}} +     
\gh_{c_{a_\smalls} \hg_{k}} \gh_{\phi_k V^{b_\smalls}} = 0 \label{1.2} \\     
&& - p^2 
\left( \gh_{c_{a_\smalls} \hg_{c_\smalls}} \gh_{V^{c_\smalls} \phi_j} \right) +     
\gh_{c_{a_\smalls} \hg_{k}} \gh_{\phi_k \phi_j} = 0 \label{1.3}      
\eea     
where the {\it reduced} functional $\gh$ defined by $\gh \equiv \g - \g^{g.f.}$ 
replace $\g$; to define $ \g^{g.f.}$ at the quantum level the 
Nakanishi-Lautrup (\ref{nl}) and the Faddeev-Popov equations (\ref{fp}) are used.  
Here $\g^{g.f.}$ is defined by the renormalized NL Eqs.~(\ref{nl}) 
and by the renormalized ghost equations (\ref{fp})-(\ref{aa}). 

By  means of these identities it is easy to prove that   
$${\rm Rank}\left({\cal M}^{L}(p) \right)     
= {\rm Rank}\left(\gh_{\phi_i \phi_j}(p) \right). $$   
This means that non-trivial physical information is only contained in   
the scalar sub-matrix $\gh_{\phi_i \phi_j}(p)$.   
Furthermore  by using the STI     
(\ref{1.1})-(\ref{1.3})     
computed at zero momentum (static STI) it is easy to show that     
\beq\label{zero} 
{\rm Rank}\left(\gh_{\phi_i \phi_j}(0) \right) = 
{\rm Rank}\left(\gh_{H_{i'} H_{j'}}(0) \right) 
\eeq    
where $i', j'$ run over the only set of physical scalar fields\footnote{for    
the SM $i'=j'=H$,     
\ie the Higgs field, for the     
2HDM \cite{2HD}, $i',j' = H, h, A, H^{\pm}$, where $H$ is the Higgs,    
$h$ is the second neutral scalar field,  $A$ is a pseudo-scalar (CP-odd) field 
and $H^{\pm}$ are the charged physical scalar fields}.     
   
For the {\it minimal} SM, where only the Higgs $H$ and the neutral 
would-be-Goldstone $G_0$ are involved, the Eq.~(\ref{zero}) coincides with  
\beq\label{zero_1} 
\gh_{G_0 G_0}(0) \gh_{H H}(0) - \gh^2_{G_0 H}(0) = 0  
\eeq 
and, therefore, $\gh_{G_0 G_0}(0)=0$ if $ \gh_{G_0 H}(0)=0$. This last  
condition is needed to cancel the IR anomalies of the STI and it is a consequence of the 
WTI. This will be shown in Sec.~\ref{IRanomalies}. Notice that 
this particular normalization condition states that the would-be-Goldstone bosons and 
the Higgs are decoupled on the mass-shell of the unphysical state. As a 
consequence this confirms that the Higgs field is the physical field of 
the present sector. In the case of extended models similar normalization 
conditions must be established in order to prevent the theory from IR 
anomalies.  

Finally we have only to fix the normalization conditions for the    
non-trivial eigenvalues $\lambda^{(n)}_{i'}(p)$,$i'=1, \dots, N_{Higgs}$.     
The complex zeros of the eigenvalues $\lambda^{(n)}_{i'}(p)$, 
namely $\lambda^{(n)}_{i'}(p^*_{i'}) =0$, can be identified  
with the masses $M_{H_{i'}}$ and the 
width $\g_{H_{i'}}$ of the physical Higgses. Clearly these 
normalization conditions fix partially the free parameters   
$\mu_{ij}, \lambda_{ijkl}$.   
  
As in the case of fermions (see next sections) two or more scalar doublets 
({\it e.g.} in the case of the 2HDM \cite{2HD} and of the MSSM \cite{MSSM}) 
can generate physical mixing angles whose renormalization must be discussed. The normalization 
conditions for the masses -- by  comparing the zeros of the eigenvalues with the physical 
masses -- is independent of the 
renormalization of the mixing angles  guaranteeing that the pole mass definitions 
can be always achieved. 

For the renormalization of the mixing angle we 
use again the WTI for the background gauge invariance. There, indeed, the mixing angles appear 
as constant parameters and, on the basis of the mass eigenstates they can be identified 
with the renormalized mixing angles.  
 
To be more precise we consider the WTI, neglecting gauge, fermion and ghost terms, 
\bea\label{zero_2}
    \mbox{\large \bf W}_{a}(\gh) = \left({\rm gauge-fermion-ghost~terms} \right) 
+ (et)^{a}_{ij} \left[ (\phi+v)_{j} \fdu{\gh}{\phi}{i} +  
(\h{\phi}+v)_{j} \fdu{\gh}{\h{\phi}}{i} \right] = 0  
\eea 
and we define the mass eigenvectors by requiring
\beq\label{zero_3}
\left.   \gh_{\phi_{i} \phi_{j}}(p) u^{(i)}_{j} \right|_{p^2 = (p^*_{i})^2} = 0 
\eeq
where $\gh_{\phi_{i} \phi_{j}}(p)$ are the two-point functions 
for scalar fields including all the counter terms computed in the 
Sec.~\ref{Generalsolution}. In the case of the Goldstone fields $p^*_{i}=0$. 
As a consequence, the eigenvectors $u^{(i)}_j$ are functions 
of the free parameters of the Higgs potential $\mu_{ij}$ and $\lambda_{ijkl}$ and of the 
allowed w.f.r. $Z^\phi_0,  Z^\phi_{a_\smalla}$ of Eq.~(\ref{tree.7}).   

Finally we can rewrite the WTI (\ref{zero_2}) in the mass eigenstates 
$\phi'_{i} = \sum_{j} u^{(i)}_{j}  \phi_{i}$ and correspondently for the 
background fields $\hat{\phi}'_{i} = \sum_{j} u^{(i)}_{j} \hat{\phi}_{i}$
\bea\label{zero_4}
    \mbox{\large \bf W}_{a}(\gh) = \left({\rm gauge-fermion-ghost~terms} \right)  
+ {\cal T}^{a}_{ij} \left[ (\phi'+v')_{j}  \fdu{\gh}{\phi}{',i} +  
(\h{\phi}'+v')_{j} \fdu{\gh}{\h{\phi}}{',i} \right] = 0  
\eea 
where ${\cal T}^{a}_{ij} = u^{(i)}_k (et)^{a}_{kl}  (u^{-1})^{(l)}_j$ 
coincide, at tree level with the physical angles among 
the scalar fields, and at the quantum level they are functions of the free parameters 
$\mu_{ij}, \lambda_{ijkl}, Z^\phi_0,  Z^\phi_{a_\smalla}$. In order to impose normalization 
conditions only on the angle the matrix $u^{(i)}_j$ should be unitary: 
$u^{(i)}_k u^{(k)}_j = \delta_{ij}$. 
This can be achieved by adjusting the free parameters 
$\mu_{ij}, \lambda_{ijkl}, Z^\phi_0,  Z^\phi_{a_\smalla}$. 

The mixing angle can now be fixed by requiring that 
at the quantum level  ${\cal T}^{a}_{ij}$ coincide with the renormalized 
mixing angles. Only by representing the WTI on the basis of the mass 
eigenstates the quantities ${\cal T}^{a}_{ij}$ can be identified with renormalized mixing angles 
avoiding the problem of w.f.r. for scalar fields. 

\subsection{Gauge Boson Sector}    
\label{sub.sec:trans_gauge_BFM}    
    
For the transverse components of the two-point functions for the  
gauge bosons we have to impose the following   
normalization conditions    
\beq\label{nor.10}    
\left. \g^{(n), T}_{A^{a'} A^{b'}}(p)\right|_{p^2=0} = 0 \hspace{2cm}    
\left. \g^{(n), T}_{A^{a'} Z^{b''}}(p)\right|_{p^2=0} = 0.     
\eeq    
in order to avoid the IR divergences. Then, recalling that the BFM leaves 
only $Z^{W}_\smalls$ as free parameters for each gauge multiplet, it seems   
almost impossible that these normalization conditions can be achieved.    
However, as we will show below, for the SM (a non-semisimple gauge model
with a single abelian factor) the Eqs.~(\ref{nor.10}) can be solved.    
 
We define  the new two-point functions by rotating, at fixed $p$, the matrices 
of two-point functions matrix of the mass eigentstates      
\bea\label{diag_two}    
\left(    
  \begin{array}{cc}    
\g^T_{{W}^{c_\smalla} W^{d_\smalla}}(p) & \g^T_{{W}^{c_\smalls} W^{d_\smalla}}(p) \\   & \\   
\g^T_{{W}^{c_\smalla} W^{d_\smalls}}(p) & \g^T_{{W}^{c_\smalls} {W}^{d_\smalls}}(p)    
  \end{array}     
\right)    
= {\cal R}^{-1}(p)      
\left(    
  \begin{array}{cc}    
\g^T_{{A}^{a'} A^{b'}}(p) & \g^T_{{Z}^{a''} A^{b'}}(p) \\   & \\   
\g^T_{{A}^{a'} Z^{b''}}(p) & \g^T_{{Z}^{a''} {Z}^{b''}}(p)    
  \end{array}     
\right)    
{\cal R}(p) \nonumber    
\eea      
where the matrix ${\cal R}(p)$ of two-point functions has the following
block structure    
\bea\label{matri}    
{\cal R}(p) = \left(    
  \begin{array}{cc}    
{\cal R}_{b' b_\smalla}(p) & {\cal R}_{b'' b_\smalla}(p) \\   & \\   
{\cal R}_{b' b_\smalls}(p) & {\cal R}_{b'' b_\smalls}(p)    
  \end{array}     
\right)    
\eea    
The indices $a',b'$ run over $1, \dots, N_\gamma$ (where $N_\gamma$ is
the number of massless gauge field),  $a'',b''$ run over $1, \dots, N_\smallz$ (where $N_\smallz$ is
the number of massive gauge field; $a_\smalla, b_\smalla, a_\smalls, b_\smalls$ run over  $1,
\dots, N_\smalla$ and  $1, \dots, N_\smalls$ where $N_\smalla,N_\smalls$ are  the number
of the abelian factors and of simple factors, respectively. 

The presence of massless particles is detected by 
computing the eigenvalues of the two-point function matrix
(\ref{diag_two}) and the STI ensure that the number of those particles,
namely $N_\gamma$ is left invariant under renormalization. This means
that 
$$
{\rm Rank }\left(    
  \begin{array}{cc}    
\g^T_{{W}^{c_\smalla} W^{d_\smalla}}(0) & \g^T_{{W}^{c_\smalls} W^{d_\smalla}}(0) \\   & \\   
\g^T_{{W}^{c_\smalla} W^{d_\smalls}}(0) & \g^T_{{W}^{c_\smalls} {W}^{d_\smalls}}(0)    
  \end{array}     
\right) = {\rm Rank } \left(\g^T_{{Z}^{a''} {Z}^{b''}}(0)\right) 
$$
which implies that if $\g^T_{{Z}^{a''} {A}^{b'}}(0)=0$, $\forall a''=1, \dots,
N_\smallz ~\cup~ \forall a''=1, \dots, N_\gamma$, then $\g^T_{{A}^{a'} {A}^{b'}}(0)=0$,
$\forall a',b'=1, \dots, N_\gamma$. Therefore the normalization conditions
(\ref{nor.10}) are not independent and they require fewer free parameters to be fixed. 
In particular, recalling that we have one free w.f.r. 
for each simple factor\footnote{we use the 
definition  $Z^W_{S}= 1 + \sum_{n \geq 1} \hbar^{(n)} \del Z^{W,(n)}_\smalls$.}
$Z^W_{S}$ and, by using the invertibility of the matrix
${\cal R} \equiv {\cal R}(0)$ computed at zero momentum, 
the normalization conditions $\g^{T,(n)}_{{Z}^{a''} {A}^{b'}}(0)=0$, $\forall a''=1, \dots,
N_\smallz ~\cup~ \forall b'=1, \dots, N_\gamma$ corresponds to 
\beq\label{nor_con_1}
\Sigma^{T,(n)}_{ {A}^{a'} {Z}^{b''}}(0) -
M^2_{Z_{b''}} \sum_{S=1}^{N_\smalls} 
\del Z^{W,(n)}_\smalls \sum_{c_\smalls} {\cal R}_{a' c_\smalls} {\cal R}^{-1}_{c_\smalls b''} = 0
\eeq
where $\Sigma^{T,(n)}_{{Z}^{a''} {A}^{b'}}(0)$ is the renormalized
amplitude up to $n-1$ order and ${\cal R}^{-1}_{c_\smalls b'} \equiv ({\cal R}^{-1})_{c_\smalls b'}$.
This equation is solved in terms for the unknown quantities $\del
Z^{W,(n)}_\smalls$ if $N_\smalls \geq N_\gamma \times N_\smallz$. The non-singularity
of $\sum_{c_\smalls} {\cal R}_{a' c_\smalls} {\cal R}^{-1}_{c_\smalls b''}$ is insured by
the the fact that ${\rm Rank } \left(\g^T_{{Z}^{a''}{Z}^{b''}}(0)\right)= N_\smallz$ 
and by the invertibility of ${\cal R}$ (existence of non-trivial tree-level mixing). 
As an example for the SM, where $N_\gamma = N_\smallz= N_{SU(2)} = 1$, 
$Z^W_{SU(2)} \equiv Z^W_3$ and $a'=A, b''=Z, c_\smalls=3$ we get 
\beq\label{nor_con_SM}
\del Z^W_3 = \frac{1}{\left(M^2_\smallz {\cal R}_{A3} {\cal R}_{3Z}^{-1} \right)} 
\Sigma^{T,(n)}_{AZ}(0)  
\eeq
where ${\cal R}_{A3} = s_\smallw, {\cal R}_{3Z}^{-1} =c_\smallw$. 

Besides the conditions  (\ref{nor.10}) we have also to impose some mass   
normalization conditions. 
Therefore, we consider the sub-matrix  $\g^{(n),T}_{Z_{a''} Z_{b''}}$ 
involving only the massive gauge fields.    
This remaining matrix is real\footnote{The two-point functions are real   
only below the particle creation thresholds.}, symmetric and non-diagonal, and it     
can be diagonalized by means of an orthogonal transformation restricted only to the     
subspace of massive fields. As a consequence, the only relevant information     
is contained in its eigenvalues $\lambda_{Z^{a''}}(p)$ and, in  
particular, in their zeros $p^*_{Z_{a''}}$.    
Notice that, for the transverse components of the two-point for gauge bosons, the    
existence of  $N_\smallz$ (number of massive gauge bosons) non-trivial   
eigenvalues $\lambda^{(n)}_\smallz(p)$     
is ensured by the STI.  
Then we compare     
the real part (in the case of unstable particles) of zeros $p^*_{Z_{a''}}$ with    
the experimental mass $M_{Z_{a''}}$    
$$    
{\cal R}e(p^*_{Z_{a''}} ) = M_{Z_{a''}}.     
$$    
This definition is independent
of the gauge parameters $\xi,\rho$  \cite{sirlin,stuart} and of the mixings. 
At the tree level the mixing of the gauge bosons in the SM concerns only the   
Weinberg angle, but, at higher orders, the radiative corrections    
generate new mixings. Those mixings carry no physical information.    
Finally we can define the mass eigenstates which appear in the LSZ
formalism. They are obtained by requiring 
\beq\label{LSZ_gau}
\left(    
  \begin{array}{cc}    
0 & 0   \\   
0 & \g^T_{{Z}^{a''} {Z}^{b''}}(0)    
  \end{array}     
\right)  \left(\ber{c} u^{A,a'}_{b'} \\ u^{A,a'}_{b''} \eer\right) =0, ~~~~
\left(    
  \begin{array}{cc}    
\g^T_{{A}^{a'} A^{b'}}(p^*_{Z_{a''}}) & \g^T_{{Z}^{a''} A^{b'}}(p^*_{Z_{a''}}) \\   & \\   
\g^T_{{A}^{a'} Z^{b''}}(p^*_{Z_{a''}}) & \g^T_{{Z}^{a''} {Z}^{b''}}(p^*_{Z_{a''}})    
  \end{array}     
\right)  \left(\ber{c} u^{Z,a''}_{b'} \\ u^{Z,a''}_{b''} \eer\right) =0.  ~~~~
\eeq
As a consequence the metric is not diagonal
\beq\label{LSZ_metric}
\left(\ber{cc}  
(u^{A,a'}, u^{A,b'}) & (u^{A,a'}, u^{Z,b''}) \\
(u^{Z,a''}, u^{A,b'}) & (u^{Z,a''}, u^{Z,b''})
\eer\right) = \left(\ber{cc}  
\delta^{a',b'} & K^{a',b''} \\
K^{a'',b'}  & H^{a'',b''}
\eer\right)
\eeq
where $(.,.)$ is the scalar product, 
and no free parameter can be adjusted in order to achieve its diagonalization. 

\subsection{Fermion Sector}    
\label{sub.sec:fermionBFM}    
    
In the fermion sector we have several free parameters (the Yukawa 
couplings $Y^i_{IJ}$, and the w.f.r. $Z^{R/L}_{IJ},\bar{Z}^{R/L}_{IJ} $
for the fermions) which can be tuned in order to satisfy the normalization conditions. 
The free parameters are selected by the BRST symmetry and, in
particular, by the invariance under the background gauge 
transformations. As a consequence the normalization conditions 
must respect those symmetries to all orders. Moreover in the 
present sector there are two kinds of physical normalization 
conditions  which must be implemented in order to compare the model 
with the experimental data: the fermion masses and the mixing
parameters. 

At the tree level those parameters can be easily expressed
in terms of the Yukawa couplings. In fact we can 
diagonalize the mass matrices (after the spontaneous symmetry breaking) 
$M_{IJ} = Y^i_{IJ} v_i$ by bi-unitary
transformations $U^{R/L}_{IJ}$ acting on the space of fermions 
and the resulting eigenvalues $\lambda^{Y,i}_{I} v_i $ must be compared 
with the fermion masses $M_I$. 
As is well known the diagonalization of the Yukawa
matrices modifies the interaction terms ($\g^{\bar{\psi}W\psi}$) 
among the fermion and the charged gauge fields
\beq\label{ckm}
\g^{\bar{\psi}W\psi} = \int \dx 
\left(\bar{\psi}^{L}_{K} U^{L,*}_{I K} T^{L,a}_{IJ} U^{L}_{J M} \not\!
  W^{a} \psi^L_M
+ \bar{\psi}^{R}_{K} U^{R,*}_{I K} T^{R,a}_{IJ} U^{R}_{J M}  \not\!W^{a}
\psi^R_M + {\rm h.c.} \right)
\eeq
and the resulting matrices $U^{L/R,*}_{I K} T^{L/R,a}_{IJ} U^{L/R}_{J M}$ 
are interpreted as the CKM angles. In the case of the minimal SM, the matrices 
$U^{L}_{IJ}$ split into up and down parts and they commute with 
the third generator of the $SU(2)$ and with $U(1)$. Therefore we recover
the usual CKM matrix. Furthermore, since the right fermions 
couple only to the $U(1)$ gauge group (and to $SU(3)$),   
the $U^{R}_{IJ}$ commute and, therefore, no CKM 
is present for the right-handed part. Notice that the only relevant 
physical CKM angles are obtained by a suitable rephasing of fermions. 

At the quantum level we have to translate those requirements in
terms of normalization conditions. First we have to define the
generations intrinsically. At the tree level this can be easily done by 
considering the masses of the fermions and organizing the fermions 
according to the hierarchy of the masses. At the quantum level 
we consider the two-point functions ${\g}_{ \bar{\psi}_{I} {{\psi}}_{J} }(p)$
and their decomposition into Lorentz-invariant terms:    
\beq\label{ferm_decomp}    
{\g}_{ \bar{\psi}_{I} {{\psi}}_{J} }(p) = \Sigma^{D}_{IJ}(p) P_L + \not\!p P_L     
\Sigma^{L}_{IJ}(p) + P_R \Sigma^{D,*}_{JI}(p) + \not\!p  P_R \Sigma^{R}_{IJ}(p)    
\eeq    
where $\Sigma^{D}_{IJ}(p), \Sigma^{L}_{IJ}(p), \Sigma^{R}_{IJ}(p)$ are     
Lorentz invariant matrices. In order to avoid heavy notation we 
suppress the labels $\psi_I$ and only the index $I$ of the flavor  
(and of the color, in the case of quarks) is used. 
Then, we can form the following   
new Lorentz invariant hermitian (below particle production thresholds)  matrix \cite{donoghue}  
\beq\label{inv_mat}     
{\cal K}_{I J}(p) \equiv p^2 \Sigma^{L}_{I J}(p)  -   
  \Sigma^{D,*}_{K I} (p) \Sigma^{R,-1}_{K L} (p) \Sigma^{D}_{L J}(p)  
\eeq    
and its eigenvalues $\lambda_{I}(p)$ correspond to    
fermion two-point functions with flavour mixing. Moreover    
the zeros of eigenvalues $\lambda_I (p^*_{I})=0$, or equivalently   
the zeros of the determinant   
\beq\label{nor_con_ferm}    
\left. {\rm Det} \left[ {\cal K}_{I K}(p) \right] \right|_{p^2 = (p^*_{I})^2}  
= 0,      
\eeq    
are gauge independent\footnote{The definition of masses has been
 taken into account in  \cite{sirl,stuart} and in the case of fermion 
 in \cite{donoghue}, where  
 the problem of gauge parameter independence is considered.   
 From a more rigorous point of view we refer to  papers \cite{kraus}. 
 An extension to flavour mixings is discussed in \cite{gg_1}. } 
and their real and imaginary parts are compared  
with the masses of fermions $M_{I}$ and their width $\g_{\psi_I}$. This  
fixes a possible choice of normalization conditions for quark and  
lepton masses. Although the QCD corrections affect the pole structure,  
 this  definition is used in the explicit computations \cite{two_loop}
of electroweak radiative corrections (switching off the strong interactions).   
As a consequence we are able to organize the fermions into generations
intrinsically by fixing the zeros of the eigenvalues of the matrix ${\cal K}_{I J}(p)$
and this choice fixes some of the free parameters $Y^{i}_{IJ}$.   

At this point we are able to define the mass eigenstates for the
fermion fields in the case of mixing. We define mass eigenstates by 
\beq\label{mass_eigen}
\left. \bar{u}^{(I)}_{K} \g_{KJ}(p) \right|_{\not p = \not p^*_I} =0, ~~~~
\left.  \g_{KJ}(p) {u}^{(I)}_{J} \right|_{\not p = \not p^*_I
} =0
\eeq 
where $ \bar{u}^{(I)}, {u}^{(I)}$ are independent eigenvectors due to
the CP violation. Those vectors depend of the free parameters 
$Y^i_{IJ}, Z^{L/R}_{IJ}, \bar{Z}^{L/R}_{IJ}$
(cf. Eqs.~(\ref{tree.13})-(\ref{tree.10.2})) and the metric
$(\bar{u}^{(I)}, u^{(J)})$ is not diagonal at the quantum level. As a 
consequence we can define the mass eigenfield by projecting the 
fermion fields $\psi_I, \bar{\psi}_I$ onto the vectors  $\bar{u}^{(I)},
{u}^{(I)}$: $f'_I =  \bar{u}^{(I)}_{K} \psi_K$ and 
$\bar{f}'_I = \bar{\psi}_J {u}^{(I)}_J$. Notice that we can introduce the matrices 
${\cal U}_{I J} = u^{(I)}_J$ and  $\bar{\cal U}_{I J} = \bar{u}^{(I)}_J$ 
which are complicate functions of the 
free parameters  $Y^i_{IJ}, Z^{L/R}_{IJ}, \bar{Z}^{L/R}_{IJ}$ (except those which are already 
fixed by the mass renormalization) and we require that the matrices are unitary in order to 
single out only the relevant mixing angles. It is easy to show that the system is invertible at tree level 
and, by the theorem of the implicit function for formal power series, this is true also at the quantum 
level. 

By following Aoki {\it et al.} we would be forced to impose
normalization conditions on the mass eigenstates $\bar{u}^{(I)},{u}^{(I)}$, 
but, due to the constraints of BFM, this can not 
be achieved without deforming the WTI. Therefore we choose to fix the 
remaining parameters, namely the CKM matrix, by using the WTI themselves. 
In fact, by recalling that the CKM angles appear as
couplings among charged gauge fields and fermions and that the couplings
among background and quantum gauge fields with 
fermions coincide in the BFM, we are able to fix the CKM by imposing
the WTI in the mass eigenstate representations ($f'_I, \bar{f}'_I$)
\bea\label{WTI_mass}
 \mbox{\large \bf W}_{a}(\g) &=& \left({\rm scalar~+~gauge~+~ghost~terms }\right) + \\
&+ & 
(eT)^{a}_{L,IJ} \left[ {\b{f}'}^{L}_{I'} \bar{\cal U}^{L,*}_{I I'}
(\bar{\cal U}^{L})_{J J'} \dms{\fdaL{\g}{\b{f}'}{R}{J'}}
+ 
\dms{\fdaR{\g}{f'}{L}{I'}} {\cal U}^{L}_{I' I} ({\cal U}^{L,*})_{J' J}  {{f}'}^{L}_{J'} 
\right] + ({\rm  L} \rightarrow {\rm R})  
\nonumber \\
&=& \left({\rm scalar~+~gauge~+~ghost~terms}\right)   +
\nonumber \\
&+& 
\left[ {\b{f}'}^{L}_{I'} \bar{V}^{L,a}_{I'J'} \dms{\fdaL{\g}{\b{f}'}{R}{J'}}
+
\dms{\fdaR{\g}{f'}{L}{I'}} V^{L,a*}_{J' I'} {f'}^{L}_{J'}
\right] + ({\rm  L} \rightarrow {\rm R})
= 0 \nonumber
\eea
where only the fermion terms are displayed and $V^{L,a}_{I'J'} = 
{\cal U}^*_{K I'} (eT)^{a}_{L, K M} {\cal U}_{M I'}$ (and analogously for the right part and 
for $\bar{V}^{L,a}_{I'J'}$) 
are the renormalized CKM matrices. The above equation fixes the free parameters
$Y^i_{IJ}$ up to unphysical rescaling and up to unphysical w.f.r.. 

Analogously to the scalar mixing the quantities $V^{L,a}_{I'J'} = 
{\cal U}^*_{K I'} (eT)^{a}_{L, K M} {\cal U}_{M I'}$ and $\bar{V}^{L,a}_{I'J'}$
can be identified with the renormalized 
mixing angles avoiding the problem of the w.f.r. and the mixing among 
physical and unphysical quantities. 

\subsection{Background Field Renormalization}    
\label{bf_renorm}
    
In this section we discuss the normalization conditions for Green's functions 
with external background fields.     
As is clear from \cite{bkg} some Green's functions with     
external background gauge and scalar fields are divergent, then     
they require their proper normalization conditions. However two orthogonal approaches      
can be assumed: the first one is to fix these normalization conditions     
independently of those of quantum fields, the second one corresponds    
to choose the normalization conditions which respect the form of WTI    
(\ref{wt_S}) and (\ref{wt}).       
The first alternative has been adopted in \cite{krau_ew}     
and it requires some ``unphysical'' normalization conditions for the    
background fields; namely the free parameters $X_\smalls,X_{ij}$ 
are fixed independently of w.f.r. $Z^W_\smalls, Z^\phi_{ij}$ of quantum fields.        
Here, we choose the second alternative in order to minimize the number    
of normalization conditions and the amount of work to renormalize the model.    
    
Furthermore, since the BFM is a powerful tool to compute S-matrix    
elements from Green's functions with external bosonic background fields \cite{ags},    
it is crucial to show that the normalization conditions imposed on those    
Green's functions directly correspond to physical normalization    
conditions on the poles of quantum Green's functions. Therefore, in the    
next paragraphs we explicitly prove the degeneracy among the zeros of two-point    
functions with external background fields and the corresponding two    
point functions with quantum fields.     
    
The parameters  $X_\smalls,X_{ij}$ are respectively related to the two-point    
functions with external BRST-sources     
${\g}_{{\gamma}^{a_\smalls}_{\mu} {\Om}^{b_\smalls}_{\nu}}(p), {\g}_{{\gamma}_{i}     
{\Om}_{j}}(p)$ which are superficially divergent;  on the contrary,    
the other two-point functions     
$\g_ {{\gamma}_{i} {\Om}^{b_\smalls}_{\nu}}(p),     
\g_ {{\gamma}^{a_\smalls}_{\mu} {\Om}_{j}}(p)$     
are superficially convergent by power counting and Lorentz covariance. To     
compare these Green's functions we firstly use the WTI to show that the    
renormalization of     
$\g_{\h{W}^{a_\smalls}_{\mu} {\gamma}^{b_\smalls}_{\nu}     
{c}^{c_\smalls}}$, $\g_{\h{\phi}_{j} {\gamma}_{k} {c}^{c_\smalls}}$     
depends on the renormalization of vertex functions with external quantum    
fields.     
In fact, by taking the derivative of (\ref{wt_S}) with respect to     
$\tilde{\gamma}^{b_\smalls}_{\nu}(p)$ and $\tilde{c}^{c_\smalls}(q)$, we    
get\footnote{In the present section we re-absorb the couplings    
  $e_\smalls,e_{a_\smalla b_\smalla}$ from WTI by rescaling the gauge fields and we impose that    
  the v.e.v. $v_i$ in the WTI coincides with its tree level value.}     
\bea\label{wti_bakg}    
\left. 
\frac{\del^{2}  \left( \mbox{\large \bf W}_{a_\smalls} \g \right) 
}{\del \tilde{\gamma}^{c_\smalls}_{\mu}(p) \del \tilde{c}^{b_\smalls}(q)}  
\right|_{\phi=0}
&=& i (p + q)^{\mu} \left(\g_{\h{W}^{c_\smalls}_{\mu}   
{\gamma}^{a_\smalls}_{\nu}\,     
{c}^{b_\smalls}}(p,q) + \g_{{W}^{c_\smalls}_{\mu} {\gamma}^{a_\smalls}_{\mu}\,     
{c}^{b_\smalls}}(p,q) \right) \nonumber \\  &+&    
v_{i}t^{c_\smalls}_{ij} \left(\g_{\h{\phi}_{j} {\gamma}^{a_\smalls}_{\mu}\,     
{c}^{c_\smalls}}(p,q) + \g_{{\phi}_{j} {\gamma}^{b_\smalls}_{\mu}\, {c}^{c_\smalls}}(p,q) \right)     
\vspace{.1cm} \\  
&+&    f^{c_\smalls a_\smalls d_\smalls} \g_{{\gamma}^{d_\smalls}_{\mu} {c}^{b_\smalls}}(-p) + 
f^{c_\smalls  b_\smalls d_\smalls} \g_{{\gamma}^{a_\smalls}_{\mu} {c}^{d_\smalls}}(q) = 0.     
\nonumber \eea    

Exploring the region of very high momenta and 
using the Weinberg's theorem \cite{wein} from the 
expression 
\bea\label{loc_wti_bakg}    
\lim_{\rho \rightarrow \infty}   \de_{p^\mu} \left. \frac{\del^{2} \left( 
\mbox{\large \bf W}_{a_\smalls} \g \right) }{\del    
  \tilde{\gamma}^{c_\smalls}_{\mu}(\rho \,p)     
\del \tilde{c}^{b_\smalls}(- \rho \, p)} \right|_{\phi=0}
= \g^{\infty}_{\h{W}^{a_\smalls}_{\mu} {\gamma}^{b_\smalls}_{\mu}     
{c}^{c_\smalls}}  + \g^{\infty}_{{W}^{a_\smalls}_{\mu} {\gamma}^{b_\smalls}_{\mu}     
{c}^{c_\smalls}} +     
f^{a_\smalls b_\smalls d_\smalls} \de_{p^{\mu}}  \g^{\infty}_{{\gamma}^{b_\smalls}_{\mu}     
{c}^{c_\smalls}},
\eea    
we can isolate the asymptotic behaviour of the Green's    
like $\g^{\infty}_{\h{W}^{a_\smalls}_{\mu} {\gamma}^{b_\smalls}_{\nu}     
{c}^{c_\smalls}}$ to be compared 
with the renormalized Green's functions with external quantum fields. 

In the same way, by considering the derivative of WTI (\ref{wt_S}) with     
respect to $\tilde{\gamma}_{i}(p)$ and $\tilde{c}^{c_\smalls}(q)$, we obtain    
\bea\label{loc_wti_bakg_2}    
\lim_{\rho \rightarrow \infty}  
\left. \frac{\del^{2}\left( \mbox{\large \bf W}_{a_\smalls} \g \right)}{\del \tilde{\gamma}^{i}(\rho \, p)     
\del \tilde{c}^{b_\smalls}(- \rho \, p)} \right|_{\phi=0} = 
v_{j}\, t^{a_\smalls}_{jk} \g^{\infty}_{\h{\phi}_{k} {\gamma}_{i}     
{c}^{b_\smalls}}  + v_{j}\, t^{a_\smalls}_{jk}    
\g^{\infty}_{{\phi}_{k} {\gamma}_{i} {c}^{b_\smalls}}  +     
t^{a_\smalls}_{ik} \g^{\infty}_{{\gamma}_{k} {c}^{b_\smalls}}  + f^{a_\smalls b_\smalls d_\smalls}    
\g^{\infty}_{{\gamma}_{i} {c}^{d_\smalls}},     
\eea    
which relate the $\g^{\infty}_{\h{\phi}_{k} {\gamma}_{i}     
{c}^{b_\smalls}}$ to Green's functions with quantum fields.     
Equations (\ref{loc_wti_bakg})  and (\ref{loc_wti_bakg_2}) are the     
all-order version of relations between     
 $X_\smalls, X_0, {\cal T}_{ij}$ and $Z_\smalls, Z_0, {\cal R}_{ij}$,     
and they reduce to them at the tree level.     
    
Furthermore, the Green's functions $ \g_{{\gamma}^{b_\smalls}_{\mu}     
{c}^{c_\smalls}}$,$\g_{{\gamma}_{k} {c}^{c_\smalls}}$,    
$\g_{{W}^{a_\smalls}_{\mu}    
  {\gamma}^{b_\smalls}_{\nu}}$, $\g_{{\phi}_{k}{\gamma}_{i} {c}^{c_\smalls}}$      
contain the radiative corrections to the     
structure constants $f^{a_\smalls b_\smalls c_\smalls}$ of the gauge group ${\cal G}$     
and to the generators $t^{a}_{ij}$. Therefore, their correct    
renormalization ensures that the model has the same gauge group and the    
same matter representation to all orders \cite{henri,henn}. 
Moreover, in the case of non-invariant regularization    
schemes, general non-invariant counterterms    
\beq\label{nor.75}    
{\g^{C.T.}_{BRST}} = \int \dx \left( x^{1}_{a_\smalls b_\smalls}     
\de_{\mu} c^{a_\smalls} \gamma^{\mu}_{b_\smalls} +     
 x^{2}_{a_\smalls b_\smalls c_\smalls}     
c^{a_\smalls} \gamma^{\mu}_{b_\smalls} W^{c_\smalls}_{\mu} +     
x^{3}_{a_\smalls i} c^{a_\smalls} \gamma^i +     
x^{4}_{a_\smalls i j} c^{a_\smalls} \gamma^i \phi^j    
\right)   
\eeq    
must be added to the action in order to restore the STI and the WTI. 
The coefficients     
$x^{1}_{a_\smalls b_\smalls}, x^{2}_{a_\smalls b_\smalls c_\smalls}$,     
$x^{3}_{a_\smalls i}, x^{4}_{a_\smalls i j}$     
are computed in terms of the breakings to the STI. This necessarily    
implies the analysis of STI for the three-points functions involving,    
at least, two ghost fields and it requires a great amount of work to    
compute these coefficients at higher orders. However,     
by using the WTI  (\ref{loc_wti_bakg}), (\ref{loc_wti_bakg_2})     
$x^{1}_{a_\smalls b_\smalls}, x^{2}_{a_\smalls b_\smalls c_\smalls}$,     
$x^{3}_{a_\smalls i} , x^{4}_{a_\smalls i j}$ can be computed from the    
breakings of WTI, and only some STI for two-point functions are    
needed \cite{amt}. Furthermore, since up to this moment, only a subset of 
two-loop electroweak    
radiative corrections have been computed \cite{two_loop}, two-loop and    
one-loop WTI and one-loop STI     
are sufficient to evaluate all possible counterterms, including those of Eq.~(\ref{nor.75}).     
    
To complete our program we have to show the relation among the     
Green's functions $ \g_{\h{\phi}_{j} {\gamma}_{k}     
{c}^{c_\smalls}}$, $\g_{\h{W}^{a_\smalls}_{\mu} {\gamma}^{b_\smalls}_{\nu}     
{c}^{c_\smalls}}$ and the two-point functions 
${\g}_{{\gamma}^{a_\smalls}_{\mu} {\Om}^{b_\smalls}_{\nu}},    
{\g}_{{\gamma}_{i} {\Om}_{j}}$. For this purpose we consider the    
STI    
\bea\label{sti_bakg}    
\g_{\h{W}^{a_\smalls}_{\mu} {\gamma}^{b_\smalls}_{\nu}     
{c}^{c_\smalls}}(p,q) & = & \g_{\Om^{a_\smalls}_{\mu} {\gamma}^{d_\smalls}_{\rho}(p+q)}     
\g_{{W}^{d_\smalls}_{\rho} {\gamma}^{b_\smalls}_{\nu}(p)     
{c}^{c_\smalls}(q)} + \nonumber \\     
& + &      
 \g_{\zeta^{d_\smalls} \Om^{a_\smalls}_{\mu}{(p)} {c}^{c_\smalls}{(q)} }     
\g_{{\gamma}^{b_\smalls}_{\nu} {c}^{d_\smalls}(q)} + 
\g_{\Om^{a_\smalls}_{\mu} {\gamma}_{k}(p+q)}     
\g_{{\phi}_{k} {\gamma}^{b_\smalls}_{\nu}(p)     
{c}^{c_\smalls}(q)}    
\nonumber \\ && \\    
\g_{\h{\phi}_{i} {\gamma}_{k}(p) {c}^{c_\smalls}(q)}     
& = & \g_{\Om_{i} {\gamma}^{d_\smalls}_{\rho}(p+q)}     
\g_{{W}^{d_\smalls}_{\rho} {\gamma}_{j}(p)     
{c}^{c_\smalls}(q)} + \nonumber \\     
& + &      
\g_{\zeta^{d_\smalls} \Om_{i}(p)\, {c}^{c_\smalls}(q) }     
\g_{{\gamma}_{j} {c}^{d_\smalls}(p)} + \g_{\Om_{i} {\gamma}_{k}(p+q)}     
\g_{{phi}_{k} {\gamma}_{j}(p) {c}^{c_\smalls}(q)}     
\nonumber    
\eea     
obtained by differentiating (\ref{st}) with respect to $\tilde{\Om}^{a_\smalls}_{\mu}(-p-q),     
\tilde{\gamma}^{b_\smalls}_{\nu}(p)$, $\tilde{c}^{c_\smalls}(q)$    
and with respect to $\tilde{\Om}_{i}(-p-q),     
\tilde{\gamma}_{j}(p)$, $\tilde{c}^{c_\smalls}(q)$.    
Since $\g_{\zeta^{d_\smalls} \Om^{a_\smalls}_{\mu}{c}^{c_\smalls} }$ and 
$\g_{\zeta^{d_\smalls}\Om_{i}{c}^{c_\smalls} }$     
are superficially convergent by power counting we immediately deduce the relations    
\bea\label{sti_bakg_loc}    
\g^{\infty}_{\h{W}^{a_\smalls}_{\mu} {\gamma}^{b_\smalls}_{\nu}     
{c}^{c_\smalls}}  = \g^{\infty}_{\Om^{a_\smalls}_{\mu} {\gamma}^{d_\smalls}_{\rho}}     
\g^{\infty}_{{W}^{d_\smalls}_{\rho} {\gamma}^{b_\smalls}_{\nu}     
{c}^{c_\smalls}},     
~~~~~\g^{\infty}_{\h{\phi}_{i} {\gamma}_{k} {c}^{c_\smalls}}     
 = \g^{\infty}_{\Om_{i} {\gamma}^{d_\smalls}_{\rho}}     
\g^{\infty}_{{W}^{d_\smalls}_{\rho} {\gamma}_{j}     
{c}^{c_\smalls}}. 
\eea     
These equations fix     
completely the local parts of Green's with background fields in terms     
of the Green's functions with quantum fields.     
Furthermore we have to recall that the renormalization of $\g_{\h{\phi}_{j} {\gamma}_{k}     
{c}^{c_\smalls}},  \g_{\h{W}^{a_\smalls}_{\mu} {\gamma}^{b_\smalls}_{\nu}     
{c}^{c_\smalls}}$ depends on the normalization conditions for     
$\g_{{\gamma}^{b_\smalls}_{\nu} {c}^{c_\smalls}}$ and    
$\g_{{\gamma}_{k}{c}^{c_\smalls}}(-p)$. The latter essentially corresponds to     
the w.f.r. for the gauge fields and for the scalar fields, and, consequently, the     
normalizations for the Green's functions with background fields depend on these     
w.f.r. as the tree level relations (\ref{tree.7}). In conclusion, the    
renormalization of the background fields $X_\smalls, X_{ij}$ are unphysical     
parameters and they are fixed to the renormalization of the quantum    
fields. However we have also to recall that the present derivation has    
been performed by rescaling the gauge fields and removing the gauge    
couplings $e_\smalls, e_{a_\smalla b_\smalla}$ from the WTI. This means that the    
renormalization of the background fields depends on the renormalization     
of the quantum fields only up to the renormalization of the gauge    
couplings and up to the vacuum expectation value $v_i$. And this     
is the well-known result presented into \cite{bkg}.    
    
In the last part of the present section, we discuss the relation among    
the zeros of the background two-point functions, $\g_{\h{W} \h{W}}$, 
and the quantum two   point functions. In particular, we prove that the zeros of both    
Green's functions coincide. This fact has a simple application in the    
computation of radiative corrections with the BFM \cite{msbkg}: by fixing the    
mass renormalization on the zeros (or on the     
complex zeros in the case of unstable particles) of  background    
two-point functions, the zeros of quantum two-point functions are also    
consistently fixed.  Needless to say, in several applications the computation of    
background two-point functions is enough to derive the complete S-matrix    
element for the process.     
    
At first,  we consider the background two-point functions for     
gauge fields and the derivatives of STI with respect     
to the fields $\tilde{\Om}^{a_\smalls}_{\mu}(p), \tilde{W}^{a_\smalla}_{\nu}(-p)$     
\beq\label{sti_bakg_two_1}    
\g_{\hat{W}^{a_\smalls}_{\mu} W^{b_\smalla}_{\nu}}      
 = \g_{\Om^{a_\smalls}_{\mu} {\gamma}^{d_\smalls}_{\rho}}     
\g_{{W}^{d_\smalls}_{\rho}  W^{b_\smalla}_\nu} +       
\g_{\Om^{a_\smalls}_{\mu} {\gamma}_{k}}      
\g_{{\phi}_{k} W^{b_\smalla}_\nu}    
\eeq     
where we have suppressed the argument since they are all two-point    
functions with same momentum. The two-point functions    
$\g_{\Om^{a_\smalls}_{\mu} {\gamma}^{d_\smalls}_{\rho}}$,  
$\g_{\Om^{a_\smalls}_{\mu} {\gamma}_{k}}$      
have been already discussed in the previous paragraphs and only the    
quantum abelian gauge fields $W^{a_\smalla}_{\mu}$ enter into the discussion,    
since their background are completely decoupled.     
In the same way, we get     
\bea\label{sti_bakg_two_2}    
\g_{\hat{W}^{a_\smalls}_{\mu} \h{W}^{b_\smalls}_{\nu}}(p)      
& = & \g_{\Om^{a_\smalls}_{\mu} {\gamma}^{d_\smalls}_{\rho}}(p)      
\g_{{W}^{d_\smalls}_{\rho}  \h{W}^{b_\smalls}_\nu}(p) +      
\g_{\Om^{a_\smalls}_{\mu} {\gamma}_{k}}(p)      
\g_{{\phi}_{k} \h{W}^{b_\smalls}_\nu}(p) \vspace{.5cm} \\    
\g_{\hat{W}^{a_\smalls}_{\mu} W^{b_\smalls}_{\nu}}(p)      
& = & \g_{\Om^{a_\smalls}_{\mu} {\gamma}^{d_\smalls}_{\rho}}(p)      
\g_{{W}^{d_\smalls}_{\rho}  W^{b_\smalls}_\nu}(p) +       
\g_{\Om^{a_\smalls}_{\mu} {\gamma}_{k}}(p)      
\g_{{\phi}_{k} W^{b_\smalls}_\nu}(p)       
\eea     
by differentiating with respect to     
$\tilde{\Om}^{a_\smalls}_{\mu}(p), \tilde{\h{W}}^{b_\smalls}_{\nu}(-p)$  and     
$\tilde{\Om}^{a_\smalls}_{\mu}(p), \tilde{W}^{b_\smalls}_{\nu}(-p)$.  Therefore, by     
decomposing the two-point functions into a transverse and a    
longitudinal part, the Eqs.~(\ref{sti_bakg_two_2}) can be put in the form     
\bea\label{matr_bak_1}    
\left(    
  \begin{array}{cc}    
\g^T_{{W}^{a_\smalla} W^{b_\smalla}}(p) & \g^T_{\hat{W}^{a_\smalls} W^{b_\smalla}}(p) \\ & \\    
\g^T_{\hat{W}^{a_\smalla} W^{b_\smalls}}(p) & \g^T_{\hat{W}^{a_\smalls} \h{W}^{b_\smalls}}(p)    
  \end{array}     
\right)     
=     
{\cal U}_{a c}(p)     
\left(    
  \begin{array}{cc}    
\g^T_{{W}^{c_\smalla} W^{d_\smalla}}(p) & \g^T_{{W}^{c_\smalls} W^{d_\smalla}}(p) \\  & \\  
\g^T_{{W}^{c_\smalla} W^{d_\smalls}}(p) & \g^T_{{W}^{c_\smalls} \h{W}^{d_\smalls}}(p)    
  \end{array}     
\right)    
{\cal U}_{d b}(p)     
\eea    
where the matrix ${\cal U}_{a b}(p) $ is     
\bea\label{def_matr}    
{\cal U}_{a b}(p) \equiv \left(    
  \begin{array}{cc}    
\del^{a_\smalla c_\smalla} & 0 \\    
0 &  \g_{\Om^{a_\smalls} {\gamma}^{c_\smalls}}(p)      
  \end{array}    
\right).    
\eea    
    
This matrix ${\cal U}_{a b}(p) $ is a non-singular matrix for $p\neq0$ and 
$\g_{\Om^{a_\smalls} {\gamma}^{c_\smalls}}(p)\neq 0$ starting at one loop. 
As a consequence, the zeros of the eigenvalues of the background    
two-point functions and the corresponding quantum partners 
coincide\footnote{Notice that, here, we used the representation $W,\hat{W}$ to denote the 
quantum field and the background field. However the S-matrix elements obtained in the 
BFM approach are constructed with quantum fields $Q^W = W-\hat{W}$ and with the 
background fields $\hat{W}$. Therefore the two-point functions with external background   
described in the literature \cite{bkg,msbkg,ags} are related to ours by the 
simple relation $\g^{Abbott}_{\hat{W}\hat{W}}(p) = 
\g_{\hat{W}\hat{W}}(p) + 2 \g_{\hat{W} {W}}(p) +  \g_{ {W} {W}}(p)$ where $\g^{Abbott}$ is 
defined in \cite{bkg}. By the Eq.~(\ref{matr_bak_1}), it is easy to show that 
$\g^{Abbott}_{\hat{W}\hat{W}}(p)$ has the same  zeros of $\g_{ {W} {W}}(p)$.}.     
This is a crucial property for the self-energies to ensure that     
the S-matrix defined with the BFM is equivalent to the conventional     
one \cite{ags}.      
    
Moreover we can immediately see     
that the diagonalization of the two-point functions matrix for the background fields     
does not imply the diagonalization of the two-point functions for     
the quantum fields. This is only achieved by a proper choice of     
normalization for the matrix ${\cal U}_{a b}(p) $. In particular the IR    
behavior might be different. Therefore, we have to study the two-point functions     
at zero momentum.     
By using the Eq.~(\ref{diag_two})
and by defining, in the same way, for the background fields, a matrix
$\h{\cal R}(p)$ 
the relation among the background two-point functions     
and the quantum two-point functions in the mass eigenstates is given by    
\bea\label{rel_mass_eigen}    
\left(    
  \begin{array}{cc}    
\g^T_{\h{A}^{a'} \h{A}^{b'}}(p) & \g^T_{\h{Z}^{a''} \h{A}^{b'}}(p) \\   & \\   
\g^T_{\h{A}^{a'} \h{Z}^{b''}}(p) & \g^T_{\h{Z}^{a''} \h{Z}^{b''}}(p)    
  \end{array}     
\right)     
= \left(\h{\cal R}^{-1} {\cal U} {\cal R} \right)(p)     
\left(     
\begin{array}{cc}    
\g^T_{{A}^{a'} A^{b'}}(p) & \g^T_{{Z}^{a''} A^{b'}}(p) \\   & \\   
\g^T_{{A}^{a'} Z^{b''}}(p) & \g^T_{{Z}^{a''} {Z}^{b''}}(p)    
 \end{array} \right)    
 \left({\cal R}^{-1} {\cal U} \h{\cal R} \right)(p)     
\nonumber    
\eea     
and we set $\widetilde{\cal U}(p)\equiv  \left({\cal R}^{-1} {\cal U}    
  \h{\cal R} \right)(p)$.      
Since the local WTI for the background gauge transformations     
ensure that    
\bea\label{mass_WTI}    
\left. \g^T_{\h{W}^{a'} \h{W}^{b'}}(p) \right|_{p^2 = 0} & = & 0~~~~a',b' = 1\dots N_{\gamma} \\    
\left. \g^T_{\h{W}^{a'} \h{W}^{b''}}(p) \right|_{p^2 = 0} & = & 
0~~~~a' = 1\dots N_{\gamma}, b''= 1 \dots N_\smallz     
\eea    
we immediately get    
\bea\label{norm_cond}    
&& \left. \g^T_{{W}^{a'} {W}^{b'}}(p) \right|_{p^2 = 0} = \sum_{a'', b''}    
\widetilde{\cal U}_{a' a''}(0)  \g^T_{\h{W}^{a''} \h{W}^{b''}}(0) \widetilde{\cal U}_{b'' b'}(0), 
~~~~a',b' = 1\dots N_{\gamma} \\    
&& \left. \g^T_{{W}^{a'} {W}^{b''}}(p) \right|_{p^2 = 0} =\sum_{a'', b''}     
\widetilde{\cal U}_{a' a''}(0)  \g^T_{\h{W}^{a''} \h{W}^{c''}}(0) \widetilde{\cal U}_{c'' b''}(0), 
~~~~a' = 1\dots N_{\gamma}, b''= 1 \dots N_\smallz \nonumber    
\eea    
from where $\widetilde{\cal U}_{a' a''}(0) = 0$ for 
$a' = 1\dots N_{\gamma}, a''= 1 \dots N_\smallz $.    
The matrices $\widetilde{\cal U}_{a' a''}$ depend on the     
already discussed Green's functions $\g^{T}_{\gamma_{a_\smalls} \Om_{b_\smalls}}(p)$   
computed at zero momentum.     
These Green's functions are superficially divergent (only the transverse parts, while the     
longitudinal part is finite) and are fixed by the normalization conditions (\ref{norm_cond}).   
However  we saw that the Green's functions  $\g^{T}_{\Om^{a_\smalls} \gamma_{b_\smalls}} $   
are related to the     
w.f.r. of the multiplets of quantum fields for each single simple sector ${\cal G}_\smalls$ of the     
group.    Hence the IR requirements given by (\ref{norm_cond}) for     
the quantum fields fix their w.f.r.s $Z_\smalls$. 

The BFM restricts the number     
of free parameters to a single w.f.r. for each simple factor and this bounds the number      
of massless particles which can be coupled to the model 
without modifying the STI and the local WTI.     
In fact increasing the number of massless particles the number of normalization     
conditions increases and the bound $N_{\gamma} N_\smallz \leq N_\smalls$ 
(where $N_\smalls$ is the     
number of simple factors) can not be respected any longer.     
    
\subsection{Renormalization of Couplings}

For what concerns the coupling renormalization it is very difficult to
discuss general properties, however we can divide the problem into 
two different categories: the couplings which are related to the
physical masses (in the case of spontaneous symmetry breaking) and those
which are true gauge couplings. For the first set, by using the gauge
invariant definition of masses we are able to provide a proper 
definition, but for the others a construction of an invariant
charge (see for example \cite{itzy} in chapter 13) is required \cite{eff}. 

The couplings $\lambda_{ijkl}, \mu_{ij}$ of the scalar sector are
related to the renormalization of the tadpoles and the masses of Higgs
fields (see Sec.~3.3). By identifying the masses of the Higgs fields
as the real part of the zeros of non-trivial eigenvalues of the
two-point function matrix some of the couplings $\lambda_{ijkl}, \mu_{ij}$ are
fixed in a gauge independent way. 

In the sector of the gauge bosons of the SM, $M_\smallz,M_\smallw$ define the
v.e.v. $v$ and the weak angle $c_\smallw$. The procedure 
of comparing the real part of the zeros of eigenvalues for the two-point 
function matrices with the physical masses is gauge parameter 
independent and therefore the weak angle  $c_\smallw$, defined by the
conditions $c^2_\smallw = \frac{M_\smallw^2}{M_\smallz^2}$ \cite{sirlin} is clearly independent of 
the gauge parameters. As discussed in the previous section, the zeros of
two-point functions with external background fields and with quantum ones coincide,
therefore, in order to fix the mass renormalization only the background
two-point functions are required. This has been already observed in reference 
\cite{msbkg} where the one loop analysis has been performed. 

However due the precision of the experimental
measurement of the $\mu$-decay, the Fermi's constant
$G_F$ is used to calculate $M_\smallw$. The definition of $G_F$ is given in terms of
the $\mu$-decay amplitude which is clearly gauge independent \cite{sirlin}. 

The QED charge $\alpha_{QED}$ is fixed in terms of the photon two-point 
function  computed at zero momentum transfer. It is easy to see that this
definition with the normalization conditions given in Sec.~3.4 is
gauge independent to all orders. 
The definition of the QCD coupling constant $\alpha_\smalls$ requires  a detailed
analysis. We refer to the literature \cite{bkg,eff,alpha_s} 
on the subject. 

Finally, concerning the charge renormalization, we have to stress again that the 
renormalization of the two-point functions for background gauge fields and 
for background scalar fields  can be immediately related to the 
charge renormalization or to the {\it v.e.v.} renormalization, respectively. 
This provides a very simple way to compute the beta functions 
of the Renormalization Group equation and the Callan-Symanzik equation. 
This advantage has been extensively used in the QCD \cite{bkg}, in the SM \cite{msbkg} 
and for extended models.  

    
\section{Anomalies}    
\label{ren_iden}   
  
The issue of algebraic renormalization of identities has been widely analyzed in the literature  
\cite{henri,bbbc_1,henn,krau_ew,pg_1,algeb}, therefore, here,  
we only recall the main results about the hard anomalies for non-semisimple 
gauge models with background fields and we discuss the problem of IR anomalies of STI 
in the minimal SM. Besides the hard anomalies and the IR anomalies, some soft anomalies can 
be present in the SM. However they are rule out by the Callan-Symanzik equation as shown 
in \cite{henri,bbbc_1,krau_ew}. 
  
\subsection{Hard Anomalies}    
   
Due to their relevance in the renormalization procedure, we describe briefly    
the hard anomalies of STI and of WTI for a non-semisimple gauge model.    

As proved in \cite{henri,henn} the structure of the general    
solution ${\cal A}$ of the consistency conditions   ${\cal S}_{0} ({\cal A}) =0$ 
\cite{brs,bec,wess,algeb} is given by    
\bea   
&& \hspace{-1cm} {\cal A}  = \dms{\int \dx} {\cal A}^{ABJ}(x) +   \dms{\int \dx}     
{\cal A}_{1}(x) +    
\dms{\int \dx}  {\cal A}_{2}(x) +   
{\cal S}_{0} \dms{\int \dx} \, {\cal B}(x) \label{ano_0} \\   
 &&\hspace{-1cm} {\cal A}^{ABJ}(x)  =  \sum_{i} r_i \epsilon^{\mu\nu\rho\sigma}    
\left[ D_{i}^{abc} c^{a} \de_{\mu} W^{b}_{\nu}  \de_{\rho} W^{c}_{\sigma} +    
\dms{\frac{F^{abcd}_i}{12}} (\de_{\mu} c^{a} ) W^{b}_{\nu} W^{c}_{\rho} W^{d}_{\sigma}    
\right]  \\   
 &&\hspace{-1cm} {\cal A}_{1}(x) =  c^{a_\smalla} {\cal R}_{a_\smalla}(x) \label{ano_1} \\   
 &&\hspace{-1cm} {\cal A}_{2}(x)  =  w^{\alp}_{a_\smalla,b_\smalla} \left(j_{\alp}^{\mu} 
W^{a_\smalla}_\mu  c^{b_\smalla} +    
\frac{1}{2}  P^{a_\smalls}_{\alp, \mu}  \gamma^{\mu}_{a_\smalls} c^{a_\smalla} c^{b_\smalla} +    
\frac{1}{2}  P^{i}_{\alp} \gamma_i c^{a_\smalla} c^{b_\smalla} +    
\frac{1}{2} \bar{P}^{I}_{\alp}  {\eta}^{I} c^{a_\smalla} c^{b_\smalla} + h.c.    
\right) 
\label{ano_2}   
\eea   
where ${\cal A}^{ABJ}(x)$ is the well known Adler-Bardeen-Jackiw anomaly \cite{abj} and     
the $r_i$ are its coefficients; the tensors $F^{abcd}_i$ are defined by    
$$    
F^{abcd}_i = D_{i}^{abx} (ef)^{xcd} + D_{i}^{adx} (ef)^{xbc} + D_{i}^{acx} (ef)^{xdb}   
$$   
with $D_{i}^{abc}$ invariant symmetric tensors of rank three    
on the algebra of the gauge group. ${\cal B}(x)$ in Eq.~(\ref{ano_0}) is a generic     
polynomial with dimension $\leq4$ and Faddeev-Popov charge zero\footnote{It provides the    
non-invariant counterterms to cancel the spurious anomalies coming    
from a non-symmetric renormalization scheme. Those counterterms which are IR 
dangerous will be discussed in the next section.}; 
${\cal R}_{a_\smalla}(x)$ are a set of    
BRST invariant polynomials. 
The latter are absent if the discrete CP invariance is not violated.  
In the expression ${\cal A}_2$ of Eq.~(\ref{ano_2})   
the coefficients $w^{\alp}_{a_\smalla,b_\smalla}$ are    
constant and antisymmetric in the abelian indices $a_\smalla,b_\smalla$ for each value of $\alp$ 
and $ P^{a_\smalls}_{\alp, \mu}, \dots, \bar{P}^{I}_{\alp}$ are defined in App. A.     
   
As observed by Barnich {et al.} \cite{henn} the anomaly terms    
(\ref{ano_2}) are trivial if and only if the conserved currents  $j_{\alp}^{\mu}$    
are trivial, that is, are equal  to an identically conserved total divergence 
when the equation of motion are satisfied; however    
as already observed, in the SM, the conserved lepton and baryon numbers     
provide four non-trivial examples of $j_{\alp}^{\mu}$ \cite{acci}. About the actual     
presence of anomalies as (\ref{ano_1}) and (\ref{ano_2}) at higher orders  there is no    
evidence from one- and two-loop calculations    
and  the only example where (\ref{ano_1}) occurs in the literature is given in    
\cite{medrano}.    
In fact by choosing an appropriate gauge fixing, one can avoid these anomalies    
as in `t Hooft gauge fixing as proved in \cite{henri} or    
as in `t Hooft-Background gauge fixing as proved in paper \cite{pg_2}.    
Generically for an arbitrary choice of the gauge fixing the abelian ghosts    
are coupled to scalars in order to protect the model against IR divergences   
of massless would-be Goldstone fields, as a consequence  anomalies as   
${\cal A}_1, {\cal A}_2$ might appear.     

To summarize we have the following anomaly cancellation mechanisms
\begin{itemize} 
\item  The coefficients $r_i$  of ${\cal A}^{ABJ}$ depend on the charges of
  fermion fields and on the gauge group structures. 
  Therefore for a proper choice of fermion content of the
  theory, the coefficients $r_i$ are zero. In addition, by the non-renormalization
  theorem (see \cite{bbbc_1} and the references therein), if $r_i$
  are zero at one loop, they will be absent to all orders.  
\item The candidates ${\cal A}_1$ are excluded by the AAE (\ref{aa}). In
  fact it is easy to show that, solving the constraints
  hierarchically, the AAE rules out the candidates which depend on the
  abelian ghost fields $c_{a_\smalla}$ \cite{pg_3}. 
\item The candidates ${\cal A}_2$ are absent because of the
  AAE.  This can also easily proved by using the WTI (\ref{wt}) for the
  abelian factors which is a consequence of the STI and the AAE. 
\item As proved in \cite{pg_1} the hard anomalies of the WTI are related to 
those of the STI. Therefore they are cancelled by the same mechanisms.  
\end{itemize}

\subsection{IR Anomalies}   
\label{IRanomalies}  
 
In this last section we present the analysis of the IR anomalies and we derive   
the conditions under which their coefficients vanish. We also prove that,   
under these conditions, the IR anomalies which depend on background fields are absent to 
all orders. Here,  we consider only the IR anomalies to STI; the IR anomalies to WTI are related to   
the STI anomalies by consistency conditions.  The IR anomalies to the Nakanishi-Lautrup 
equations (\ref{nl}), to the Faddeev-Popov equations (\ref{fp}) and to 
the Abelian Antighost equation (\ref{aa}) will be discussed in 
App.~B and in App.~C, respectively. 
   
As a consequence of the Quantum Action Principle (QAP) \cite{brs,libro,QAP}, 
a non-invariant renormalization scheme breaks the STI 
\beq\label{ren.20.1}    
{\cal S}(\g) = \Delta^{3}_{S,4} \cdot \g    
\eeq     
by local (integrated) terms $\Delta^{3}_{S,4}=\int \dx \Delta^{3}_{S,4}(x)$ 
whose invariance properties (${\cal S}_{0} \Delta^{3}_{S,4} =0$), UV and IR power 
counting degrees\footnote{We underline that we use the UV and IR power counting 
degrees defined within the BPHZL scheme 
\cite{zimm}, but our considerations remain valid for any renormalization scheme.} 
($d_{UV} \Delta^{3}_{S,4}=4, d_{IR} \Delta^{3}_{S,4}=3$)
and Faddeev-Popov charge are fixed by the algebraic properties of 
the functional operator ${\cal S}_0$ (cf. App.~(\ref{sub.sec:functio.symm})).
If the theory is renormalizable, it must be possible to 
re-absorb the breaking terms $\Delta^{3}_{S,4}$ by suitable non-invariant counterterms. 
  
However some of the anomaly candidates which can be written as a variation   
of a local counterterms   
\beq\label{ren.20.2}    
 \Delta^{3}_{S,4} = {\cal S}_{0} \int \dx \left(x^1_{a' b''} A^{a'}  Z^{b''} +     
x^2_{a' i} c^{a'}_{A} \gamma^{i}   
\right) + \h{\Delta}^{3}_{S,4},     
\eeq    
whose IR degree is smaller than four, cannot be  removed by counterterms. 
Indeed, these counterterms would generate IR divergences 
at the next order. As a consequence, the STI is anomaly-free only if the coefficients    
$x^1_{a' b''},\, x^2_{a' i}$ of the breaking terms vanish to all orders. The 
breaking terms $\h{\Delta}^{3}_{S,4}$ contain other candidates 
which are removed by counterterms. 
Notice that also some terms containing one or two background fields  
({\it e.g.} $\int \dx A^{a'}_\mu  \h{Z}^{b''}_\mu$) 
might appear as IR anomalies to STI; however, 
due to the fact that the background fields do not propagate, 
the corresponding counterterms do not produce IR divergences.  Finally, 
we do not assume the CP symmetry as an invariance of the model and  the 
basis of monomials for $\Delta^{3}_{S,4}$ contains both CP-odd and CP-even terms.  
   
To derive the minimal set of conditions under which the IR dangerous part of $\Delta^{3}_{S,4}$
vanishes we have, first, to differentiate the STI with respect to those fields entering in the 
monomials ${\cal S}_{0} \int \dx A^{a'}_\mu  Z^{b''}_\mu$ and  
${\cal S}_{0} \int \dx c^{a'}_{A} \gamma^{i}$, and, then,
to choose a convenient configuration for momenta to compute the coefficients  
$x^1_{a' b''},\, x^2_{a' i}$. It turns out that the most convenient point 
is at zero momentum and the requirement that  $x^1_{a' b''}= x^2_{a' i}=0$ 
implies sensible normalization conditions for two-point Green's 
functions at zero momentum. In particular, among these conditions there are 
independent equations which must be solved  in terms of the free parameters of the 
model. The existence of a solution to those equations 
has been already covered in Sec.~\ref{normalizations}. 

As an example, the two conditions 
$\g_{\h{\gamma}_i c^A_{a'}}(0) = 0$ and $\g_{A^{a'}  \phi_i}(0) =0$  are required to cancel 
the IR dangerous parts of $\Delta^{3}_{S,4}$, 
but they cannot be solved by adjusting free parameters. 
It will be shown that from the explicit form of the breaking terms  
(\ref{ren.20.2}), by using the nilpotency of the Slavnov-Taylor operator  ${\cal S}_0$ 
and the WTI they are not independent and they are implemented simultaneously. 
It is important to stress that this particular cancellation of anomalies works only   
for the SM which contains one neutral massless field.  

Let us consider the STI in terms of the mass eigenstates  \cite{aoki,krau_ew}
\bea\label{ir.1} 
{\cal S}(\g) &=& 
\int \dx \left[ 
\left(c_\smallw \de_\mu c_\smalla - s_\smallw \de_\mu c_\smallz \right) 
\left(c_\smallw \frac{\del \g}{\del A_\mu} - s_\smallw \frac{\del \g}{\del Z_\mu} \right) 
+ 
\frac{\del \g}{\del \gamma^3_\mu}
\left(s_\smallw \frac{\del \g}{\del A_\mu} + c_\smallw \frac{\del \g}{\del Z_\mu} \right) 
\nonumber \right. \\ 
&+& \left. 
\frac{\del \g}{\del \zeta^3} 
\left(s_\smallw \frac{\del \g}{\del c_\smalla} + c_\smallw \frac{\del \g}{\del c_\smallz} \right) 
+ 
\frac{\del \g}{\del \gamma^0} 
\frac{\del \g}{\del G_0} 
+ 
\frac{\del \g}{\del \gamma^H} 
\frac{\del \g}{\del H} 
+ \dots \right] = \Delta^{3}_{S,4} 
\eea 
where the ellipsis collect those terms which are not involved in the present 
analysis. Here the symbol $\g$ stands for the {\it reduced} functional 
given in Eq.~(\ref{re_1}) and the sources $\gamma^3_\mu, \gamma_0$ -- 
for the third component of the triplet gauge field $W^3_\mu$ and for the Goldstone --
coincide with the modified ones defined in (\ref{re_2}). This choice is very convenient, 
although not strictly necessary, since it automatically takes into account the Faddeev-Popov 
Eqs.~(\ref{fp}) and the abelian antighost Eq.~(\ref{aa}). 

The IR dangerous breaking terms are given by the following local monomials 
\bea\label{ir.2} 
\Delta^{3}_{S,4} &=& 
{\cal S}_0 \int \dx \left(x_1 A_{\mu} Z^{\mu} +  x_2 \gamma_0 c_\smalla +  
x_3 \gamma_H c_\smalla  \right) = \\  
&=& \int \dx 
\left[ x_1 
\left(\de_\mu c_\smalla Z_\mu + A_\mu \de_\mu c_\smallz + \dots  
\right) 
+ x_2 
\left(\frac{\del \g_0}{\del G_0} c_\smalla - \gamma_0 s_\smallw 
\frac{\del \g_0}{\del \zeta_3} \right) \nonumber \right. \\
&+& \left.  x_3 
\left(\frac{\del \g_0}{\del H} c_\smalla - \gamma_H s_\smallw 
\frac{\del \g_0}{\del \zeta_3} \right) 
\right]. 
\eea 
where terms with higher powers of fields are not shown. 

In order to compute the coefficients $x_{i}, i=1,2,3$
we  differentiate both sides of the 
Eq.~~(\ref{ir.1}) with respect to $\tilde{c}_\smalla(p)$ and 
$\tilde{Z}_\mu (-p)$ 
and with respect to the ordinary derivative $\de_{p_{\mu}}$. 
Then, we evaluate the entire expression at zero momentum 
\bea\label{ir.3} 
x_1 + v x_2 
&=&\left.  i \de_{p_\mu} \frac{\del^2}{\del \tilde{c}_\smalla(p) \del \tilde{Z}_\mu (-p)} 
{\cal S}(\g)\right|_{\phi=0,p=0} = 
 \\ 
&=&
\vspace{1cm} \left(c_\smallw^2  + s_\smallw \g_{\gamma^3 c_\smalla}(0) \right) \g_{A Z}(0) + 
\left(- s_\smallw  + \g_{\gamma^3 c_\smalla}(0) \right) c_\smallw \g_{Z Z}(0) 
\nonumber  \\ 
&-& \g_{\gamma_0 c_\smalla}(0) \g_{G_0 Z}(0) 
- \g_{\gamma_H c_\smalla}(0) \g_{H Z}(0) \nonumber 
\eea 
The conventional decomposition in terms of 
longitudinal and transverse parts of the gauge 
two-point functions and the definition 
for the scalar-gauge mixed two-point functions (Sec.~\ref{normalizations}), 
below the Eq.~(\ref{new_form_matr})) are used. 
   
Notice that the two-point functions $\g_{\gamma_0 c_\smalla}, \g_{\gamma_H c_\smalla}$ 
must vanish at zero momentum transfer because of the IR degrees 
of the fields and the BRST sources. 
The first one, namely  $\g_{\gamma_0 c_\smalla}(0)=0$ 
is ensured by the ghost equations (see Sec.~\ref{normalizations}), but 
the second one is a consequence of the STI and the AAE as we will show below. 
Furthermore to ensure the correct IR behavior of the 
$Z-A$ mixing, as discussed in Sec.~\ref{normalizations}, 
$\g_{A Z}^L(0)=0$ and, finally, the vanishing 
of $x_1 + v x_2$ is achieved only if $\g_{\gamma^3 c_\smalla}(0) = s_\smallw$. 
 
The other anomaly coefficients are determined by the 
following STI 
\bea\label{ir.4} 
 -p^2\, x_2 &=&\left.  
\frac{\del^2}{\del \tilde{c}_\smalla (p) \del \tilde{G}_0 (-p)} 
{\cal S}(\g)\right|_{\phi=0} = \\ 
&=& - p^2 \left[ 
\left(c_\smallw^2  + s_\smallw \g_{\gamma^3 c_\smalla}(p) \right) \g_{A G_0}(p) + 
\left(- s_\smallw  + \g_{\gamma^3 c_\smalla}(p) \right) c_\smallw \g_{Z G_0}(p) 
\right] 
 \nonumber\\
&-& \g_{\gamma_0 c_\smalla}(p) \g_{G_0 G_0}(p) 
- \g_{\gamma_H c_\smalla}(p) \g_{H G_0}(p) 
 \nonumber  \\
\nonumber
\\ 
\label{ir.4.1}
(M_H^2 - p^2) x_3 &=& \left.  
\frac{\del^2}{\del \tilde{c}_\smalla (p) \del \tilde{H} (-p)} 
{\cal S}(\g)\right|_{\phi=0}  = \\
&=& -p^2 \left[ 
\left(c_\smallw^2  + s_\smallw \g_{\gamma^3 c_\smalla}(p) \right) \g^L_{A H}(p) + 
\left(- s_\smallw  + \g_{\gamma^3 c_\smalla}(p) \right) c_\smallw \g^L_{Z H}(p) 
\right] 
\nonumber \\
&-& \g_{\gamma_0 c_\smalla}(p) \g_{G_0 H}(p) 
- \g_{\gamma_H c_\smalla}(p) \g_{H H}(p)  \nonumber
\eea 
where $M_H$ is the Higgs mass.  
From the second one computed at zero momentum, we immediately 
get $x_3 =0$ as a consequence of $\g_{\gamma_0 c_\smalla}(0)=\g_{\gamma_H c_\smalla}(0)=0$. 
However the Eqs.~(\ref{ir.4}) are not sufficient for our purposes, 
in fact, in order to verify that also $x_2=0$, it is necessary the differentiate the 
Eq.~(\ref{ir.4}) with respect to $p^2$ and to study the 
system of scalar fields. This has been already analyzed 
in Sec.~\ref{normalizations}, however the corresponding identities are 
derived in the gauge eigenstates. In particular here we show that the WTI 
and the STI implies that vanishing of the mixed Green's function $\g_{G_0 H}(p)$ at zero 
momentum. 

Here we need the following two equations 
\bea\label{ir.5} 
&&\left.  
\frac{\del^2}{\del \tilde{c}_\smallz(p) \del \tilde{G}_0 (-p)} 
{\cal S}(\g)\right|_{\phi=0,p=0} =  
- \g_{\gamma_0 c_\smallz}(0) \g_{G_0 G_0}(0) 
- \g_{\gamma_H c_\smallz}(0) \g_{H G_0}(0) = 0 
\nonumber \\ 
&&
\\&&\left.  
\frac{\del^2}{\del \tilde{c}_\smallz(p) \del \tilde{H} (-p)} 
{\cal S}(\g)\right|_{\phi=0,p=0} =  
- \g_{\gamma_0 c_\smallz}(0) \g_{G_0 H}(0) 
- \g_{\gamma_H c_\smallz}(0) \g_{H H}(0) = 0 
\nonumber
\eea 
which can be spoiled by breaking terms reabsorbable 
without any IR obstruction. As a consequence of these 
equations, the determinant of 
the matrix of two-point functions for the scalars (\ref{mat_scal}) 
vanishes (see Eq.~ \ref{zero}). Finally, by the normalization condition 
$\g_{G_0 H}(0)=0$, the two-point function for the neutral 
Goldstone  $\g_{G_0 G_0}(0)=0$ vanishes at zero momentum. 
Going back to Eqs.~(\ref{ir.4}), 
we can compute the derivative with respect to $p^2$ and, setting $p=0$, 
we deduce that $\g_{A G_0}(0)=0$ is a necessary and sufficient conditions 
to guarantee that the coefficients $x_2$ is zero. 
To complete the proof we have to derive $\g_{G_0 H}(0)=\g_{A G_0}(0)=0$. 

The condition $\g_{G_0 H}(0)=0$ is satisfied as a consequence of the 
invariance under the background gauge transformations and the relations among  
the two-point function with external background fields and the two-point functions with 
external quantum fields. In particular by the STI
\bea\label{ir.9.1} 
\left.  
\frac{\del^2}{\del \tilde{G}_0 (p) \del \tilde{\Om}_0 (-p)} 
{\cal S}(\g)\right|_{\phi=0,p=0} =
- \g_{\gamma_0 \Om_0}(0) \g_{G_0 H}(0) 
- \g_{\gamma_H \Om_0}(0) \g_{H H}(0) + \g_{\h{G}_0 H}(0) = 0,  
\eea 
where $\Om_0$ is the BRST variation of the background Goldstone
$\h{G}_0$, 
and the WTI\footnote{Notice that both the eqs. (\ref{ir.9.1})-(\ref{ir.9.2}) can be spoiled 
by breaking terms, however they can be removed by counterterms without introducing any IR 
divergences.} 
\beq\label{ir.9.2}
\left(\g_{\h{G}_0 H}(0) + \g_{{G}_0 H}(0) \right) =0,  
\eeq
obtained by differentiating the functional identities~(\ref{wt_S}) with respect to 
$\widetilde{H}(0)$, we immediately get 
\beq\label{ir.9.3}
\left(1 + \g_{\gamma_0 \Om_0}(0) \right) \g_{{G}_0 H}(p) = 
\g_{\gamma_H \Om_0}(0) \g_{{H} H}(0). 
\eeq
This implies that, if the CP-violation is induced only by means of the CKM matrix 
for fermions,  $\g_{G_0 H}(0)=0$. In fact this equation relates the CP-odd 
Green's function $\g_{G_0 H}(0)$ with external quantum fields which are directly coupled to 
the fermions to the CP-odd Green's function $\g_{\gamma_H \Om_0}(0)$ whit external 
classical fields which do not couple directly to fermions. Therefore we can conclude that 
$\g_{G_0 H}(0)=\g_{\gamma_H \Om_0}(0)=0$. Notice that in the case of CP-violation 
Higgs potential (see for example \cite{2HD}) this conclusion cannot be achieved any longer. 

To prove that $\g_{A G_0}(0)=0$ we consider the following 
WTI\footnote{Notice that for convenience we use the notation $\hat{A}_\mu$ 
to denote the background field for the photon. However in our approach without 
an explicit background field for the $U(1)$ gauge boson we have that 
$\g_{\hat{A}_\mu A_\nu} = - s_\smallw \g_{\hat{W}^3_\mu A_\nu}$ where 
$\hat{W}^3_\mu$ is the background field for the $SU(2)$ factor. Using the 
background field for the photon the following equations stay unchanged.}  
\beq\label{ir.1.new}
-i p_\mu \left(\g_{\h{A}_\mu Z_\nu}(p) + \g_{{A}_\mu Z_\nu}(p) \right) =0 
\eeq
which expresses the 
transversality of the mixed two-point functions. Notice that this identity can be 
broken by the renormalization procedure by terms of the form 
$\Delta_{\lambda_\smalla Z_\nu}(p) = y_1 p_\mu + y_2 p^2 p_\mu$ 
which must be removed by the following counterterms
\beq\label{ir.2.new}
\int \dx  \left(y_1 \h{A}_\mu Z^\nu + y_2 \de^\mu \h{A}_\mu \de_\nu Z^\nu \right).  
\eeq
Only the first terms has IR degree equal to 3. However, since  this term is linear in the 
quantum field $Z_\nu$, it does not contribute to the 
irreducible Green's functions and it never produces IR divergences. 

From Eq.~(\ref{ir.1.new}), by using the decomposition for two-point 
functions into longitudinal and transverse part, we arrive to 
\beq\label{ir.3.new}
\left(\g^L_{\h{A} Z}(p) + \g^L_{{A} Z}(p) \right) =0 
\eeq
and the normalization condition $ \g^L_{{A} Z}(0)=0$  implies that  
$\g^L_{\h{A} Z}(0)=0$. Finally, by analyticity of Green's functions, we 
obtain that  $\g^T_{\h{A} Z}(0)=0$. 

Furthermore we know from Sec.~\ref{bf_renorm} that the two-point functions
with external background fields are in general related, by means of the STI, to 
the two-point functions with quantum fields. and in particular 
we have 
\beq\label{ir.4.new}
\g^T_{\h{A} Z}(p) \propto  \g^T_{\h{W}_3 Z}(p) = 
 \g^T_{\Om_3 \gamma_3}(p) \left( s_\smallw \g^T_{{A} Z}(p) + c_\smallw \g^T_{{Z} Z}(p)  \right) 
\eeq
where $\g^T_{\Om_3 \gamma_3}(p)$ 
is the transverse part of $\g_{\Om^\mu_3 \gamma^\nu_3}(p)$. This equation 
implies that   
$\g^T_{\Om_3 \gamma_3}(0) \g^T_{{Z} Z}(0) = 0$ to all orders, and, since 
$ \g^{T,(0)}_{{Z} Z}(0) \neq 0$ already at tree level, we deduce that 
$\g^T_{\Om_3 \gamma_3}(0)$ must vanish. Notice that this is a consequence of the 
normalization conditions for the transverse part of the two-point functions 
for the gauge fields and, it follows from imposing the background gauge invariance as in its tree 
level form. 
Removing this essential constraint we cannot deduce this result and the 
STI must be deformed as discussed in \cite{krau_ew}.  Notice that the STI (\ref{ir.4.new}) 
can be spoiled by breaking terms, however as proved in \cite{pg_1} these breaking 
terms are compensated by suitable counterterms dependent on the background fields. This 
means that, as in the case of (\ref{ir.1.new}), the corresponding counterterms do not 
produce any IR divergence. 

Analogously to the mixed two-point function we have also the following STI 
for mixed Goldstone-photon two-point function
\beq\label{ir.5.new}
 \g_{\h{A} G_0}(p) \propto  \g_{\h{W}_3 G_0}(p) = 
 \g^L_{\Om_3 \gamma_3}(p) \g_{{W}_3 G_0}(p) + 
 \g^L_{\Om_3 \gamma_0}(p) \g_{{G}_0 G_0}(p)  +
 \g^L_{\Om_3 \gamma_H}(p) \g_{{H} G_0}(p). 
\eeq
Notice that if the CP invariance were a symmetry of the model 
the last terms would be absent. By recalling that $ \g_{{H} G_0}(0)= \g_{{G}_0 G_0}(0)=0$, that 
$\g^T_{\Om_3 \gamma_3}(0)=\g^L_{\Om_3 \gamma_3}(0)=0$ by analyticity and 
$\g^{(0)}_{{W}_3 G_0}(p)\neq 0$, we deduce that  $\g_{\h{A} G_0}(0)=0$. 

To establish that $\g_{{A} G_0}(0)=0$ we can use again the WTI for the 
background gauge invariance
\beq\label{ir.6.new}
\left(\g^L_{\h{A} G_0}(p) + \g^L_{{A} G_0}(p) \right) =0 
\eeq
which immediately implies our claim. Notice that also in this case the 
WTI can be broken by local terms 
$\Delta_{\lambda_\smalla G_0}(p) = y_3 p^2$ (a constant breaking 
terms is excluded by IR power counting) which is 
removed by using counterterms of the form
\beq\label{ir.7.new}
\int \dx  \left(y_3 \h{A}_\mu \de^\mu G_0\right)  
\eeq
which is not IR dangerous and has no effect in irreducible Green's functions. 

Finally we have to show that $\g^{(n)}_{\gamma_H c_\smalla}(0) =
0,~~ \forall n$. This can be achieved by considering the AAE and the STI.  
In particular, by taking the functional derivative of AAE (\ref{aa}) 
with respect to $\tilde{\gamma}_H(p)$ at zero momentum, we have 
\bea\label{ir.8}
c_\smallw \g_{\gamma_H c_\smalla}(0) - s_\smallw 
\g_{\gamma_H c_\smallz}(0) - v \g_{\gamma_H
  \Om_0}(0)  = 0.  
\eea
By considering Eq.~(\ref{ir.9.1}) and 
$\g_{G_0 H}(0) = 0$, $\g_{\gamma_0 c_\smalla}(0) = 0$, from the second equation 
of~(\ref{ir.5}) we get $\g^{(n)}_{\gamma_H c_\smallz}(0)=0,~~ \forall n$, and finally 
from eq.~(\ref{ir.8}) and $\g_{\gamma_H \Om_0}(0)=0$ we obtain 
$\g_{\gamma_H c_\smalla}(0)=0$ which is our claim.

Finally, by this minimal set of normalization conditions we are 
able to exclude any IR anomalies and,  
by considering the new identity 
\bea\label{ir.6} 
&& \left.  
- i \de_{p_\mu} 
\frac{\del^2}{\del \tilde{c}_\smalla (p) \del \tilde{A}_\mu (-p)} 
{\cal S}(\g)\right|_{\phi=0,p=0} = \\
&=& \left[ - s_\smallw \left(c_\smallw  + \g_{\gamma^3 c_\smallz}(0) \right) \g^L_{A A}(0) + 
\left(s^2_\smallw  -  c_\smallw \g_{\gamma^3 c_\smallz}(0) \right) \g^L_{Z A}(0) \right] 
\nonumber \\
&-& 
\g_{\gamma_0 c_\smallz}(0) \g_{G_0 A}(0) 
- \g_{\gamma_H c_\smallz}(0) \g_{H A}(0) = 0 
\nonumber
\eea 
we obtain the well-known result \cite{aoki} 
\beq\label{ir.7} 
\left(
\g^L_{AA}(0) \g^L_{ZZ}(0) - (\g^L_{AZ})(0)^2 
\right) = 0 . 
\eeq 
for the $Z-A$ mixing. Notice that this can be only achieved in 
restricted set of models. Therefore, by imposing $\g_{AZ}(0)=0$, 
we immediately get $\g_{AA}(0)=0$ from (\ref{ir.7}) and $\g_{AH}(0)=0$ 
from (\ref{ir.6}). 

We conclude the section reminding the 
reader that this particular cancellation 
of the IR anomaly can be only achieved in the {\it minimal} SM, 
because there is only one abelian factor.  
 
The analysis of the IR anomalies cancellation in the case of 
extended models as the 2HDM and in the 
MSSM requires a more involved proof  because 
any kind of mixing among the scalar fields are admitted by 
symmetries and CP violation. In the same way 
the corresponding BRST sources $\gamma_i$ can mixed among themselves and 
we have to prove that $\g_{c_\smalla \gamma_i}(0)=0$ for all $i$. The 
analysis of these model will be presented in a forthcoming paper.

\section{Conclusions}  
  
We have shown that the BFM allows for  a consistent 
renormalization of non-semisimple gauge theories and the Standard  
Model. We have discussed the relation between the background gauge invariance,  
expressed in terms of Ward-Takadashi Identities on the Green's function  
and the normalization conditions. In particular,  we have analyzed the IR problems  
and we have defined a ``minimal'' set of normalization conditions ({\it partially on-shell scheme}) 
compatible  with the background gauge invariance which avoids the (off-shell) IR  
singularities.   
  
Furthermore we have studied the anomalies of STI and WTI   
with special attention to their IR anomalies. We have shown that the IR  
anomalies for the SM can be excluded by a suitable choice of free parameters.   
This implies that, in the case of extended models, the STI and the WTI   
must be deformed.   
  
In addition, we have considered the problem of mixings in all  
sectors of non-semisimple gauge models and we discuss the renormalization  
conditions for the CKM matrix elements and the mixing angles. We are able to formulate the  
renormalization independently of the  
CP-invariance, showing that new counterterms are 
fixed by the WTI for the background gauge invariance 
and by a proper renormalization of tadpoles. 

In conclusion, we have analyzed the renormalization of the background fields and  
we have proved that the normalizations of their two-point functions are related to the  
renormalizations of quantum fields. This  
ensures that there is no independent renormalization of the background  
fields and the S-matrix elements built by background external Green's  
functions, as prescripted by the Background Field Method, have the same poles as the S-matrix   
constructed in the conventional way.    
    
\section*{Acknowledgments}  
  
The author is deeply indebted to D. Maison for discussions, suggestions and  
for a careful reading of the manuscript. I also want to thank C. Becchi and R. Ferrari   
for useful remarks and comments. Moreover, I would like to acknowledge  
P. Gambino for useful clarifications on the electroweak radiative   
calculations and, finally,  E. Kraus and K. Sibold concerning the algebraic 
renormalization.


\section*{Appendix A: Functional Operators and their Algebra}     
\label{sub.sec:functio.symm}  
 \renewcommand{\theequation}{A.\arabic{equation}}  
\label{Invariantoperators}   
 
In the present section we will  present the functional operators and
their algebra in the gauge eigenfield formalism. The expressions of
those operators in the mass eigenfield formalism is provided in \cite{aoki,krau_ew}. 

Starting from the STI (\ref{st}), from the WTI (\ref{wt_S}) for simple factors, from the
choice of the gauge fixing and from the AAE discussed in Sec.~\ref{sub.sec:symmetries}, 
and by requiring the closure of the algebra of the functional operators we
derive the complete set of identities.   
 
The choice of the gauge fixing  (\ref{gaugefixing}) is implemented at the quantum level  
as equations of motion for the $b_a$ fields
\bea\label{nl}  
  \dms{\frac{\del \g}{\del b_{a}}} & = & \left[\del^{a b_{S}} \nabh  
  ^{b_{S}c_{S}}_{\mu} (W-\h{W})^{\mu}_{c_{S}} + \del^{a b_{A}}  
  \partial_{\mu} (W -\h{W})^{\mu}_{b_{A}} + \right. \nonumber \\ &&  
  \left.  \rho^{a b} (\h{\phi}+v)_{i} (et)^{b}_{ij} (\phi+v)_{j} +  
    \Lambda^{a c} b_{c} 
  \right] \equiv \Delta^{Cl}_{b_a}.   
\eea   
For deriving the consistency conditions among the STI and other  
    functional identities we introduce the linearized Slavnov-Taylor  
    operator:   
\bea\label{linst} \ber{l} \oop{.3cm} {\cal S}_0 =  
    \dms{\int} \dx \left[ \dms{\fd{\g_0}{\gamma}{a_\smalls}{\mu}}  
    \dms{\dfa{W}{a_\smalls}{\mu}} + \dms{\fd{\g_0}{W}{a_\smalls}{\mu}}  
    \dms{\dfa{\gamma}{a_\smalls}{\mu}} + \de_{\mu}c^{a_\smalla}  
    \dms{\dfa{W}{a_\smalla}{\mu}} + \dms{\fdd{\g_0}{\gamma}{i}}  
    \dms{\dfu{\phi}{i}} + \dms{\fdu{\g_0}{\phi}{i}}  
    \dms{\dfd{\gamma}{i}} + \right. \\ \left. \hspace{1cm}  
    \dms{\fdL{\g_0}{\b{\eta}}{R}{I}} \dms{\dfaR{\psi}{R}{I}} +  
    \dms{\fdR{\g_0}{{\psi}}{L}{I}} \dms{\dfaL{\b{\eta}}{L}{I}} +  
    \dms{\fdR{\g_0}{\eta}{R}{I}} \dms{\dfaL{\b{\psi}}{R}{I}} +  
    \dms{\fdL{\g_0}{\b{\psi}}{L}{I}} \dms{\dfaR{\eta}{L}{I}} +  
    \dms{\fdd{\g_0}{\zeta}{a_\smalls}} \dms{\dfu{c}{a_\smalls}} +  
    \dms{\fdd{\g_0}{c}{a_\smalls}} \dms{\dfu{\zeta}{a_\smalls}} + \right. \\   
    \left. \hspace{1cm} b_{a} \dms{\dfd{\bc}{a}} + \Om^{a_\smalls}_{\mu}  
      \dms{\dfa{\hat{W}}{a_\smalls}{\mu}} + \Om_{i} \dms{\dfd{\h{\phi}}{i}}  
    \right]   
\eer   
\eea   
where $\g_0$ is the tree level action (\ref{setti.5}). 
Notice that because of invariance of $\g_0$ under  
    BRST transformations the operator ${\cal S}_0$ is nilpotent: $  
    {\cal S}^2_0=0 $.  
  
Due to the presence of several multiplets  
of scalars and fermions, the classical action turns out to be     
invariant under some global symmetries (here labeled by $\alp$)\cite{acci}.  
The following WTI for the global symmetries implement  
  those symmetries at the quantum level   
\bea\label{glob_symm} {\cal  
    W}_{\alp} \g = \dms{\int \dx }\left[ P^{a_\smalls}_{\alp, \mu}  
  \fd{\g}{W}{a_\smalls}{\mu} + P^{i}_{\alp} \fdu{\g}{\phi}{i} +  
  \bar{P}^{I}_{\alp} \fdu{\g}{\bar{\psi}}{I} + \fdu{\g}{\psi}{I}  
  P^{I}_{\alp} \right] =0 \eea 
where 
$P^{a_\smalls}_{\alp, \mu}, P^{i}_{\alp},  
  \bar{P}^{I}_{\alp}, P^{I}_{\alp}$ are local polynomials for each  
  $\alp$ and $P^{a_\smalls}_{\alp, \mu}, \dots,   P^{I}_{\alp}$ are the generators of the 
  global symmetry in the field representation.   
  
In the SM there are four different global currents 
which correspond to  conserved quantum numbers (at the level of perturbation 
theory),   
namely the individual lepton numbers $L_e, L_\mu, L_\tau$ and the  
baryon number $B$.  The corresponding currents are given by   
\begin{eqnarray}\label{boh_0}   
&&\hspace{-1cm} j^{B}_{\mu}  =  \sum_{\beta} \left(\b{Q}_{L,\beta} \gamma_{\mu} Q_{L,\beta} +   
 \b{u}_{R,\beta} \gamma_{\mu} u_{R,\beta} +  \b{d}_{R,\beta} \gamma_{\mu} d_{R,\beta}    
\right)  \\  
&&\hspace{-1cm} j^{L_{\alp}}_{\mu} =  \left( 
\b{L}_{L,\alp} \gamma_{\mu} L_{L,\alp} + \b{e}_{R,\alp} \gamma_{\mu} e_{R,\alp} \right), \nonumber
\end{eqnarray}  
where $\alp=e,\mu,\nu$ and $\beta$ counts  the generations,
which can mix with the current of the hypercharge ${\cal Y}$ 
\beq\label{boh_1}
j^{\cal Y}_{\mu} = \sum_{\alp} \left(\frac{1}{6}  
\b{Q}_{L,\alp}   
\gamma_{\mu} Q_{L,\alp} +   
\frac{2}{3} \b{u}_{R,\alp} \gamma_{\mu} u_{R,\alp} -  \frac{1}{3}   
\b{d}_{R,\alp} \gamma_{\mu} d_{R,\alp} - \frac{1}{2} \b{L}_{L,\alp}   
\gamma_{\mu} L_{L,\alp} - \b{e}_{R,\alp} \gamma_{\mu} e_{R,\alp} \right) \nonumber   
\eeq  
as discussed in \cite{henri,henn,krau_ew,pg_1}. 
The currents (\ref{boh_0}) provide the four corresponding  WTI  
\bea\label{glob_sym.2}  
{\cal W}_{B} \g = \dms{\int \dx} \de^{\mu}  j^{B}_{\mu}, ~~~~  
{\cal W}_{L_\alp} \g = \dms{\int \dx} \de^{\mu}  j^{L_\alp}_{\mu}   
\eea  
which commute which the other functional equations and among themselves. 

Among the functional identities we have the following commutation relations   
(here expressed for a generic functional ${\cal F}$)  
\def\F{{\cal F}}  
\bea\label{fi.6}  
&&   
\vspace{.5cm}   
{\cal S}_{\F}{\cal S}(\F) = 0, ~~~~  
{\cal S}_{\F} \mbox{\large \bf W}_{a_\smalls}(\F) - \mbox{\large  
  \bf W}_{a_\smalls} {\cal S}(\F) = 0   
\nonumber  
\\   
&&   
\vspace{.5cm} \dms{\frac{\del}{\del  
    b_{b}}} \left({\cal G}_{a_\smalla}(\F) - \Delta^{Cl}_{c_{a_\smalla}}  
\right) - {\cal G}_{a_\smalla} \left(\dms{\frac{\del}{\del  
    b_{b}}}(\g) - \Delta^{Cl}_{b_b}\right) = 0   
\nonumber   
\\  
&&  
\vspace{.5cm}   
\left[\mbox{\large \bf W}_{a_\smalls},  
\mbox{\large \bf W}_{b_\smalls} \right] (\F) = (ef)^{a_\smalls b_\smalls c_\smalls} \mbox{\large \bf  
  W}_{c_\smalls}(\F)   
\nonumber  
\\  
&&  
\left[\mbox{\large \bf W}_{a}, \left(\dms{\frac{\del}{\del b_{b}}} -  
\Delta^{Cl}_{b_b}\right) \right] (\F) = (ef)^{abc} \left( 
\dms{\frac{\del}{\del b_{c}}} - \Delta^{Cl}_{b_c}\right) (\F)  
\eea  
which respectively express the nilpotency of the   
Slavnov-Taylor operator, the invariance of the Slavnov-Taylor operator  
under the background gauge transformations of the simple factors, the   
compatibility between the AAE and the gauge fixing and, finally, the covariance of  
the Eq.~(\ref{nl}) with respect to the background gauge transformations.  
  
However the system of the above equations does not close under the  
(anti-)commutation relations and in particular the { Faddeev - Popov} equations of  
motion ($\fp$ equations) are derived by requiring the involution of the system:   
\bea\label{fp} &&  
  \overline{\cal G}_{a}(\g) \equiv \dms{\frac{\del}{\del b_{b}}} {\cal  
    S}(\g) - {\cal S}_{\g} \left(\dms{\frac{\del}{\del b_{b}}}(\g) -  
  \Delta^{Cl}_{b_b} \right) =   
\nonumber   
\\ && = \fdu{\g}{\bc}{a} +  
  \del^{aa_{S}} \nabh ^{a_{S}b_{S}}_{\mu} \fd{\g }{\gam}{{b_{S}}}{\mu}  
  + \rho^{ab} (\h{\phi}+v)_{i} (et)^{b}_{ij} \fdu{\g}{\gam}{j} = \\ &&  
  = - \del^{aa_{A}} \de^{2} c_{a_{A}} - \del^{aa_{S}}  
  \nabla_{\mu}^{a_{S}b_{S}} \Om^{b_{S}}_{\mu} - \rho^{ab} \Om_{i}  
  (et)^{b}_{ij} (\phi+v)_{j} \equiv \Delta^{Cl}_{\bc_a} \nonumber \eea  
Although these equations are not independent of the NL Eqs.~(\ref{nl}) 
and of STI, they provide a direct method for eliminating the ghost fields  
from the vertex functional. 
  
The anti-commutation relation between   
the STI (\ref{st}) and the AAE (\ref{aa}) is more interesting, since
this implies the following WTI  
\bea\label{wt}   
&& \mbox{\large \bf W}_{a_\smalla}(\g)   
\equiv  
  {\cal G}_{a_\smalla}{\cal S}(\g) - {\cal S}_{\g} \left({\cal G}_{a_\smalla}  
  (\g) - \Delta^{Cl}_{c_{a_\smalla}} \right) = -  
  \de_{\mu} \left(\fd{\g}{W}{{a_{A}}}{\mu}+  
  \fd{\g}{\h{W}}{{a_{A}}}{\mu} \right) + 
\nonumber \\ \nonumber \\ && 
+ (et)^{a_\smalla}_{ij}  
  \left[ (\Phi+v)_{j} \fdu{\g}{\Phi}{i} + (\h{\Phi}+v)_{j}  
    \fdu{\g}{\h{\Phi}}{i} + \Om_{j} \fdu{\g}{\Om}{i}+ \gam_{j}  
    \fdu{\g}{\gam}{i} \right] +   
\\ &&   
+ (et)^{a_\smalla}_{R,IJ} \left[ {\b{\eta}}^{R}_{I}  
\dms{\fd{\g}{\b{\eta}}{R}{J}} + {{\psi}}^{R}_{I}  
\dms{\fd{\g}{\psi}{R}{J}} + {\eta}^{R}_{I} \dms{\fd{\g}{\eta}{R}{J}} +  
{\b{\psi}}^{R}_{I} \dms{\fd{\g}{\b{\psi}}{R}{J}} \right] +  ({\rm  
R} \rightarrow {\rm L})   
\nonumber   
\eea   
which describe the  background gauge invariance for the abelian factors ${\cal G}_\smalla$.   
By eliminating the abelian background gauge fields we get   
\beq\label{wt_2}  
\mbox{\large \bf W}_{a_\smalla}(\g) = \de^2 b_{a_\smalla}   
\eeq and, since the  
l.h.s. is linear in quantum fields $b_{a_\smalla}$ it does not require any  
new external source to fix its renormalization.  
  
At this point it is easy to check that the functional   
operators generates an algebra over the space of local and   
integrated functionals, in particular we have   
to supply also the following remaining relations   
\bea\label{algebra_2}   
&&   
\vspace{.5cm}   
{\cal S}_{\F} \left(\overline{\cal G}_{a}(\F) -   
\Delta^{Cl}_{\bc_a} \right) + \overline{\cal G}_{a}  
{\cal S}(\F) = 0,   ~~~~
\dms{\frac{\del}{\del b_{b}}}   
\left(\overline{\cal G}_{a}(\F) - \Delta^{Cl}_{bc_a}  
\right) - \overline{\cal G}_{a} \left(\dms{\frac{\del}{\del  
    b_{b}}}(\g) - \Delta^{Cl}_{b_b}\right) = 0,   
\nonumber   
\\ &&  
\vspace{.5cm}   
{\cal S}_{\F} \mbox{\large \bf W}_{a_\smalla}(\F) - \mbox{\large  
  \bf W}_{a_\smalla} {\cal S}(\F) = 0,   ~~~~~~
\left[ \mbox{\large \bf W}_{a_\smalla},  
\mbox{\large \bf W}_{b} \right] (\F) = 0,   
\nonumber   
\\ &&   
\vspace{.5cm}   
\left[ \mbox{\large \bf  
  W}_{a}, \left(\overline{\cal G}_{b} - \Delta^{Cl}_{\bc_b} \right)  
\right](\F) = (ef)^{abc} \left(\overline{\cal G}_{c} -  
\Delta^{Cl}_{\bc_c} \right)(\F),   
\nonumber   
\\ &&   
\vspace{.5cm}  
\left[  
\mbox{\large \bf W}_{a}, \left(\dms{\frac{\del}{\del b_{b}}} -  
\Delta^{Cl}_{b_b}\right) \right] (\F) = (ef)^{abc} \left( 
\dms{\frac{\del}{\del b_{c}}} - \Delta^{Cl}_{b_c}\right) (\F), 
\\ &&  
\vspace{.5cm}\left[ \mbox{\large \bf W}_{a}, \left({\cal G}_{a_\smalla} -  
\Delta^{Cl}_{c_{a_\smalla}} \right) \right] (\F) = 0, ~~~~~~~~  
\left\{ \left({\cal G}_{a_\smalla} -  
\Delta^{Cl}_{c_{a_\smalla}} \right), \left({\cal G}_{b_\smalla} -  
\Delta^{Cl}_{c_{b_\smalla}} \right) \right\}(\F) = 0,   
\nonumber   
\\ &&  
\vspace{.5cm} \left\{ \left(\overline{\cal G}_{a} -  
\Delta^{Cl}_{\bc_a} \right), \left({\cal G}_{b_\smalla} -  
\Delta^{Cl}_{c_{b_\smalla}} \right) \right\}(\F) = 0, ~~~~   
\left\{ \left(\overline{\cal G}_{a} - \Delta^{Cl}_{\bc_a} \right),  
  \left(  \overline{\cal G}_{b} - \Delta^{Cl}_{\bc_b} \right)  
  \right\}(\F) = 0. \nonumber   
\eea    
This ensures the integrability of the   
system (in the sense of the Fr\"obenius' theorem).   

\section*{Appendix B: Renormalization of NL Equations}  
\setcounter{equation}{0}      
\renewcommand{\theequation}{B.\arabic{equation}}  
\label{app:ren_b}  
  
In this section we will discuss the renormalization of the NL Eqs.~(\ref{nl}) 
by means of the algebraic   
technique with special care to the problems of IR divergences. In fact,   
although the renormalization of the NL equations does not present any   
algebraic obstruction, we have to be sure that the   
introduction of non-invariant counter term does not introduce some   
IR anomalies. In particular we do not analyze the complete algebraic renormalization   
of the NL Eqs.(\ref{nl}), but we look only for the lower dimension terms.   
The renormalization of these equations   
is also studied in book \cite{libro} and in recent paper \cite{krau_ew}.  
  
By using the relation among the $b_{a}$ and the corresponding   
anti-ghost fields $\bc^{a}$, we introduce the ``massless''   
$b_{A}^{a'}$ and the ``massive'' $b_{Z}^{a''}$   
Nakanishi-Lautrup which are the BRST variation of the anti-ghost fields:  
$s \bc^{a'}_\smalla = b_{A}^{a'},~ s \bc^{a''}_\smallz = b_{Z}^{a''} $ 
with the same IR and UV degree.   
  
Then the quantum extension of the NL equations to the all orders gives   
\bea\label{nomass_nl_sy}  
\frac{\del \g}{\del  b_{A}^{a'}} =   
{\Delta}^{Cl, a'}_{b_\smalla} + \left[{Q}^{a'}_{b_\smalla} \cdot \g \right]^2_2,~~~~   
 \frac{\del \g}{\del b^{a''}_{Z}} = {\Delta}^{Cl, a''}_{b_\smallz}   
+ \left[ {Q}^{a''}_{b_\smallz} \cdot \g \right]^1_2   
\eea  
where the classical terms ${\Delta}^{Cl, a'}_{b_\smalla}, {\Delta}^{Cl, a''}_{b_\smallz}$   
are specified in Eqs.~(\ref{nl}).    
  
Recursively assuming that the lower order breaking terms  
of the Eqs.~(\ref{nomass_nl_sy}) up to order $\hbar^{n-1}$  
are compensated by means of counterterms we deduce  
\bea\label{sy_0}  
 \left[ {Q}^{a'}_{b_\smalla} \cdot \g \right]^2_2 = \hbar^n {Q}^{a'}_{b_\smalla}   
+ O(\hbar^{n+1} {Q}_{b_\smalla}),~~~~
 \left[ {Q}^{a''}_{b_\smallz} \cdot \g \right]^1_2 = \hbar^n  
{Q}^{a''}_{b_\smallz} + O(\hbar^{n+1} {Q}_{b_\smallz})   
\eea  
where $\bar{Q}^{a'}_{b_\smalla}, {Q}^{a''}_{b_\smallz}$   
are local polynomials in terms of fields and  
their derivatives with zero Faddeev-Popov charge.   
By UV and IR power counting, by covariance and by Faddeev-Popov charges, the   
possible candidates for $\bar{Q}^{a'}_{b_\smalla}, {Q}^{a''}_{b_\smallz}$ are   
\bea\label{nl_sy_q}  
 \bar{Q}^{a''}_{b_\smallz} & = & {Y}^{a'' b}_{Z,1} \partial_{\mu} W^{b}_{\mu} +   
{Y}^{a'' i}_{Z,2} \phi_i + \bar{Y}^{a'' i}_{Z,3} \h{\phi}_i   
 + {Y}^{a'' i j}_{Z,4} {\phi}_i  \phi_j  
 + {Y}^{a'' i j}_{Z,5} \h{\phi}_i  \phi_j \\   
&& + {Y}^{a'' i j}_{Z,6} \h{\phi}_i  \h{\phi}_j   
 + {Y}^{a'' b'}_{Z,7} b^{a'}_{A}  + {Y}^{a'' b''}_{Z,8} b^{a'}_{A} +   
{Y}^{a'' b d}_{Z,9} c_b \bc_d  
\nonumber \\  
 \bar{Q}^{a'}_{b_\smalla} & = & {Y}^{a' b}_{A,1} \partial_{\mu} W^{b}_{\mu} +   
{Y}^{a' i}_{A,2} \phi_i + \bar{Y}^{a' i}_{A,3} \h{\phi}_i   
 + {Y}^{a' i j}_{A,4} {\phi}_i  \phi_j  
 + {Y}^{a' i j}_{A,5} \h{\phi}_i  \phi_j \\   
&& + {Y}^{a' i j}_{A,6} \h{\phi}_i  \h{\phi}_j   
 + {Y}^{a' b'}_{A,7} b^{a'}_{A}  + {Y}^{a' b''}_{A,8} b^{a'}_{A} +   
{Y}^{a' b d}_{A,9} c_b \bc_d  \nonumber
\eea  
where the coefficients ${Y}_{A,\alpha}, {Y}_{Z,\alpha}$ are constant. The only dangerous   
terms are the $ {Y}^{a' i}_{A,2} \phi_i, \bar{Y}^{a' i}_{A,3} \h{\phi}_i$ if   
the scalar fields $\phi_i$ are massless (at least one of them). Nevertheless the   
non-invariant counter terms $b^{a'}_{A} {Y}^{a' i}_{A,2} \phi_i,   
b^{a'}_{A} \bar{Y}^{a' i}_{A,3} \h{\phi}_i$   
are IR convergent because of the IR degree of the $b^{a'}_{A}$ fields.   
  
Moreover no other obstruction occurs to the compensation of the breaking terms   
to the NL equations by means of local counter terms. In the next appendix we will   
see that for the ghost equations (\ref{aa}) and (\ref{fp}), this possibility   
occurs and a suitable rotation among massless and massive fields has to be taken   
into account in order to avoid the IR divergences.   
   
We have to remind that the presence of the   
non-linear breaking terms $ {Y}^{a'' i j}_{Z,4}, {Y}^{a'' b c}_{Z,9},   
{Y}^{a' i j}_{A,4}, {Y}^{a' b c}_{A,9}$ introduce   
new Feynman rules that have to be considered at the   
higher orders.

\section*{Appendix C: Renormalization of $\Phi\Pi$E and AAE}  
\setcounter{equation}{0}      
\renewcommand{\theequation}{C.\arabic{equation}}  
\label{app:ren_ghost}  
  
In this appendix we  briefly discuss the renormalization of the functional  
equations for the ghost and anti-ghost fields (\ref{aa}) and (\ref{fp}). Although in the main   
text we underlined the implication of the AAE, here both  
the Faddeev-Popov equation and the AAE turn out to be necessary to compute   
ghost dependent counterterms.  
The present discussion follows essentially the lines of the proof of absence of  
anomalies for the Faddeev-Popov equation and AAE in  \cite{henri} and   
in the paper by T.Clark \cite{cla}.   
  
By using the QAP the Faddeev-Popov equation and the AAE  
at higher order are given by  
\bea\label{fp_sy}  
 \bar{\cal G}^{a} \g = \bar{\Delta}^{Cl, \bc}_{a} +  
\left[\bar{Q}^{a} \cdot \g \right]^1_2,~~~~   
{\cal G}^{a_\smalla} \g =  
{\Delta}^{Cl, c}_{a_\smalla} + \left[ {Q}^{a_\smalla} \cdot \g \right]^3_4   
\eea  
where $\bar{\Delta}^{Cl, \bc}_{a}$  
is the left hand side of Eq.~~(\ref{fp}) and ${\Delta}^{Cl, c}_{a_\smalla}$   
is given in Eq.~(\ref{aa}).   
Recursively, by assuming that lower order breaking terms  
$\bar{Q}^{a}, {Q}^{a_\smalla}$  
are compensated by means of suitable 
counterterms up to order $\hbar^{n-1}$ the 
r.h.s. of Eq.~(\ref{fp_sy}) become 
\bea\label{fp_sy_q_0}    
\left[\bar{Q}^{a} \cdot \g \right]^1_2 = \hbar^n \bar{Q}^{a} + O(\hbar^{n+1} \bar{Q}),~~~~  
\left[{Q}^{a_\smalla} \cdot \g \right]^3_4 = \hbar^n {Q}^{a_\smalla} + O(\hbar^{n+1} {Q})   
\eea  
where $\bar{Q}^{a}, {Q}^{a_\smalla}$ are local polynomials in terms of fields and  
their derivatives with Faddeev-Popov charge +1 and -1 respectively.   
By UV and IR power counting, by the Lorentz covariance and by
Faddeev-Popov charges conservation,  the   
possible candidates for $\bar{Q}^{a}, {Q}^{a_\smalla}$ are   
\bea\label{fp_sy_q}  
&& \bar{Q}^{a} = \bar{X}^{a b}_1 c_b + \bar{X}^{a[b c]d}_2 c_b c_c \bc_d +  
\bar{X}^{a i}_3 \Om_i + \bar{X}^{a b_\smalls}_{\mu,4} \Om_{b_\smalls}^{\mu}   
\\ \label{aae_sy_q}  
&& {Q}^{a_\smalla} = X^{a_\smalla b}_1 \bc_b + X^{a_\smalla [b c]d}_2 \bc_b \bc_c c_d +  
X^{a_\smalla i}_3 \gamma_i + X^{a_\smalla b_\smalls}_{\mu,4} \gamma_{b_\smalls}^{\mu} +   
X^{a_\smalla I}_{\mu,5} \bar{\eta}_I + h.c.   
\eea  
where $\bar{X}^{a b}_1, \bar{X}^{a i}_3, \bar{X}^{a b_\smalls}_{\mu,4}$,  
${X}^{a_\smalla b}_1, {X}^{a_\smalla i}_3, 
{X}^{a_\smalla b_\smalls}_{\mu,4}, X^{a_\smalla I}_{\mu,5}$   
are polynomials of quantum fields while the constant coefficients   
$\bar{X}^{a[b c]d}$ are totally antisymmetric tensors of the Lie algebra
of ${\cal G}$.  
This follows from the consistency conditions derived by the commutation
relations  (\ref{algebra_2}):  
\bea\label{con_co}  
\bar{\cal G}^{a}_{(x)} \bar{\cal G}^{b}_{(y)} +  
\bar{\cal G}^{b}_{(y)} \bar{\cal G}^{a}_{(x)}  = 0, ~~~~~~  
{\cal G}^{a_\smalla}_{(x)}  {\cal G}^{b_\smalla}_{(y)}  +  
{\cal G}^{b_\smalla}_{(y)}  {\cal G}^{a_\smalla}_{(x)} = 0, ~~~~~~  
{\cal G}^{a_\smalla}_{(x)}  \bar{\cal G}^{b}_{(y)}  +  
\bar{\cal G}^{b}_{(y)}  {\cal G}^{a_\smalla}_{(x)} = 0  
\eea   
which imply relations among the coefficients of $\bar{Q}^{a},{Q}^{a_\smalla}$.  
  
As is well known (see for further details the papers \cite{henri} and  
\cite{libro}) the breaking terms (\ref{fp_sy_q})-(\ref{aae_sy_q}) can be  
removed by means of counterterms and no anomaly appears for equations  
(\ref{fp_sy}). But, although from an algebraic point  
of view there are local counterterms which cancel the apparent breaking terms,  
we have to be sure that those counterterms do not introduce any IR  
divergences. To this purpose we have to check the structure of the lower  
dimensional terms in the explicit decomposition  (\ref{fp_sy_q})-(\ref{aae_sy_q}),  
that is $\bar{X}^{a b}_1, {X}^{a_\smalla b}_1$ and verify if the corresponding counter  
terms could give IR problems.  
It is easy to see that the only dangerous candidates are  
\beq\label{da_cou}  
{\cal L}^{c.t.}_{\rho_{IR} \leq 3}(x) =  
\bar{c}^{a''}_\smallz K_{a'',b'}^1 c^{b'}_\smalla +  
\bar{c}^{a'}_\smalla K_{a',b''}^2 c^{b''}_\smallz + \bar{c}^{a'}_\smalla K_{a',b'}^3(\phi,\h{\phi})   
c^{b'}_\smalla   
\eeq  
with IR degree $\rho_{IR} \leq 3$. Only $K_{a',b'}^3(\phi,\h{\phi})$ could  
depend on the scalars if they are massless, otherwise the coefficients are  
constant and represent mass terms for the massless ghost fields
$c_\smalla^{a'}, \bar{c}_\smalla^{a'}$.  
However the last term of the ${\cal L}^{c.t.}_{\rho_{IR} \leq 3}$   
is not necessary; in fact  
IR power counting implies that  
\bea  
\label{fp_sy_q_1}  
\bar{Q}^{a}_{\rho_{IR} \leq 1} =  \bar{X}^{a b''}_{1,con} c^{b''}_\smallz,~~~~  
{Q}^{a_\smalla}_{\rho_{IR} \leq 3}  =  X^{a_\smalla b''}_{1,con} \bar{c}^{b''}_\smallz  
\eea  
where $\bar{X}^{a b''}_{1,con}, X^{a_\smalla b''}_{1,con}$ are the field independent  
constant parts of the polynomials $\bar{X}^{a b''}_{1}, X^{a_\smalla b''}_{1}$ and the  
index $b''$ runs only over the indices of massive ghost fields $c^{b''}_\smallz$.  
Furthermore the consistency conditions (\ref{con_co}) impose the  
constraint $\bar{X}^{a a_\smalla}_{1,con} + X^{a_\smalla a}_{1,con} = 0$,  
reducing  the only free coefficients  to $\bar{X}^{a b''}_{1,con}$.   
In the case $\bar{X}^{a b''}_{1,con}\neq 0$ the corresponding  
counterterms (\ref{da_cou}) cannot be introduced in the tree level  
action. However another solution can be found. We can use  
the matrix ${G}^{ab}$ introduced above in order to remove the anomaly terms  
(\ref{fp_sy_q_1}), or equivalently, to fix the normalization  
conditions  
\beq\label{no_con}  
\g_{\bar{c}^{a'}_\smalla c^{b''}_\smallz}(p^2 = 0) = 0, \hspace{.5cm}  
\g_{\bar{c}^{a''}_\smallz c^{b'}_\smalla }(p^2 = 0) = 0, \hspace{.5cm}  
\g_{\bar{c}^{a'}_\smalla c^{b'}_\smalla}(p^2 = 0) = 0  
\eeq  
ensuring the correct normalization properties of massless ghost fields.  
As is well known the IR problems arise when radiative corrections  
mix massive and massless fields. Therefore the anomalies in the  
functional equations for the ghost fields can be removed by rotating  
the anti-ghost fields. Finally we would like to stress that the  
coefficients $\bar{X}^{a b''}_{1,con}$ computed within BPHZL  
scheme, as a consequence, are zero and the normalization conditions  
(\ref{no_con}) are automatically satisfied. On the other side  
the choice of other normalization conditions for physical fields  
(as for Standard Model with on-shell normalization conditions)  
might spoil the (\ref{no_con}) and spurious anomalies as  
(\ref{fp_sy_q})-(\ref{aae_sy_q}) might appear.

%
%
\def\ap#1#2#3{{\it Ann. Phys.} (NY) #1 (19#3) #2}  
\def\jmp#1#2#3{{\it  J. Math. Phys.} #1 (19#3) #2}  
\def\np#1#2#3{{\it Nucl. Phys.} {\bf B#1} (19#3) #2}  
\def\pl#1#2#3{{\it Phys. Lett.} {\bf #1B} (19#3) #2}  
\def\pr#1#2#3{{\it Phys. Rev.} {\bf D#1} (19#3) #2}  
\def\prep#1#2#3{{\it Phys. Rep.} #1 (19#3) #2}  
\def\prl#1#2#3{{\it Phys. Rev. Lett.} #1 (19#3) #2}  
\def\rmp#1#2#3{{\it Rev. Mod. Phys.} #1 (19#3) #2}  
\def\sj#1#2#3{{\it Sov. J. Nucl. Phys.} #1 (19#3) #2}  
\def\zp#1#2#3{{\it Zeit. Phys.} {\bf C#1} (19#3) #2}  
\def\cmp#1#2#3{{\it Comm. Math. Phys.} {\bf #1} (19#3) #2}  
\def\nc#1#2#3{{\it Il Nuovo Cimento} #1A (19#3) #2}

\vfill  
\eject  
\end{document}